\documentclass[11pt,a4paper,english]{article}

\usepackage{epsfig}
\usepackage{amsfonts}
\usepackage{amssymb,amsmath}
\usepackage[ps,matrix,arrow]{xy}
\usepackage{eucal}

\makeatletter

\catcode`@=11
\renewcommand{\section}
{\@startsection{section}{1}{0pt}{\medskipamount}{\medskipamount}{\large\bf}}
\makeatletter\renewcommand{\subsection}
{\@startsection{subsection}{2}{\z@}{-3.25ex plus -1ex minus -.2ex}
{1.5ex plus .2ex}{\it }}

\numberwithin{equation}{section}
\catcode`@=12

\newcommand{\Tr}[1]{\:{\rm Tr}\,#1}

\newcommand{\mbf}[1]{{\boldsymbol {#1} }}
\newcommand{\complex}{{\mathbb C}} 
\newcommand{\zed}{{\mathbb Z}} 
\newcommand{\nat}{{\mathbb N}} 
\newcommand{\real}{{\mathbb R}} 
\newcommand{\torus}{{\mathbb T}}
\newcommand{\sphere}{{\mathbb S}}

\def\e{{\,\rm e}\,}

\newcommand{\alg}{{\cal A}}

\newcommand{\Hcal}{{\cal H}}
\newcommand{\Ecal}{{\cal E}}
\newcommand{\ch}{{\rm ch}}

\newcommand{\K}{{\rm K}}
\def\ii{{\,{\rm i}\,}}
\def\dd{{\rm d}}

\newcommand{\Hom}{\mathrm{Hom}}

\newcommand{\End}{\mathrm{End}}
\newcommand{\sign}{\mathrm{sgn}}

\def\beq{\begin{equation}}
\def\bee{\begin{equation}}
\def\eeq{\end{equation}}
\def\bea{\begin{eqnarray}}
\def\eea{\end{eqnarray}}
\def\bd{\begin{displaymath}}
\def\ed{\end{displaymath}}

\newcommand{\Cint}{\int\kern-10.5pt-\kern7pt}

\newcommand{\PP}{{\mathbb{P}}}

\newcommand{\be}{\begin{equation}}
\newcommand{\ee}{\end{equation}}

\newcommand\fverb{\setbox\pippobox=\hbox\bgroup\verb}
\newcommand\fverbdo{\egroup\medskip\noindent%
                        \fbox{\unhbox\pippobox}\ }
\newcommand\fverbit{\egroup\item[\fbox{\unhbox\pippobox}]}
\newbox\pippobox

{

\def\pa{\partial}

\def\eps{\epsilon}

\def\th{\theta}

\def\w{\wedge}

\def\be{\begin{equation}}
\def\ee{\end{equation}}
\def\bea{\begin{eqnarray}}
\def\eea{\end{eqnarray}}
\def\nn{\nonumber\\}

\begin{document}

\begin{titlepage}
\setcounter{page}{0}
\begin{flushright}
ITP--UU--08/14 , SPIN--08--13\\
DAMTP--2008--24\\
HWM--08--1 , EMPG--08--01\\
NI08016--SIS , ESI--2019
\end{flushright}

\vskip 1.8cm

\begin{center}

{\Large\bf Cohomological gauge theory, quiver matrix models \\[5pt] and
   Donaldson--Thomas theory}

\vspace{15mm}

{\large\bf Michele Cirafici$^{(a)}$, Annamaria Sinkovics$^{(b)}$ and
 Richard~J.~Szabo$^{(c)}$}
\\[6mm]
\noindent{\em $^{(a)}$ Institute for Theoretical Physics and Spinoza
  Institute \\
Utrecht University, 3508 TD Utrecht, The Netherlands \\Email: {\tt
  M.Cirafici@uu.nl}}\\[3mm]
\noindent{\em $^{(b)}$ Department of Applied Mathematics and
  Theoretical Physics
\\ Centre for Mathematical Sciences, University of
Cambridge \\ Wilberforce Road, Cambridge CB3 0WA, UK\\ Email: {\tt
  A.Sinkovics@damtp.cam.ac.uk}}\\[3mm]
\noindent{\em $^{(c)}$ Department of Mathematics, Heriot--Watt
  University and\\ 
Maxwell Institute for Mathematical Sciences\\
Colin Maclaurin Building, Riccarton, Edinburgh EH14 4AS, UK\\ Email: {\tt
  R.J.Szabo@ma.hw.ac.uk}}

\vspace{15mm}

\begin{abstract}
\noindent

We study the relation between Donaldson--Thomas theory of Calabi--Yau
threefolds and a six-dimensional topological Yang--Mills theory.
Our main example is the topological $U(N)$ gauge theory on flat
space in its Coulomb branch. To evaluate its partition function we
use equivariant localization techniques on its noncommutative
deformation. As a result the gauge theory localizes on noncommutative
instantons which can be classified in terms of $N$-coloured
three-dimensional Young diagrams. We give to these noncommutative
instantons a geometrical description in terms of certain stable framed
coherent sheaves on projective space by using a higher-dimensional
generalization of the ADHM formalism. {}From this formalism we construct
a topological matrix quantum mechanics which computes an index of BPS
states and provides an alternative approach to the six-dimensional
gauge theory.

\end{abstract}




\end{center}
\end{titlepage}

{\baselineskip=12pt
\tableofcontents
}

\newpage

\allowdisplaybreaks

\section{Introduction}

Topological field and string theories that arise from a topological
twist of a physical model capture the BPS sector of their physical
 counterparts. They have been used successfully in the last decades 
to compute certain classes of nonperturbative effects, thereby
considerably improving our understanding of the quantum world.
Topological theories also have a deep relation with mathematics as
they relate BPS quantities to geometrical invariants of the
underlying manifold. For example, topological string theory is
equivalent to certain enumerative problems for Calabi--Yau
threefolds and it computes invariants such as the Gopakumar--Vafa,
Gromov--Witten and Donaldson--Thomas invariants. These invariants 
play a key role in the mirror symmetry conjecture
which is by now one of the best understood examples of duality in string
theory (see \cite{Hori:2003ic} for a review). On the other hand, these
invariants enter directly in the 
computation of the entropy of a class of supersymmetric black
holes that arises in string theory compactifications on a
Calabi--Yau manifold via the Ooguri--Strominger--Vafa (OSV) conjecture
and provide some computational control on the enumeration of quantum
gravity microstates~\cite{Ooguri:2004zv}.

Donaldson--Thomas invariants count bound states of D0--D2 branes
with a single D6-brane wrapping a Calabi--Yau threefold. The geometrical
object that describes this configuration is called an ideal sheaf and
the Donaldson--Thomas invariants can be roughly thought of as
``volumes'' of the moduli spaces of ideal sheaves. When the
Calabi--Yau space is toric the topological string theory has a
reformulation in terms of the classical statistical mechanics of a
melting crystal~\cite{Okounkov:2003sp}. In this setting the
Donaldson--Thomas invariants enumerate atomic configurations in the
melting process. Just like the Gopakumar--Vafa invariants,
Donaldson--Thomas theory computes certain F-terms in the low energy
effective action of the string theory
compactification~\cite{Bershadsky:1993cx}--\cite{Gopakumar:1998jq}. Indeed,
it is conjectured that the  Donaldson--Thomas invariants are
equivalent to the Gopakumar--Vafa and Gromov--Witten invariants as
they arise from different expansions of the same topological string
amplitude. This conjecture is known to be true in the toric
setting~\cite{MNOP}. The melting crystal picture also has an
interpretation as a sum over fluctuating K{\"a}hler
geometries~\cite{Iqbal:2003ds}. This sum can be explicitly rewritten
as a path integral of an auxiliary six-dimensional topological gauge
theory. This theory localizes on the moduli space of solutions of
the Donaldson--Uhlenbeck--Yau equations which can be
thought of as generalized higher-dimensional instantons. The
proposal of~\cite{Iqbal:2003ds} consists in identifying the
instanton counting problem associated to this gauge theory with
the Donaldson--Thomas enumeration of ideal sheaves.

Accordingly, Donaldson--Thomas theory is reduced to six-dimensional
instanton calculus. To this end one can adopt the approach
envisaged by Nekrasov in the four-dimensional case and apply the
techniques of equivariant localization in topological field
theories~\cite{sw}. Roughly, Nekrasov's formalism consists of a
noncommutative deformation of the gauge theory that resolves
certain singularities of the instanton moduli space and a
particular redefinition of the BRST charge that localizes the
instanton measure on pointlike configurations. The BRST charge is
now an equivariant differential with respect to the natural toric
action on the $\complex^2$ target space and as a result
integration over the instanton collective coordinates is reduced
to a sum over pointlike instantons via the equivariant
localization formula. These ideas were applied to
the six-dimensional gauge theory under consideration on a generic
toric Calabi--Yau threefold in~\cite{Iqbal:2003ds}, with the main
difference being the appearance of degenerate gauge field
configurations which wrap rational curves. The gauge theory partition
function can in this way be successfully matched with the melting
crystal partition function.

The full picture is thus consistent, and gives strong support to the
equivalence between the six-dimensional gauge theory on the D6-brane and
Donaldson--Thomas theory. Nevertheless, a direct link is still
lacking. The equivalence becomes clear only after the gauge theory
partition function is matched with the topological vertex
amplitude. It would be desirable to have a direct understanding of
Donaldson--Thomas theory in the gauge theoretic picture and in 
particular to investigate the ideal sheaf counting problem in terms of
the gauge theory variables.
Moreover, the fact that we are now dealing with a gauge theory
poses two pressing questions. Firstly, it is natural to consider a
higher rank generalization of the problem which corresponds to
introducing an arbitrary number $N$ of D6-branes in the picture. While
this is conceptually simple the relation (if any) with the usual
topological string theory is rather obscure. We expect that a
resolution of this puzzle may shed some new light on the black
hole microstate counting problem. Secondly, topological string
amplitudes have modular transformation properties that allows one to
carry the information about the enumerative invariants along the
various regions of the Calabi--Yau moduli space. In particular,
knowledge of the Gromov--Witten invariants at the large radius point
is enough to compute them, say, at orbifold
points~\cite{Aganagic:2006wq}. One may wonder how this property is
seen in Donaldson--Thomas theory given that the gauge theory is
naturally defined at the large radius point.

The aim of this paper is to investigate in detail the
aforementioned proposals and conjectures, and to make a first step
towards the resolution of the outlined problems. In particular, we
will explore the random partition combinatorics of the nonabelian
theory and its formulation in terms of a topological vertex. Partial
results in this direction were described in~\cite{Sduality}, as well
as in~\cite{jafferis}. We work mainly on
flat space in the Coulomb branch of a $U(N)$ gauge theory. In this
case we are able to set up a general formalism that we hope could
be applied to more general settings. We attack the problem in two
independent but intimately connected ways. As a start we define
and evaluate the partition function of the noncommutative
deformation of the gauge theory. The noncommutative gauge theory
localizes onto a sum over critical points that are classified in
terms of $N$-coloured three-dimensional Young diagrams. The
fluctuation factor around each critical point has the form of a
ratio of functional determinants that can be computed by direct
evaluation. The result is a simple generalization of the MacMahon
function which captures the counting of BPS states for a configuration
of D6-branes widely separated in the transverse directions.

We give a purely geometric interpretation of these noncommutative
instantons. Under a certain set of plausible assumptions we can relate
the noncommutative gauge field configurations with the stable framed
moduli space of certain coherent sheaves on $\PP^3$, a
compactification of the target space $\complex^3$. The main technical
tool is the use of Beilinson's spectral sequence to parametrize a
coherent sheaf on $\PP^3$ with a set of matrix equations. We show that
a special class of such sheaves arise as the cohomology of certain
nonlinear monads on $\PP^3$. The matrix equations are naturally
interpreted as a higher-dimensional generalization of the ADHM
formalism. This construction provides a direct relationship between
the gauge theory and Donaldson--Thomas theory, and should be compared
to the recent construction by Diaconescu~\cite{Diaconescu} for the
local Donaldson--Thomas theory of curves in terms of ADHM quiver
sheaves.

{}From this generalized ADHM language we construct a topological matrix
quantum mechanics that dynamically describes the stable
coherent sheaves. In the string theory language this model
corresponds to the effective action on the gas of D0-branes that
are bound to the D6-branes (in the presence of a suitable
$B$-field)~\cite{CIMM}--\cite{Witten:2000mf}. Equivalently, we
may think of the matrix model as
arising from the quantization of the collective coordinates around
each instanton solution. In either way the matrix model recovers
the classification of the critical instanton configurations in
terms of $N$-coloured three-dimensional Young diagrams via the
special properties of the linear maps in the generalized ADHM
equations. The computation of the instanton fluctuation factors is
reduced to the evaluation of an appropriate equivariant index
associated with the topological matrix model. The results agree
with the direct computation done in the original noncommutative
gauge theory.

As an application of our formalism we compute the partition function
of the rank $N$ gauge theory in the Coulomb branch on a generic toric
Calabi--Yau manifold. The result is the $N$-th power of the
abelian partition function with an $N$ dependent sign shift. This
shift is consistent with the expected shift in the topological
string coupling constant when the amplitude is related to the D-brane
charges at the attractor point of the BPS moduli space. In particular,
our results should provide a testing ground for the OSV conjecture
with many D6-branes. Although the gauge theory provides essentially
the same Donaldson--Thomas invariants, the Gromov--Witten theory is
new and involves an additional expansion parameter, the rank $N$ of
the gauge group.

It is instructive to compare at the outset with the analogous problem
in four dimensions. Since the six-dimensional gauge theory we consider
is maximally supersymmetric, it is tempting to believe that its
dynamics are qualitatively similar to the well-studied $\mathcal{N}=4$
supersymmetric $U(N)$ Yang-Mills theory in four dimensions which is a
topological twist of the $\mathcal{N}=2$ Seiberg--Witten theory. In
that case the partition function is independent of the vevs of the
scalar fields, and hence of the gauge symmetry breaking pattern,
because it always computes the Euler characteristic of the
(compactified) instanton moduli space. More precisely, for the
$\mathcal{N}=2$ Seiberg--Witten theory, which computes the low energy
effective action, one can use holomorphy properties to argue that the
prepotential is a generic solution~\cite{sw}, even when some of the
Higgs field eigenvalues coincide. By using the equivalence between
Donaldson--Witten theory and Seiberg--Witten
theory~\cite{Moore:1997pc}, this shows (modulo subtleties associated
to wall-crossing phenomena) that it is enough to localize the gauge
theory onto the Cartan subgroup, and indeed a direct computation of
Donaldson--Witten invariants for $SU(N)$ gauge group gives no newer
information than the $U(1)^{N-1}$ Seiberg--Witten theory computation.

However, there is no reason to expect that the infrared dynamics of
the six-dimensional gauge theory is similar. Moreover, at present it
is not known what is the rigorous stability condition to place on
higher-rank sheaves over three-dimensional Calabi--Yau manifolds in
order to construct a well-behaved compactification of the moduli
space. Due to our lack of understanding of this moduli space, there is
no guarantee that the instanton configurations that we localize onto
span the appropriate moduli space of nonabelian gauge field
configurations. For the gauge theory on $\complex^3$ we have good
control over the instanton moduli space and an explicit construction
(with the caveat that we have not rigorously defined the appropriate
compactification of the instanton moduli space). We argue in the
following that in this case the allowed sheaves are those provided by
the noncommutative deformation, and give an explicit algebro-geometric
description including appropriate stability conditions. For more
general toric threefolds, our computations merely provide a
localization of the full nonabelian Donaldson--Thomas theory onto
those sheaves which are invariant under the equivariant action of the
maximal torus of the gauge group. For more general symmetry breaking,
the sheaves should carry a nontrivial framing (as proposed
in~\cite{Diaconescu}) and the fixed point locus need not consist of
isolated points. Usage of the localization formulas in this case would
require the much more complicated integration over degenerate fixed
point submanifolds of moduli space weighted by the Euler class.

This paper is organized as follows. In Section 2 we
review the geometrical setting of the six-dimensional gauge theory
and its equivariant deformation. Section 3 is devoted to the
noncommutative deformation of the gauge theory and the evaluation of
the partition function on flat (noncommutative) space. Section 4 is
devoted to geometrical aspects and contains the derivation of
our ADHM-like formalism. This formalism is used in Section 5 to
construct a topological matrix quantum mechanics that computes an
index of BPS states. In Section 6 we compute the partition
function of the rank $N$ gauge theory in the Coulomb branch on a
generic toric Calabi--Yau manifold. Section 7 contains some further
discussion and applications of our results.

\bigskip

\section{Topological $\mbf{\mathcal{N}_T = 2}$ Yang--Mills theory in
  six dimensions}

In this section we will collect some known results concerning the
six-dimensional gauge theory which captures the physics of the
K{\"a}hler quantum foam. This theory has been studied in the
literature both as a cohomological field theory~\cite{Acharya:1997gp}
and in the balanced formalism~\cite{Dijkgraaf:1996tz}, i.e. with
$\mathcal{N}_T=2$ \emph{topological} charges.
In order to keep the discussion completely general we will
consider the $U(N)$ gauge theory defined on a toric threefold, but
for concrete applications we will often specify to an abelian
$U(1)$ or $U(1)^N$ theory on $\complex^3$.
Some of the results reviewed here are also valid in the more
general context of a generic K{\"a}hler manifold.

\subsection{The cohomological gauge theory}

The gauge theory in question can be defined through a topological
twist of the maximally supersymmetric Yang--Mills theory in six
dimensions. One starts with $\mathcal{N}=1$ $U(N)$ supersymmetric
Yang--Mills theory in ten dimensions and dimensionally reduces on
a six-dimensional K{\"a}hler manifold $X$ with $U(3)$ holonomy.
After the twist the bosonic spectrum of the theory consists of a
gauge field $A_{i}$, a complex Higgs field $\Phi$ and $(3,0)$-form
$\rho=\rho^{3,0}$ along with their complex conjugates, all taking
values in the adjoint representation of the $U(N)$ gauge group. In the
fermionic sector we can twist the spin bundle with the canonical
line bundle over $X$ and obtain explicitly an isomorphism between
fermions and differential forms~\cite{Blau:1997pp}. The spectrum
consists of a complex scalar $\eta$, one-forms $\psi^{1,0}$ and
$\psi^{0,1}$, two-forms $\psi^{2,0}$ and $\psi^{0,2}$, and finally
three-forms $\psi^{3,0}$ and $\psi^{0,3}$ for a total of $16$
degrees of freedom matching those of the bosonic sector. The
resulting supersymmetric gauge theory is cohomological and has two
topological charges. It can therefore be studied as a balanced
topological field theory as in~\cite{Hofman:2000yx}. The bosonic
part of the action has the form
\begin{eqnarray}
S  &=& \frac{1}{2}\, \int_X\, \Tr \left( \dd_A \Phi \wedge * \dd_A
\overline{\Phi} + \big[\Phi \,,\, \overline{\Phi}~\big]^2 +
\big|F_A^{2,0} + \overline{\pa}\,_A^{\dagger} \rho\big|^2 +
\big|F_A^{1,1}\big|^2  \right) \nonumber\\  &&+\,
\frac{1}{2}\, \int_X \,\Tr \Big(   F_A \w F_A \w k_0 +
\mbox{$\frac\vartheta3$}\, F_A
\w F_A \w F_A \Big) \ ,
\label{bosactionTGT}\end{eqnarray}
where $\dd_A=\dd+\ii[A,-]$ is the gauge-covariant derivative, $*$ is
the Hodge operator with respect to the K\"ahler metric of $X$,
$F_A=\dd A+A\wedge A$ is the gauge field strength, $k_0$ is the
background K{\"a}hler two-form of $X$, and $\vartheta$ is the
six-dimensional theta-angle which will be identified later on with the
topological string coupling~$g_s$.

In~\cite{Iqbal:2003ds} the $U(1)$ gauge theory has been given a
suggestive interpretation as a K{\"a}hler quantum foam. One starts
with a path integral which represents a sum over K{\"a}hler
geometries with quantized K{\"a}hler class $k = k_0 + g_s\, F_A$.
Expanding the K{\"a}hler gravity action $\int_X\, k \wedge k
\wedge k$ and gauge fixing the residual symmetry gives precisely
the $\mathcal{N}_T = 2$ topological gauge theory. In this
interpretation it is essential that the gauge field curvature
$F_A$ is a representative of the Chern class of a complex line
bundle over $X$. For higher rank gauge groups $U(N)$, $N>1$ a
characterization of the gauge theory as a gravitational theory is not 
known.

The gauge theory has a BRST symmetry and hence localizes 
onto the moduli space of solutions of the fixed point
equations
\begin{eqnarray} \label{inste} F_A^{2,0} &=&
  \overline{\pa}\,_A^{\dagger} \rho
  \ , \nonumber\\[4pt]
F_A^{1,1} \w k_0 \w k_0 + \big[\rho\,,\, \overline{\rho}\,\big] &=&
l~k_0 \w k_0 \w k_0 \ , \nonumber\\[4pt] \dd_A \Phi &=& 0 \ .
\end{eqnarray}
The right-hand side of the second equation is a quantum correction
coming from the magnetic charge of the gauge bundle, where $l$ is a
constant. The solutions of these equations minimize the action
(\ref{bosactionTGT}) and we will therefore call them generalized
instantons or just instantons. We will be interested in the set of
minima where the field $\rho$ is set to zero. This is always possible
on a Calabi--Yau background, because of the uniqueness of the
holomorphic three-form in that case. Then the first two equations
reduce to the Donaldson--Uhlenbeck--Yau (DUY) equations which are
conditions of stability for holomorphic bundles over~$X$ with finite
characteristic classes.

We will generically denote an appropriate compactification of the
moduli space of solutions to the first two equations of
(\ref{inste}) with $\mathcal{M}$, or with $\mathcal{M}(N, c_1 ,
\mathrm{ch}_2 , \mathrm{ch}_3)$ labelling the solutions by their
rank and Chern classes when we want to be more explicit about
their topology, or by $\mathcal{M}(X)$ when we want to stress the
role of the underlying variety $X$ on which the gauge theory is
defined. We will also include in $\mathcal{M}$ those
configurations that solve (\ref{inste}) where the gauge field is
possibly singular on $X$. More precisely, we will not restrict
attention to holomorphic bundles but in principle consider also
coherent sheaves of $\mathcal{O}_X$-modules over $X$ of rank $N$.
It is not clear that this compactification always exists for a
generic threefold as very little is known about these moduli
spaces, and we may expect the singular loci to be intractable. This
lack of a geometrical understanding is a major obstacle in
carrying out the localization program. We will see in the rest of
this paper how some progress can be made in specific situations.

\subsection{Local geometry of the instanton moduli
  space\label{Localmodsp}}

The moduli space $\mathcal{M}$ is a highly singular and badly
behaved complex variety. When $N=1$ and $c_1 = 0$ this problem has
been cured in~\cite{MNOP,Iqbal:2003ds}, and with an appropriate
compactification there is an isomorphism $\mathcal{M}( 1 , 0,
\mathrm{ch}_2 ,\mathrm{ch}_3) \cong\mathcal{I}_k( X , \beta)$ with
the moduli space of ideal sheaves over $X$ with fixed two-homology
class $\beta$ and holomorphic Euler characteristic $k$ as defined
for example in~\cite{MNOP}. This moduli space is also isomorphic
to the Hilbert scheme of points and curves in $X$,
$\mathsf{Hilb}^k( X , \beta)$.
In the following we will keep the rank $N$ arbitrary as we can
formally consider a nonabelian gauge theory. In principle one
could try to formulate the Donaldson--Thomas theory as a
localization problem for a gauge theory in arbitrary rank. However,
one should face the difficult technical problem of integrating
over the appropriately compactified moduli space. We can hope to
make computational progress in the Coulomb phase of the gauge
theory where the gauge symmetry is completely broken down to the
maximal torus $U(1)^N$
by the Higgs field vacuum expectation values and the moduli space
essentially reduces to $N$ copies of the Hilbert scheme where
localization techniques can and have been successfully applied.
A precise geometric characterization of $\mathcal{M}$ is not known
and lies beyond the scope of this paper. In particular,
integration over $\mathcal{M}$ is ill-defined and requires
appropriate homological tools to be dealt with. For our practical
purposes we will ignore these (important) issues and base our
analysis on gauge theory techniques. It will be enough to know
that for $N=1$ a perfect obstruction theory can be developed and a
virtual fundamental class of $\mathcal{M}$ has been
defined~\cite{Donaldson:1996kp}. In Section~\ref{Matrixsheaves} we
will explore a sheaf theoretical description of the moduli space
$\mathcal{M}$ for $N\geq1$ in the case $X=\complex^3$.

{}From the gauge theory perspective we are dealing with holomorphic
bundles and pairs $(A, \rho)$. The first step in understanding the
moduli space $\mathcal{M}$ is to characterize its local geometry. For
this, we introduce the instanton deformation
complex~\cite{Iqbal:2003ds}
\begin{equation} \label{defcomplex}
\xymatrix@1{0 \ar[r] & \Omega^{0,0} ( X , \mathrm{ad}\, \frak g)
\ar[r]^{\hspace{-1.3cm} C} & ~\Omega^{0,1} ( X , \mathrm{ad}\, \frak
g) \oplus \Omega^{0,3} ( X , \mathrm{ad}\, \frak g)
\ar[r]^{\hspace{1.3cm} D_A} & \Omega^{0,2} ( X , \mathrm{ad}\, \frak
g) \ar[r] & 0 } \ ,
\end{equation}
where $\Omega^{\bullet,\bullet} (X , \mathrm{ad}\, \frak g)$ denotes
the bicomplex of complex differential forms taking values in the
adjoint gauge bundle over $X$, and the maps $C$ and $D_A$
represent a linearized complexified gauge transformation and the
linearization of the first equation in (\ref{inste}) respectively.
A precise definition can be found
in~\cite{Iqbal:2003ds,Baulieu:1997jx}. This complex is elliptic and
its first cohomology represents the holomorphic tangent space to
$\mathcal{M}$ at a point $(A,\rho)$, $T_{(A,\rho)} \mathcal{M}$. The
degree zero cohomology represents reducible pairs $(A,\rho)$, whereby
the gauge field $A$ is a reducible connection which we assume
vanishes.

On the other hand, in general there is also a finite-dimensional
second cohomology that measures obstructions and is customarily called
the obstruction or normal bundle $\mathcal{N}$. It is associated with
the kernel of the conjugate operator $D_A^\dag$. Since this is the
operator that enters in the kinetic term for the antighost fields, the
terminology ``antighost bundle'' is used for $\mathcal{N}$ in the
physics literature as its fibres are spanned by the antighost zero
modes. It is precisely this bundle which provides an ``integration
measure'' on the moduli space. On general grounds the fermionic BRST
symmetry localizes the partition function of the topological gauge
theory onto the moduli space $\mathcal{M}$ (once a topological sector
is chosen by fixing a critical point). There are nevertheless
remaining fermionic terms to be integrated over. In particular, there
is a four-fermion interaction schematically of the form $\mathcal{R}\,
\psi\, \psi\,\psi\, \psi$ where $\psi$ collectively denotes the Fermi
fields. A careful analysis of the BRST transformations and of the
on-shell action shows that this term produces an integral
representative of the Euler characteristic of the antighost bundle
over the moduli space~\cite{Iqbal:2003ds,Hofman:2000yx}. Equivalently,
matching the Fermi zero modes with the fermionic path integral measure
brings down the pfaffian of the curvature $\mathcal{R}$.

As the detailed analysis is quite involved, we will just denote this
integral symbolically as
\begin{equation} \label{eulerobstr}
\int_{\mathcal{M}}\, \textsf{e} (\mathcal{N}) \ .
\end{equation}
It is difficult to give a precise definition of this integral
and to evaluate it. For this one needs rather sophisticated tools such
as a perfect obstruction theory and to deal with the relevant virtual
fundamental class of $\mathcal{M}$. Throughout this paper we will \emph{assume} the existence of a perfect obstruction theory and a corresponding virtual cycle. Then the localization formulas of Section~\ref{Eqmodel} below will be used to \emph{define} the integrals~(\ref{eulerobstr}). A special
case where we can make extensive computational progress in the evaluation of
(\ref{eulerobstr}) is when the ambient variety is $X = \complex^3$. Since $\beta$ is necessarily trivial in this case
(equivalently $\mathrm{ch}_2=0$), the relevant moduli space for $N=1$
can be identified with the Hilbert scheme
$\mathsf{Hilb}^k(\complex^3)=(\complex^3)^{[k]}$ consisting of
zero-dimensional subschemes in $\complex^3$. For $k>3$ this moduli
space generically contains branches of varying dimension and so is not
even a manifold.

\subsection{The equivariant model\label{Eqmodel}}

If the moduli space $\mathcal{M}$ were smooth and compact, then we
could proceed to evaluate the integrals (\ref{eulerobstr}) by
choosing an appropriate representative of the cohomology class
$\textsf{e} (\mathcal{N})$. However, this is not the case as
moduli spaces of instantons suffer from non-compactness problems
arising both from singularities where instantons shrink to zero
size as well from the non-compactness of the ambient space $X$ on
which the gauge theory is defined. In field theoretical terms, we
can think of divergences coming from the first problem as
associated with small distances while the second problem reflects
the need for an infrared regularization. The ultraviolet behaviour
improves substantially by introducing a noncommutative deformation
of the gauge theory that provides a natural compactification of the
instanton moduli space.
We will define and study the deformed instanton calculus in the
ensuing sections, but for the time being we will assume that this
problem has been solved.

A direct evaluation of the integrals (\ref{eulerobstr}) is still a
formidable problem. A particularly powerful approach is to use
equivariant localization. In our problem we will assume that the gauge
theory is defined on a toric threefold $X$. It carries the action
of a non-compact three-torus $\torus^3$, and we further assume that
this action lifts to the moduli space $\mathcal{M}$. Working
$\torus^3$-equivariantly means that we restrict our attention to
critical gauge field configurations that are $\torus^3$-invariant. A
practical way to implement this scheme is to modify the gauge theory
so that the BRST differential on the space of fields becomes an
equivariant differential, so that an infinitesimal $\torus^3$
rotation can be undone by a gauge transformation.

When $X=\complex^3$, a physical realization of the equivariant
modification with respect to the toric isometry is to put the gauge
theory on the ``$\Omega$-background''. For the present gauge theory
this procedure was developed in~\cite{Iqbal:2003ds}. If we think of
our cohomological gauge theory as arising from a topological twist of
maximally supersymmetric Yang--Mills theory in six dimensions, then
working equivariantly is equivalent to replacing the original scalar
BRST operator $Q$ used for the twist with a linear combination of the
scalar and vector supercharges $Q^{i}$ given by
\begin{equation}
Q_\Omega= Q + \epsilon_a\, \Omega^a_{ij}\, x^{i} \,
Q^{j} \ .
\end{equation}
Here $\epsilon_a$, $a=1,2,3$ are formal parameters of the $\torus^3$
action and $\Omega^a =\Omega^a_{ij}\, x^{j}~\frac{\partial}{\partial
  x_i}$ are vector fields which generate the $SO(6)$ rotational
isometries of $\real^6\cong\complex^3$. We will restrict to the $U(3)$
holonomy subgroup that preserves the natural,
translationally-invariant K{\"a}hler two-form of $\complex^3$. Then
the toric symmetry group $\torus^3$ is the maximal torus of this
$U(3)$. The result is a topological deformation of the gauge theory,
augmenting the action by a $Q_\Omega$-exact term, which depends
explicitly on the vector fields $\Omega^a$. Since the BRST charge
$Q_\Omega$ is a linear combination of supercharges, perturbative
bosonic and fermionic contributions cancel, and the partition function
is saturated by instantons. The key point here is that, in the Coulomb
phase, the $\Omega$-deformation localizes the instanton measure onto
point-like instanton configurations which are invariant under
$\torus^3$ rotations. The critical points of the deformed gauge theory
action are thus isolated. Due to localization, the semiclassical
approximation is exact and the full path integral reduces to a sum
over contributions from isolated point-like instantons.

This deformation turns the evaluation of the integrals
(\ref{eulerobstr}) into a problem that is both well-defined and
computationally accessible by equivariant localization techniques.
In addition, the $\Omega$-background acts as an infrared
regularization and the equivariant volumes of the instanton moduli
spaces are finite. In the gauge theory action (\ref{bosactionTGT}) the
$\Omega$-background modifies, among other things, the kinetic term for
the Higgs field to
\begin{equation}
\Tr \big( \dd_A
\Phi - \imath_{\Omega} F_A \big) \wedge * \big( \dd_A \overline{\Phi}-
\imath_{\overline{\Omega}} F_A\big)
\end{equation}
where $\imath_{\Omega}$ is contraction with the vector field that
generates the toric isometries. This modifies the third equation of
(\ref{inste}) to
\begin{equation}
\dd_A \Phi = \imath_{\Omega} F_A \ ,
\end{equation}
which manifestly minimizes the action on the $\Omega$-background.
The Higgs potential and the fermionic terms are also affected by
the $\Omega$-background, but we will not need their precise forms.

The discussion of Section~\ref{Localmodsp} above can be carried
out in an equivariant setting with minor modifications and the
suitable version of the integral (\ref{eulerobstr}) can be now
evaluated with equivariant localization techniques. This will be done
explicitly in the following sections for the Coulomb phase of the
gauge theory. For the moment let us just
sketch the idea behind the procedure, glossing over many details. In
the equivariant model on $X= \complex^3$ with the natural $\torus^3$
action, the Euler class ${\sf e}(\mathcal{N})$ is an element of the
$U(N)\times\torus^3$ equivariant cohomology of the moduli space
$\mathcal{M}$, where the gauge group $U(N)$ acts by rotating the
trivialization of the instanton gauge bundle at infinity in
$\complex^3$. Its equivariant integral $\oint_{\mathcal{M}}\,{\sf
  e}(\mathcal{N})$ is the localization of the pushforward of ${\sf
  e}(\mathcal{N})\in H^\bullet_{U(N)\times\torus^3}(\mathcal{M})$ in
$H^\bullet_{U(N)\times\torus^3}({\rm
  pt})=\complex\big[{\rm Sym}^N(\complex);\epsilon_a\big]$ under the
collapsing map $\mathcal{M}\to{\rm pt}$ onto a point, where the
symmetric product ${\rm Sym}^N(\complex)=\complex^N/S_N$ parametrizes
the complex conjugacy classes of $U(N)$. These classes can be labelled
by elements $\mbf a=(a_1,\dots,a_N)$ of the Cartan subalgebra
$\mathfrak{u}(1)^{\oplus N}\cong\real^N$. Denoting $\xi=(\mbf a,\epsilon_a)$,
there is a moment map
\be
\mu[\xi]=\mu_{U(N)}[\mbf
a]+\mu_{\torus^3}[\epsilon_a]\,:\,\mathcal{M}~\longrightarrow~
\big(\mathfrak{u}(1)^{\oplus N}\big)^*\oplus\mathsf{Lie}\big(\torus^3
\big)^*
\label{momentmap}\ee
and an invariant symplectic two-form $\omega$ on $\mathcal{M}$ such
that the $U(N)\times\torus^3$ action on $\mathcal{M}$ is hamiltonian,
\be
\mathrm{d} \mu[\xi]= - \imath_{V_\xi} \omega \ ,
\label{hamactionmod}\ee
where $V_\xi$ is the vector field on $\mathcal{M}$ representing the
toric action generated by $\xi$.

Using (\ref{hamactionmod}) we can represent ${\sf e}(\mathcal{N})$ by
a cohomologically equivalent equivariant differential form, and hence
replace $\oint_{\mathcal{M}}\,{\sf e}(\mathcal{N})$ by an ordinary
integral over the moduli space, $\int_{\mathcal{M}}\,{\sf
  e}(\mathcal{N})\wedge\exp(\omega+\mu[\xi])$. We may then appeal to
the Duistermaat--Heckman localization formula
\begin{equation}
\int_{\mathcal{M}}\, \frac{\omega^n}{n!} ~\e^{-\mu[\xi]} =
\sum_{\stackrel{\scriptstyle f\in\mathcal{M}}{\scriptstyle V_\xi(f)=0}}~
\frac{\e^{-\mu[\xi](f)}}{\prod\limits_{i=1}^n \,w_i[\xi](f)} \ ,
\label{DHformula}\end{equation}
where $n=\dim_{\complex}\mathcal{M}$ and the sum on the right-hand
side runs over the fixed points of the toric action. The parameters
$w_i [\xi]$ are the weights of the toric action on the tangent
space to the critical points on $\mathcal{M}$. The
Duistermaat--Heckman theorem assumes that the toric action on the
ambient space $X$ lifts to the moduli space and that its critical
points are isolated. This is precisely what happens in the case where
the gauge theory is defined on the $\Omega$-background. One just needs
to evaluate the integrand at each critical point
and sum over all critical points with the appropriate weights.
Thus in practice the evaluation of (\ref{eulerobstr}) in the
equivariant model simply relies on being able to exhibit a complete
classification of the critical points of the toric action. This will
be explored in the following sections.

The assumptions made above on the instanton moduli space are not
generally satisfied. To deal with such situations there is a powerful
generalization of the Duistermaat--Heckman formula, the Atiyah--Bott
localization formula for virtual classes. This formalism can be
applied in the case at hand, and has been done in~\cite{MNOP} for rank
one sheaves. In the gauge theory setting, however, it is easier to
deal with the equivariant localization as if the Duistermaat--Heckman
formula and its supersymmetric generalizations were still reliable. In
fact, the general story is a bit more involved than what we have been
discussing thus far. When dealing with gauge theory on a toric manifold
one must deal with the localization procedure more carefully. Now
the $\torus^3$-invariant configurations are not necessarily point-like
and the partition function localizes into a sum of contributions
coming from the points and curves which are fixed by the toric
action. Each of these contributions can be expressed as an integral
over the instanton moduli space, or more precisely over one of its
connected components. Then one can deal with the integrals over the
component moduli spaces as outlined above.

In~\cite{Iqbal:2003ds} the cohomological gauge theory on a toric
threefold $X$ was expressed as a sum over contributions coming from the
vertices of the underlying toric graph $\Delta(X)$, which correspond
to point-like generalized instantons with action $\int_X\,\Tr F_A
\wedge F_A \wedge F_A$. These contributions are connected by gauge
field configurations fibred over the rational curves which connect the
fixed points in the toric diagram, with action $\int_X\, \Tr k_0
\wedge F_A \wedge F_A$. This procedure can be summarized as a set of
gluing rules which are completely analogous to those of the melting
crystal reformulation of the topological A-model string theory,
whereby the framing conditions to be imposed when matching the
contributions from two different vertices of the crystal get mapped
into a set of asymptotic boundary conditions for the gauge field
configurations, or equivalently a framing of the generalized
instanton gauge bundle at infinity.

\bigskip

\section{Noncommutative gauge theory\label{NCGT}}

In this section we will consider another deformation of the
six-dimensional maximally supersymmetric gauge theory which
simplifies explicit computations. This deformation resolves the
small instanton singularities of the moduli space $\mathcal{M}$
and enables explicit construction of instanton solutions.
Moreover it provides a compactification of the instanton moduli
space that is very natural from a gauge theoretical perspective.
Its geometric interpretation will be elucidated in the next
section. For the time being we consider the $U(N)$ gauge theory
defined on $X=\complex^3 \cong\real^6$, and deform it on the
noncommutative space $\real^6_{\theta}$. This is equivalent to
regarding the gauge theory as an infinite-dimensional matrix model
where the fields are replaced by operators acting on a separable
Hilbert space~\cite{Konechny}--\cite{Szabo}. In this case there is
a single patch in the geometry and only six-dimensional point-like
instantons contribute to the partition function. In particular,
there is no contribution from four-dimensional instantons
stretched over rational curves $\PP^1$, which are absent in this
geometry. As the detailed calculations in the following are
somewhat technical, let us start by summarizing the main results
of this section.

\subsection{Statement of results}

We localize the noncommutative gauge theory with respect to the
equivariant action of the abelian group $\torus^3 \times U(1)^N$ on
$\complex^3$. The resulting instanton expansion of the partition
function
is a generating function for Donaldson--Thomas
invariants given by
\begin{eqnarray}
\mathcal{Z}_{\rm DT}^{U(1)^N}\big(\complex^3\big) =\sum_{\vec\pi}\,
(-1)^{(N+1)\,|\vec\pi|}~q^{|\vec\pi|}
\label{genfnnonabinvs}\end{eqnarray} where $q = - \e^{\ii
\vartheta}=\e^{-g_s}$. The sum runs through $N$-coloured,
three-dimensional random partitions $\vec\pi=(\pi_1,\dots,\pi_N)$ with
$|\vec\pi|=|\pi_1|+\dots+|\pi_N|$ boxes. The set of components
$\pi_l$ of fixed size $k_l=|\pi_l|$ is in correspondence with the
Hilbert scheme ${\sf Hilb}^{k_l}(\complex^3)$ of $k_l$ points in
$\complex^3$, which consists of ideals of codimension $k_l$ in the
polynomial ring $\complex[z^1,z^2,z^3]$. The generating function
(\ref{genfnnonabinvs}) is a sum over torus invariant
configurations of $U(N)$ noncommutative instantons, which in the
equivariant model consists of contributions from $k=|\vec\pi|$
instantons on top of each other at the origin. In the abelian case
$N=1$ one recovers the anticipated MacMahon function \be
M(q)=\prod_{n=1}^\infty\,\frac1{\big(1-q^n\big)^n} \ ,
\label{MacMfn}\ee the generating series for ordinary
three-dimensional partitions. Whenever $N$ is odd, the sign is
independent of the partitions and the partition function closely
resembles the abelian result. On the other hand, when $N$ is even
the partition function counts invariants with an alternating sign.
In Sections~\ref{Matrixsheaves} and~\ref{TopMQM} we will connect this
result to the counting of torsion-free sheaves on $\complex^3$, while
some physical applications will be described in Section~\ref{Appl}.

\subsection{Noncommutative instantons\label{NCinsts}}

The coordinates $x^i$ of $\real_\theta^6$ satisfy the Heisenberg
algebra
\be
\big[x^i\,,\,x^j\big] = \ii \theta^{ij} \ , \quad i,j=1,\ldots,6 \ ,
\label{NCalg}\ee
where $\theta=(\theta^{ij})$ is a constant real antisymmetric
$6\times6$ matrix which we assume is nondegenerate. Using an $SO(6)$
rotation, we can choose coordinates such that $\theta$ assumes its
Jordan canonical form
\be
\theta = \left(
\begin{tabular}{cccccc}
0 &$\th_1$ &&&& \\
$-\th_1$ & 0 &&&& \\
&&0& $\th_2$&&\\
&&$-\th_2$&0&&\\
&&&&0& $\th_3$\\
&&&&$-\th_3$&0\\
\end{tabular} \right) \ .
\label{thetaJordan}\ee
The algebra of ``functions'' on $\real_\theta^6$ will be denoted
$\alg$.

The instanton equations on the noncommutative background can be
simplified by introducing the covariant coordinates
\be
X^i = x^i + \ii \theta^{ij}\, A_j \ ,
\label{Xcovcoords}\ee
and their complex combinations
\be
Z^{i} = \mbox{$\frac1{\sqrt{2
\theta_i}}$}\,\big(X^{2i-1} + \ii X^{2i}\big) \qquad \mbox{for} \qquad
i=1,2,3 \ .
\label{Zcovcoords}\ee
Using the Heisenberg commutation relations (\ref{NCalg}) to represent
derivatives as inner derivations of the algebra $\alg$, the instanton
equations (\ref{inste}) can then be rewritten in the form
\bea
\big[Z^{i}\,,\, Z^{j}\big] + \eps^{i
j k} \,\big[Z^{\dagger}_{k}\,,\, \rho\big] &=& 0 \ , \nonumber\\[4pt]
\big[Z^{i}\,,\, Z^{\dagger}_{i}\big] + \big[\rho\,,\, \rho^{\dagger}
\big] &=& 3~1_{N\times N} \ , \nonumber\\[4pt]
\big[Z^{i}\,,\, \Phi\big] &=& 0
\label{adhmform} \eea
where $i,j,k=1,2,3$. The combination of the noncommutative deformation
with the $\Omega$-background parametrized by $\eps_{i}$, $i=1,2,3$
changes the last equation in (\ref{adhmform}) to
\be
\big[Z^{i} \,,\, \Phi\big] = \eps_{i}\, Z^{i} \qquad
(\mbox{no~sum~on}~i) \ .
\label{eqvar}\ee
For the remainder of this paper we will always set the $(3,0)$-form
field $\rho$ to zero, as we work on a Calabi--Yau geometry. Then
$\torus^3$-invariance of the (unique) holomorphic three-form
constrains the equivariant parameters of the $\Omega$-background by
the equation
\be
\eps_1+\eps_2+\eps_3=0 \ .
\label{epsconstr}\ee

These sets of equations can be solved by three-dimensional harmonic
oscillator algebra. Defining
$\alpha_i=\frac1{\sqrt{2\theta_i}}\,(x^{2i-1}+\ii x^{2i})$, the
commutation relations (\ref{NCalg}) are equivalent to
\be
[\alpha_i,\alpha_j]=0 \qquad \mbox{and} \qquad
\big[\alpha_i\,,\alpha_j^\dag\big]=\delta_{ij} \ .
\label{oscalg}\ee
The unique irreducible representation of this algebra is provided by
the Fock module
\be
\Hcal=\complex\big[\alpha_1^\dag\,,\,\alpha_2^\dag\,,\,
\alpha_3^\dag\big]|0,0,0\rangle=\bigoplus_{n_1,n_2,n_3\in\nat_0}\,
\complex|n_1,n_2,n_3\rangle \ ,
\label{Fockspdef}\ee
where $|0,0,0\rangle$ is the Fock vacuum with
$\alpha_i|0,0,0\rangle=0$ for $i=1,2,3$, and the orthonormal basis
states $|n_1,n_2,n_3\rangle$ are
connected by the action of the creation and annihilation operators
subject to the commutation relations (\ref{oscalg}). The operators
(\ref{Zcovcoords}) may then be taken to act on the Hilbert space
\be
\mathcal{H}_W =W\otimes \mathcal{H}
\label{FockWdef}\ee
where $W\cong\complex^N$ is a Chan--Paton multiplicity space of
dimension $N$, the number of D6-branes (and the rank of the gauge
theory). The space $W$ carries the nonabelian degrees of
freedom and we understand $Z^i$ and $\Phi$ as $N \times N$ matrices of
operators acting on $\mathcal{H}$, i.e. as elements of the algebra
$M_{N\times N}(\complex)\otimes\alg$. In the supersymmetric matrix
model where the matrices are operators acting on the Hilbert space
(\ref{FockWdef}), we only need to focus on the operators $Z^i$, $\Phi$
and their hermitean conjugates since we are interested in the class of
minima where all other fields, including the twisted fermions, are
set to zero. For example, the vacuum solution with $F_A=0$ is given by
\bea
Z^{i} &=& \alpha_{i}~1_{N\times N} \ , \nonumber\\[4pt]
\Phi&=& \sum_{i=1}^{3}\, \epsilon_i \,\alpha_i^{\dagger}\, \alpha_i~
1_{N\times N} \ .
\label{vacinstsol}\eea

For $U(1)$ gauge theory, other solutions are found with the solution
generating technique, described for example
in~\cite{kraus,nekrasov}. Fix an integer $n\geq1$ and consider a
partial isometry $U_n$ which projects all states $|i,j,k\rangle$ with
$i+j+k<n$ out of the Fock space $\Hcal$. It obeys
\be
U_n^{\dagger} \,U_n = 1 - \Pi_n
\qquad \mbox{and} \qquad U_n \,U_n^{\dagger} =1
\label{partialisoeqs}\ee
where $\Pi_n$ is the projector
\be
\Pi_n=\sum_{i+j+k<n}\, |i,j,k \rangle \langle
i,j,k| \ .
\label{Projndef}\ee
We may then build a solution from the vacuum (\ref{vacinstsol}) of the
form
\bea
Z^{i} &=& U_n\,\alpha_{i}\,
f(\mathcal{N}\,) \,U_n^{\dagger} \ , \nonumber\\[4pt]
\Phi &=& U_n \, \sum_{i=1}^{3}\, \epsilon_i\, \alpha_i^{\dagger}\,
\alpha_i\,U_n^{\dagger} \ ,
\label{U1ansatz}\eea
where $f$ is a real function of the total number operator
\be
\mathcal{N}= \sum_{i=1}^3\, \alpha_i^{\dagger}\, \alpha_i \ .
\label{numberop}\ee
The function $f(\mathcal{N}\,)$ is found by substituting this ansatz
into the instanton equations to generate a recursion
relation. With the initial condition $f(0)=f(1)=\dots=f(n-1)=0$ and
the finite action condition $f(r)\to1$ as $r\to \infty$, one then
finds the solution
\be
f(\mathcal{N}\,) = \sqrt{1 - \frac{n\, (n+1)\,
    (n+2)}{(\mathcal{N}+1)\,
(\mathcal{N}+2)\,(\mathcal{N}+3)}}~(1-\Pi_n) \ .
\label{fnumberopsol}\ee
The topological charge of the noncommutative $U(1)$ instantons defined
by
\be
k:={\rm ch}_3=
-\mbox{$\frac\ii6$}\,\theta_1\,\theta_2\,\theta_3\,
\Tr_\Hcal\,F_A\wedge F_A\wedge F_A=\mbox{$\frac16$}\,n\,(n+1)\,(n+2)
\label{NCtopchargeU1}\ee
is the number of states in $\Hcal$ with $\mathcal{N}<n$, {i.e.},
the number of states removed by the operator~$U_n$, or equivalently
the rank of the projector (\ref{Projndef}).

\subsection{Nonabelian solutions\label{Nonabsols}}

We will now make some comments on how to generalize the $U(1)$
solutions described above to generic $U(N)$ gauge group. One can
start, for example, from the noncommutative $\mathfrak{u}(3)$-valued
instanton gauge field configuration constructed
in~\cite{Ivanova:2006ek}, which is a smooth deformation of the
canonical connection on the Stiefel bundle over $\PP^3$, written in
local coordinates on a patch $\complex^3$ of the projective space
$\PP^3$. To describe this solution, we set all $\theta_i:=\theta$,
$i=1,2,3$ for simplicity, and consider the exterior derivative $\dd$
as a vector space morphism $\dd:\alg\to\Omega^1_\alg$, where the
sheaf of one-forms $\Omega^1_\alg$ over $\alg$ is the bimodule
$\alg^{\oplus6}$. Introduce elements
\be
\psi_i=\sqrt{6\theta\,k}\,\left(\dd\alpha_i-2\theta\,\alpha_i\,\big(
\gamma^2+\gamma\,\sqrt{1+3\theta}~\big)^{-1}\,\alpha_j^\dag~
\dd\alpha^j\right)\,\gamma^{-1}
\label{psiiu3def}\ee
of $\Omega^1_\alg$ for $i=1,2,3$, where $\gamma:\Hcal\to\Hcal$ is the
invertible operator
\be
\gamma=\sqrt{2\theta\,\mathcal{N}+1+3\theta} \ .
\label{gammainvop}\ee

{}From these operators one can construct a gauge field regarded as a
morphism of $\alg$-modules
\be
F_A\,:\,\Hcal_{W_0}~\longrightarrow~\mathcal{H}_{W_0}\otimes_\alg
\Omega^2_\alg \ ,
\label{FAU3morph}\ee
where $\mathcal{H}_{W_0}$ is the module $W_0\otimes\Hcal$ with
$W_0\cong\complex^3$ the fibre space of the Stiefel bundle, and the
sheaf of two-forms over $\alg$ is the bimodule
$\Omega^2_\alg=\alg^{\oplus15}$. It is given by the $3\times3$ matrix
of Fock space operators
\be
F_A=\big(\psi_i\wedge\psi_j^\dag\big)
\label{FApsidef}\ee
which, in the basis of $\Omega^2_\alg$ generated by
(\ref{psiiu3def}), has components
\be
F_A^{2,0}=0 \qquad \mbox{and} \qquad \big(F_A^{1,1}\big)_{ij}=e_{ij}
\label{FAu3comps}\ee
where $e_{ij}$ are the standard $3\times3$ matrix units. The
corresponding covariant coordinates $Z_0^i$ thus obey the commutation
relations
\be
\big[Z_0^i\,,\,Z_0^j\big]=0 \qquad \mbox{and} \qquad
\big[Z_{0,i}\,,\,Z_{0,j}^\dag\big]=3e_{ij} \ ,
\label{Z0commrels}\ee
and consequently they solve the noncommutative DUY equations in
(\ref{adhmform}) for $U(3)$ gauge group with instanton charge
$k$. See~\cite{Ivanova:2006ek} for the detailed expressions for the
covariant coordinates $Z_0^i$.

The desired $N\times N$ covariant coordinates $Z^i$ may be realized in
terms of the $U(3)$ solution $Z_0^i$ above by appealing to the Hilbert
hotel argument, following~\cite{Popov:2005ik}. For this, we introduce
a lexicographic ordering $\nat_0^3\sim\nat_0$ on the Fock space
$\mathcal{H}$ so that $|n_1,n_2,n_3\rangle=|q\rangle$ with
$q\in\nat_0$, and fix an orthonormal basis $\vec\rho_a$,
$a=0,1,2$ of the fibre space $W_0$. Then
$\vec\rho_{a}\otimes|q\rangle$ is an orthonormal basis for
$\Hcal_{W_0}$ and there is a one-to-one correspondence
$\vec\rho_{a}\otimes|q\rangle\leftrightarrow|3
q+a\rangle$ of basis states. Similarly, by fixing an
orthonormal basis $\vec\lambda_a$, $a=0,1,\dots,N-1$ of the $U(N)$
representation space $W\cong\complex^N$, there is a one-to-one
correspondence $\vec\lambda_{a}\otimes|q\rangle\leftrightarrow
|N\,q+a\rangle$ for the corresponding orthonormal basis of
$\Hcal_W$. Let us now introduce the rectangular $N\times3$ unitary
isomorphism $U:\Hcal_{W_0}\to \Hcal_W$ by the formula
\be
U=\sum_{a=0}^{2}~\sum_{b=0}^{N-1}~\sum_{\stackrel{
\scriptstyle q,r=0}{\scriptstyle 3q+a=N\,r+b}}^\infty\,
|N\,r+b\rangle\langle 3q+a|= \sum_{a=0}^{2}~
\sum_{b=0}^{N-1}~\sum_{\stackrel{
\scriptstyle q,r=0}{\scriptstyle 3q+a=N\,r+b}}^\infty\,
\vec\lambda_{b}\,
{\vec\rho_{a}}{}^\dag\otimes|b\rangle\langle a| \ .
\label{Uimdef}\ee

Starting from the $U(3)$ solution of the noncommutative gauge theory
above, one then constructs the $U(N)$ solutions
\be
Z^i=U\,Z_0^i\,U^\dag  \ .
\label{UNsolfromvac}\ee
Of course, these nonabelian solutions only constitute a subset of the
full BPS solution space. More general solutions will contain
appropriate versions of the function $f(\mathcal{N})$ which
featured into the solution of the noncommutative DUY equations when
$N=1$, and which reflect extra free moduli such as the size and
relative orientation of the D0-branes on top of each other at the
origin of $X=\complex^3$ (as exhibited explicitly in~\cite{kraus} for
the four-dimensional case). However, in the following we will not need
the details of this solution. We will only need to know that, even in
the nonabelian case, the solution is parametrized by specific partial
isometry operators as in Section~\ref{NCinsts} above. Some further
aspects of the generic $U(N)$ noncommutative instantons in six
dimensions will be described in the next section.

\subsection{Three-dimensional coloured partitions}

A key feature of the solution of the six-dimensional cohomological
gauge theory will be its interpretation as a statistical theory of
three-dimensional random partitions. To see random partitions
emerging, let us diagonalize the field $\Phi$ using the $U(N)$ gauge
symmetry to get
\be
\Phi = \left(
\begin{tabular}{cccc}
$\Phi_1$&&& \\
&$\Phi_2$ && \\
&&$\ddots$&\\
&&&$\Phi_N$\\
\end{tabular} \right) \ .
\label{Phidiag}\ee
This transformation induces a Vandermonde determinant $\det({\rm
  ad}\,\Phi)$ in the path integral measure (\ref{fluctratioC3})
below. One can now classify the fixed points of the nonabelian gauge
theory by generalizing the arguments of~\cite{Iqbal:2003ds,swpart}. We
are prescribed to compute the path integral over configurations of the
Higgs field whose asymptotic limit is $\mbf{a} = \mathrm{diag}(a_1 ,
\dots , a_N)\in\mathfrak{u}(1)^{\oplus N}$. With this choice of boundary
condition the noncommutative field $\Phi$ has the form
\begin{equation} \label{Phi}
\Phi = \mbf{a} \otimes {1}_{\mathcal{H}} +
{1}_{N\times N} \otimes \Phi_{\mathcal{H}}
\end{equation}
on (\ref{FockWdef}), up to a function that goes to zero at infinity
faster that any power of $x^i$. Note that this boundary condition only
holds on $X=\mathbb{C}^3$. Nontrivial geometries require a more
involved gluing together of different $\complex^3$ patches, as we
discuss in Section~\ref{Nonabpartfn}. The degeneracies of the
asymptotic Higgs vevs breaks the gauge group $U(N)\to\prod_l\,U(k_l)$
with
\be
\sum_l\,k_l=N \ .
\label{kisumN}\ee
Correspondingly, the Chan--Paton multiplicity space $W$ decomposes into
irreducible representations $W=\bigoplus_l\,W_l$ with $\dim_\complex
W_l=k_l$.

With our choice of the equivariant action of $\torus^3 \times U(1)^N$,
the gauge theory will localize onto the maximal torus.
Consequently the Higgs vevs $a_l$ may all be assumed distinct and
the vacuum is the one with maximal symmetry breaking $k_l=1$ for
$l=1,\dots,N$.
This means that the full Hilbert space (\ref{FockWdef}) splits into a
sum of $N$ ``abelian'' Hilbert spaces $\mathcal{H}$, each one
``coloured'' by the Higgs vev $a_l$.
The theory localizes on noncommutative $U(1)$ instantons that are
in correspondence with maps of the full Hilbert space onto the
subspace
\begin{equation}
\mathcal{H}_{\mathcal{I}} = \bigoplus_{l=1}^N\, \mathcal{I}_{a_l}
\big[\alpha_1^{\dagger} \,,\, \alpha_2^{\dagger} \,,\,
\alpha_3^{\dagger} \big]| 0 ,0 , 0 \rangle \ ,
\label{idealsubsp}\end{equation}
where $\mathcal{I}_{a_l}$ are ideals in the polynomial ring
$\complex[z^1,z^2,z^3]$. Each partial isometry $U_n$ satisfying
(\ref{partialisoeqs}) identifies the Fock space (\ref{Fockspdef}) with
a subspace of the form in (\ref{idealsubsp}), with $\mathcal{I}$ the
ideal of codimension $k$ consisting of polynomials
$f\in\complex[z^1,z^2,z^3]$ for which
\be
\Pi_n\cdot f\big(\alpha_1^\dag\,,\,\alpha_2^\dag\,,\,\alpha_3^\dag
\big)|0,0,0\rangle=0 \ .
\label{Pinideal}\ee
These ideals are generated by monomials $z^i\,z^j\,z^k$ and are in
one-to-one correspondence with three-dimensional partitions. Thus in
complete analogy with the four-dimensional
case~\cite{sw,swpart,fucito}, the solutions correspond to
three-dimensional (plane) partitions with the triples $(i,j,k)$
in (\ref{Projndef}) corresponding to boxes of the partition. More
precisely, the solution can be found in terms of coloured partitions
\be
\vec\pi=(\pi_1, \ldots, \pi_N) \ ,
\label{vecpidef}\ee
which are rows of $N$ ordinary three-dimensional partitions $\pi_l$
labelled by $a_l$. We will explain this correspondence in more detail
in Section~\ref{TopMQM}.

A specific class of observables of the gauge theory is given by
the trace of powers of the Higgs field $\Phi$. Since the gauge theory
is cohomological all the interesting observables (including the
integral of the Chern character over $\mathbb{C}^3$) can be
expressed through these quantities by means of descendent
relations. In particular, for the solution associated to the
sum of ideals
$\mathcal{I}=\mathcal{I}_{a_1}\oplus\cdots\oplus\mathcal{I}_{a_N}$
corresponding to the three-dimensional coloured partition
(\ref{vecpidef}), this produces the normalized character
\begin{eqnarray}\label{nabcharacter}
\chi_{\mathcal{I}}(t)&=&\big(1-\e^{t
  \,\epsilon_1}\big)\,\big(1-\e^{t\,
\epsilon_2}\big) \,\big(1-\e^{t \,\epsilon_3}\big)\,
\Tr_{\mathcal{H}_{\mathcal{I}}} \,\e^{t\, \Phi} \\[4pt] &=&
\sum_{l=1}^N\, \e^{t\, a_l} \,\Big( 1 - \big(1-\e^{t
  \,\epsilon_1}\big)\,\big(1-\e^{t\,
\epsilon_2}\big) \,\big(1-\e^{t \,\epsilon_3}\big)\, \sum_{(i,j,k)\in\pi_{l}}\,
\e^{t\,(\epsilon_1\,(i-1)+\epsilon_2\,(j-1)+\epsilon_3\,(k-1))} \Big)
\ . \nonumber
\end{eqnarray}

This construction has a heuristic interpretation in terms of
D-branes. Localizing the gauge theory onto the Cartan subalgebra is
equivalent to displacing the $N$ D6-branes in the four-dimensional
extended space outside the Calabi--Yau manifold $X$. The vevs of the
Higgs field correspond to the positions of neighbouring D-branes
relative to one another. This means that there are $N$ separated
D6-branes, each one with its own bound state of D0-branes
corresponding to a six-dimensional instanton through anomalous
couplings to Ramond--Ramond fields on the D6-brane worldvolume. The
D0-branes are indexed by the boxes of the three-dimensional partition
$\pi_l$ and the ``colour'' of the partition is the information
relative to which D6-brane they are bound to. $k_l$ D0-branes bound to
a D6-brane labelled by $a_l$ are described by a three-dimensional
partition with $k_l$ boxes in the $l$-th sector of the Hilbert space
and correspond to a 
charge $k_l$ instanton on the worldvolume of the D6-brane in position
$a_l$. It should be stressed though that this geometric interpretation
is somewhat naive and if the instanton measure has a nontrivial
dependence on the Higgs vevs $a_l$, then this could reflect open
string degrees of freedom stretching between different D6-branes. 

\subsection{Instanton weight}

We will now examine the noncommutative instanton contributions to
the partition function of the $U(N)$ topological gauge theory on
$\mathbb{C}^3$ in the vacuum of maximal symmetry breaking.
Proceeding with localization the contribution of an instanton
associated to a collection of ideals $\mathcal{I}$ comes with the
weight factor \be \exp{\Big( - {\vartheta \over 48 \pi^3}
\,\Tr_{\Hcal_{\mathcal{I}}}\,
  F_A \w F_A \w
  F_A\Big)} = \exp{\Big(  {\ii \vartheta \,{\chi}_{\mathcal{I}}^{(3)}
    \over \eps_1\, \eps_2 \,\eps_3}\Big)} \ ,
\label{calEweight}\ee
where ${\chi}_{\mathcal{I}}^{(3)}$ is the coefficient of $t^3$ in
the power series expansion of the character (\ref{nabcharacter}) about
$t=0$. Using (\ref{epsconstr}) one finds
\be
{\chi}_{\mathcal{I}}^{(3)} = \sum_{l=1}^N\,
\Big( -\frac{a_l^3}{6} + \epsilon_1\, \epsilon_2\, \epsilon_3\,
\sum_{(i,j,k)\in\pi_l}\, 1 \Big) \ .
\label{calEideal3}\ee
The first term is independent of the partitions $\pi_l$ and can be
dropped as a universal perturbative contribution. The second term
yields, for each $l$, the total number of boxes $k_l=|\pi_l|$ in the
partition $\pi_l$, which coincides with the topological charges
(\ref{NCtopchargeU1}) of the corresponding noncommutative $U(1)$
instantons. Thus for the weight of an instanton we obtain
\be
\e^{\ii \vartheta\, |\vec\pi|}
\label{instweight}\ee
in the sector of instanton charge
\be
k=|\vec\pi|=\sum_{l=1}^N\,|\pi_l|
\label{instchargepim}\ee
given by the total number of boxes in the coloured partition.

\subsection{Instanton measure}

In addition to the weight there is also a contribution from the
determinants representing quantum fluctuations around the instanton
solutions. In the noncommutative gauge theory the ratio of fluctuation
determinants is represented by~\cite{Iqbal:2003ds}
\begin{equation}
\frac{\det \left(\mathrm{ad}\, \Phi \right)\, \det \left(\mathrm{ad}\,
\Phi + \epsilon_1 + \epsilon_2 \right)\, \det \left(\mathrm{ad}\, \Phi
+ \epsilon_1 + \epsilon_3 \right)\, \det \left(\mathrm{ad}\, \Phi +
\epsilon_2 + \epsilon_3 \right)}{\det \left(\mathrm{ad}\, \Phi+
\epsilon_1 + \epsilon_2 + \epsilon_3 \right)\, \det
\left(\mathrm{ad}\, \Phi + \epsilon_1 \right)\, \det \left(\mathrm{ad}\,
\Phi + \epsilon_2 \right)\, \det \left(\mathrm{ad}\, \Phi + \epsilon_3
\right)}
\label{fluctratioC3}\end{equation}
where $\Phi$ is in general nonabelian and given by (\ref{Phi}). The
origin of this ratio will be explained in more detail in
Section~\ref{TopMQM}. In the following we will compute the instanton
measure explicitly. The ratio of determinants can be written as
\begin{equation}
\exp \Big(-\int_0^\infty\, \frac{\dd t}{t}~ \Tr_{\Hcal_{\mathcal{I}}}
\e^{t\, \Phi}\, \Tr_{\Hcal_{\mathcal{I}}} \e^{-t \,\Phi}\, \big(1-\e^{t\,
\epsilon_1}\big)\, \big(1-\e^{t\, \epsilon_2}\big)\, \big(1-\e^{t\,
\epsilon_3}\big)\Big)
\end{equation}
which can be recast as an index-like quantity
\begin{equation} \label{ratioindex}
Z_I^{U(1)^N}(\vec\pi)=\exp \Big(-\int_0^\infty\, \frac{\dd t}{t}~
\frac{\chi_{\mathcal{I}} (t)\, \chi_{\mathcal{I}} (-t)}
{\big(1-\e^{t\, \epsilon_1}\big)\, \big(1-\e^{t\, \epsilon_2}\big)\,
  \big(1-\e^{t \,\epsilon_3}\big)}\Big) =:\e^{-I^{U(1)^N}(\vec\pi)} \ ,
\end{equation}
where $\vec\pi$ is the coloured partition corresponding to the given
BPS state.

First, we review the abelian case $N=1$ in some detail. Let us break
up the integral $I^{U(1)}(\pi)$ in (\ref{ratioindex}) into three
contributions $I^{U(1)}(\pi)=I^{U(1)}_{\rm vac} + I^{U(1)}_{1}(\pi) +
I^{U(1)}_{2}(\pi)$ given by
\bea
I^{U(1)}_{\rm vac} &=& \int_{0}^{\infty}\, {\dd t
\over t}~ {1 \over  \big(1 - \e^{t \,\eps_1} \big)\,\big(1 - \e^{t\,
  \eps_2} \big) \, \big(1 -
\e^{t\, \eps_3} \big) } \ , \nonumber\\[4pt]
I^{U(1)}_{1}(\pi) &=& \sum_{(i,j,k) \in \pi}~
\int_{0}^{\infty}\, {\dd t \over t}~ \left( - \e^{t\, (\eps_1\, i +
\eps_2\, j+ \eps_3\, k) }
+ \e^{-t\, (\eps_1\, i + \eps_2\, j + \eps_3\, k) }  \right) \ ,
\\[4pt]
I^{U(1)}_{2}(\pi) &=& \sum_{\scriptstyle (i,j,k) \in
  \pi\atop\scriptstyle
(i',j',k'\,) \in \pi} ~ \int_{0}^{\infty} \,{\dd t \over t}~\e^{t
\,(\eps_1 \,
(i-i'\,)  + \eps_2\, (j-j'\,) + \eps_3\, (k-k'\,) )} \nonumber\\
&& \qquad\qquad\qquad\qquad\quad \times\,\left( \e^{t \,\eps_1} -
\e^{-t\, \eps_1} + \e^{t \,\eps_2} - \e^{-t\, \eps_2} + \e^{t\,
  \eps_3} - \e^{-t\, \eps_3} \right) \ . \nonumber
\eea
The integral $I^{U(1)}_{\rm vac}$ is the universal vacuum contribution
from the empty partition and will be dropped in the following. 
Evaluating the remaining integrals we obtain
\bea
I^{U(1)}_{1}(\pi) &=& \ii \pi \,\sum_{(i,j,k) \in \pi}\,
\sign(\eps_1\, i
+ \eps_2\, j + \eps_3\, k) \ , \\[4pt] I^{U(1)}_{2}(\pi) &=&
\ii \pi \,\sum_{\scriptstyle
(i,j,k) \in \pi\atop\scriptstyle (i',j',k'\,) \in \pi}\, \Big(
\sign\big(
\eps_1\, (i - i' +1) + \eps_2\, (j - j'\,) + \eps_3\, (k -k'\,)\big)
\nonumber\\ && \qquad\qquad\qquad\quad +\, \sign\big(
\eps_1\, (i - i' \,) + \eps_2\, (j - j' +1 ) +
\eps_3\, (k -k'\,)\big) \nonumber\\ && \qquad\qquad\qquad\quad +\,
\sign\big( \eps_1\, (i - i'\,) + \eps_2\, (j - j'\,) +
\eps_3\, (k -k' +1)\big)\Big) \ . \nonumber \eea

Since we are interested only in the exponentials of these integrals,
the value $\pm\,1$ of the signum function is irrelevant (we count
$\e^{\pm \ii\pi}=-1$). Thus we have
\bea
\e^{-I^{U(1)}_{1}(\pi)} &=& (-1)^{|\pi| - {\rm diag}} \ , \nonumber
\\[4pt]
\e^{-I^{U(1)}_{2}(\pi)} &=& (-1)^{3 |\pi|^2 -{\rm diag}} \ ,
\eea
where diag denotes the ``diagonal'' terms for which the signum
functions combine to give zero and have to be subtracted since they
give a factor $+1$ instead of $-1$. The direct counting of how many
diagonal terms are present for a given partition $\pi$ is a bit
complicated, and moreover one has to take into account the constraint
(\ref{epsconstr}). Instead, we will use induction to prove the result
\be
\e^{-I^{U(1)}_{1}(\pi) - I^{U(1)}_{2}(\pi)} = (-1)^{|\pi|}
\label{measure} \ee
for any three-dimensional partition $\pi$.

Consider first the partition $\pi=(1,1,1)$ consisting of a single box, 
so that $|\pi|=1$. In this case one has $\e^{I^{U(1)}_1}(\pi)=1$,
$\e^{I^{U(1)}_2(\pi)}=-1$ and $\e^{-I^{U(1)}_1(\pi)
  -I^{U(1)}_2(\pi)}=-1$, so this partition satisfies
(\ref{measure}). Suppose now that for some partition $\pi$,
eq.~(\ref{measure}) is true. In the induction step we increase the
number of boxes in this partition by one (in such a way that we still
obtain a valid partition). Denote the new partition with the extra box
at a given position $(a,b,c)$ by $\pi^*$. One has $|\pi^*| - |\pi|=1$,
and so we have to prove that
\be
\e^{-I_1^{U(1)}(\pi^*) - I_2^{U(1)}(\pi^*) + I^{U(1)}_1(\pi) +
  I^{U(1)}_2(\pi)} = -1 \ .
\label{eI12starminus}\ee
For the first integral we have
\be
I_1^{U(1)}(\pi^*) - I^{U(1)}_1(\pi)  = \ii \pi ~\sign (\eps_1\, a
+ \eps_2\, b + \eps_3\, c) \ .
\label{firstintab}\ee
For the second integral, after some algebra we obtain
\bea
&&I^{U(1)}_2(\pi^*)  - I^{U(1)}_2(\pi)  \nonumber\\
&& \qquad\qquad ~=~ 6 |\pi| + 3 - \min(a,b-1,c-1) -
\min(a-1, b,c) - \min(a-1,b,c-1) \nn && \qquad\qquad\qquad
- \, \min(a,b-1,c)  -
\min(a-1,b-1,c) - \min(a,b,c-1) \ .
\label{secondintab}\eea
{}From these results it is easy to establish (\ref{eI12starminus}),
proving the induction step.

Putting everything together, the abelian instanton measure is given by
\be
Z^{U(1)}_{I}(\pi) = Z^{U(1)}_{\rm vac} ~ (-1)^{|\pi|} \ .
\ee
Dropping the perturbative vacuum contribution and combining this
result with the weight (\ref{instweight}), the full instanton
contribution to the $U(1)$ partition function is given by
\be
\mathcal{Z}^{U(1)}_{\rm DT}\big(\complex^3\big)= \sum_{\pi}\,
\big(-\e^{\ii \vartheta}\big)^{|\pi|} = M(q) \ .
\label{DTMq}\ee
This is the known Donaldson--Thomas partition function on
$\complex^3$.

This construction can be easily generalized to the case where the full
nonabelian gauge symmetry is broken to $U(1)^N$. Using the
character~(\ref{nabcharacter}), the exponent $I^{U(1)^N}(\vec\pi)$
of the fluctuation determinant (\ref{ratioindex})
can again be written as the sum of three integrals \bea I_{\rm
vac}^{U(1)^N} &=& \int_{0}^{\infty}\, {\dd t \over t}~ {1 \over
\big(1 - \e^{t\, \eps_1} \big) \,\big(1 - \e^{t\,
  \eps_2} \big)\, \big(1 - \e^{t\,
\eps_3} \big) }~ \sum_{l,n=1}^{N} \,\e^{t\,(a_l -a_n)} \ , \\[4pt]
I_1^{U(1)^N}(\vec\pi) &=&   \int_{0}^{\infty}\, {\dd t \over t} ~
\sum_{l,n=1}^{N} \e^{t\,(a_l -a_n)} \nonumber\\ &&
\qquad\qquad\qquad \times\, \Big( -  \sum_{(i,j,k) \in \pi_l}
\,\e^{t\, (\eps_1 \,i + \eps_2\, j + \eps_3\, k) } +
\sum_{(i,j,k) \in \pi_n} \,\e^{-t\, (\eps_1\, i + \eps_2\, j +
  \eps_3\, k) }\Big) \ ,  \nonumber \\[4pt]
I_2^{U(1)^N}(\vec\pi) &=& \int_{0}^{\infty}\, {\dd t \over t}
~\sum_{l,n=1}^{N} \,\e^{t\,(a_l -a_n)}~\sum_{\scriptstyle (i,j,k)
\in \pi_l \atop\scriptstyle (i',j',k'\,) \in \pi_n} \,\e^{t\,
(\eps_1\,
(i-i'\,)  + \eps_2\, (j-j'\,) + \eps_3\, (k-k'\,) }\nonumber\\
&& \qquad\qquad\qquad \times\, \left( \e^{t\, \eps_1} -
\e^{-t\, \eps_1} + \e^{t\, \eps_2} - \e^{-t \,\eps_2} + \e^{t
  \,\eps_3} - \e^{-t\, \eps_3} \right)\nonumber
\eea where as before $\pi_{n}, \pi_{l}$ denote components of the
coloured partition (\ref{vecpidef}) and $a_{n}, a_{l}$ are
components of the classical value of the Higgs field
(\ref{Phidiag}). Evaluating the integrals,
the non-trivial contributions can be written as \bea
I_1^{U(1)^N}(\vec\pi)_{\rm diag} &=& \ii \pi\, \sum_{l=1}^{N}~
\sum_{(i,j,k) \in \pi_l} \,\sign(\eps_1\, i + \eps_2\,
j + \eps_3\, k ) \ , \nonumber\\[4pt] I_1^{U(1)^N}(\vec\pi)_{\rm
  offdiag} &=& \ii \pi \, \sum_{\scriptstyle l,n=1 \atop \scriptstyle
l\neq  n}^{N}~ \sum_{(i,j,k) \in \pi_l} \,\sign(\eps_1\, i + \eps_2\,
j + \eps_3\,
k + a_{ln}) \ , \nonumber\\[4pt] I_2^{U(1)^N}(\vec\pi)_{\rm diag} &=&
\ii \pi\, \sum_{l=1}^N~
\sum_{\scriptstyle (i,j,k) \in \pi_l \atop\scriptstyle (i',j',k'\,)
\in \pi_l}\, \Big( \sign\big( \eps_1\, (i - i' +1) + \eps_2\, (j -
j'\,) +
\eps_3\, (k -k'\,)\big) \nonumber\\ && \qquad\qquad\qquad\qquad\quad +
\, \sign\big( \eps_1\, (i - i' \,) +
\eps_2\, (j - j' +1 ) + \eps_3\, (k -k'\,)\big) \nonumber\\ &&
\qquad\qquad\qquad\qquad\quad  + \,  \sign\big( \eps_1\, (i - i'\,) +
\eps_2\, (j - j'\,) + \eps_3\, (k -k' +1)\big)  \Big) \ , \nonumber
\\[4pt]
I_2^{U(1)^N}(\vec\pi)_{\rm offdiag} &=& \ii \pi\,
\sum_{\scriptstyle
  l,n=1 \atop  \scriptstyle l<n }^N~
\sum_{\scriptstyle (i,j,k) \in \pi_l \atop\scriptstyle (i',j',k'\,)
\in \pi_n}\, \Big( \sign\big( \eps_1 \,(i - i' +1) + \eps_2\, (j -
j'\,) + \eps_3 \,(k -k'\,) + a_{ln}\big) \nonumber\\ &&\qquad\qquad
+\, \sign\big( \eps_1\, (i - i' \, ) +
\eps_2\, (j - j' +1 ) + \eps_3\, (k -k'\,) + a_{ln}\big) \nonumber\\
&&\qquad\qquad +\, \sign\big(
\eps_1\, (i - i'\,) + \eps_2\, (j - j'\,) + \eps_3\, (k -k' +1) +
a_{ln}\big) \nonumber\\ &&\qquad\qquad -\, \sign\big( \eps_1 \,(i - i'
-1) + \eps_2\, (j - j'\, ) + \eps_3 \,(k
-k'\,) + a_{ln}\big) \nonumber\\ && \qquad\qquad-\, \sign\big(
\eps_1\, (i - i'\, ) + \eps_2  \,(j - j' -1
) + \eps_3\, (k -k'\,) + a_{ln}\big) \nonumber\\ && \qquad\qquad-\,
\sign\big( \eps_1 \,(i - i'\,)
+ \eps_2\, (j - j'\,) + \eps_3\, (k -k' -1) + a_{ln}\big) \Big) \ ,
\eea
where we have broken up the integrals into diagonal and
off-diagonal components in the summations over the colour indices, and
denoted $a_{ln}:=a_l-a_n$.

In the $U(1)^N$ phase one has $a_{ln} \neq 0$ for all $1\leq
l\neq n \leq N$, for otherwise the gauge symmetry would be enhanced.
Then one can always choose the infrared regulators $\eps_1$ and
$\eps_2$ such that $\eps_1\, M + \eps_2 \,L  +a_{ln} \neq 0$ for
any pair of integers $M,L$, in addition to the Calabi--Yau
condition (\ref{epsconstr}). This ensures that the signum
functions containing these vevs are never zero. It is easy to see
that in this case one has \bea \e^{-I_1^{U(1)^N}(\vec\pi)_{\rm
diag}} &=& \e^{-
  \sum_l\,I_{1}^{U(1)}(\pi_l)} \ , \nonumber\\[4pt]
\e^{-I_1^{U(1)^N}(\vec\pi)_{\rm offdiag}} &=& \prod_{\scriptstyle
l,n=1
  \atop \scriptstyle l \neq n}^N\,
(-1)^{|\pi_l|} \ , \nonumber\\[4pt] \e^{-I_2^{U(1)^N}(\vec\pi)_{\rm
    diag}} &=& \e^{-\sum_l\,I_{2}^{U(1)}(\pi_l)} \ , \nonumber\\[4pt]
\e^{-I_2^{U(1)^N}(\vec\pi)_{\rm offdiag}} &=& 1 \ , \eea and hence
the full contribution is given by \be
\e^{-I_1^{U(1)^N}(\vec\pi)-I_2^{U(1)^N}(\vec\pi)} =\e^{-
\sum_l\,(I_{1}^{U(1)}( \pi_l)+ I_{2}^{U(1)}(\pi_l))}\, \Big(~
\prod_{\scriptstyle l,n=1 \atop \scriptstyle l\neq n}^N
(-1)^{|\pi_l|} ~\Big) \ . \ee Using the abelian result
(\ref{measure}) to get \be
\e^{-I_{1}^{U(1)}(\pi_l)-I_{2}^{U(1)}(\pi_l)} = (-1)^{|\pi_l|} \ ,
\ee we arrive finally at the instanton measure \be
\e^{-I_1^{U(1)^N}(\vec\pi)-I_2^{U(1)^N}(\vec\pi)}= (-1)^{N
\,|\vec\pi|} \ . \ee Combining this result with the weights
(\ref{instweight}) leads to the partition function
(\ref{genfnnonabinvs}).

\bigskip

\section{Matrix equations and moduli of stable coherent
  sheaves\label{Matrixsheaves}}

The purpose of this section is two-fold. Firstly, we will recast the
BPS solutions of the noncommutative gauge theory explicitly in purely
commutative terms, by relating the counting of noncommutative
instantons to the counting of a special class of stable coherent
sheaves in three dimensions. This gives a geometrical explanation for
and generalizes the natural gauge theoretical identification between
the compactified instanton moduli space $\mathcal{M}(X)$ and the
Hilbert scheme of points ${\sf Hilb}^k(X)$ in the abelian case $N=1$
when the underlying variety is $X =\complex^3$. Secondly, we will
exhibit an ADHM-like parametrization of the equations defining
$\mathcal{M}(\complex^3)$, analogous to the four-dimensional
case. This will naturally bridge the noncommutative gauge theory
formalism with the topological matrix quantum mechanics that we will
introduce in the next section. This matrix model is equivalent to a
quantization of the collective coordinates of the gauge theory around
an instanton solution, and it will provide an alternative means for
analysing the six-dimensional cohomological gauge theory. As before,
we begin by summarizing the main results of this section before
plunging into the detailed technical calculations.

\subsection{Statement of results}

There are natural mappings between isomorphism classes of the
following three objects:
\begin{itemize}
\item[(A)] $U(N)$ noncommutative instantons on $\complex^3$ of
  topological charge ${\rm ch}_3=k$.
\item[(B)] Linear maps $B_i\in\End(V)$, $i=1,2,3$, $I\in\Hom(W,V)$ and
  $J,K\in\Hom(V,W)$ which solve the ``ADHM-type'' matrix equations
\bea
[B_1,B_2]+I\,J&=&0 \ , \nonumber\\[4pt]
[B_1,B_3]+I\,K&=&0 \ , \nonumber\\[4pt]
[B_2,B_3]&=&0 \ , \nonumber\\[4pt]
\sum_{i=1}^3\,
\big[B_i\,,\,B_i^\dag\,\big]+I\,I^\dag-J^\dag\,J-K^\dag\,K&=&3~1_V
\label{ADHMeqs}\eea
on finite-dimensional hermitean vector spaces $V\cong\complex^k$ and
$W\cong\complex^N$, modulo the natural action of $U(V)$ given by
\be
B_i~\longmapsto~g\,B_i\,g^{-1} \ , \quad
I~\longmapsto~g\,I \ , \quad J~\longmapsto~J\,g^{-1} \quad
\mbox{and} \quad K~\longmapsto~K\,g^{-1} \ .
\label{Ukactionmatrixeqs}\ee
\item[(C)] Rank $N$ torsion-free sheaves $\Ecal$ on $\PP^3$ 
  with $\ch_3(\Ecal)=k$, fixed trivializations on three lines at
  infinity, vanishing $H^1(\PP^3,\Ecal(-2))$, and satisfying certain
  stability conditions.
\end{itemize}
These three classes are not equivalent but rather represent
alternative characterizations of one another. We conjecture isomorphisms between the moduli spaces of data (A), (B) and (C) of fixed topological type, after appropriate specifications of additional structures. This requires modification of some of the natural maps defined below, which are generically not isomorphisms but merely birational morphisms at best.

The map from (A) to (B) follows from a rewriting of the
noncommutative DUY equations in (\ref{adhmform}) using special
properties of projective modules over the noncommutative space
$\real_\theta^6$~\cite{Konechny}. The map from (C) to
(B) follows from detailed calculations in sheaf cohomology which
rewrites the sheaves $\Ecal$ as the cohomology of certain complexes of
sheaves on $\PP^3$. It canonically identifies
both the D0-brane and D6-brane Chan--Paton spaces $V\subset\Hcal$ and
$W$ as certain sheaf cohomology groups. In particular, $W$ is
associated with the framing of $\Ecal$. The last equation of
(\ref{ADHMeqs}) is a stability condition. Combined with the linear
maps $\phi\in\End(V)$, $P\in\Hom(W,V)$ solving the matrix equations
\bea
[B_1,\phi]&=&\epsilon_1\,B_1 \ , \nonumber\\[4pt]
[B_2,\phi]-P\,J&=&\epsilon_2\,B_2 \ , \nonumber\\[4pt]
[B_3,\phi]-P\,K&=&\epsilon_3\,B_3 \ ,
\label{matrixphieqs}\eea
which come from the noncommutative Higgs fields
$\Phi$ obeying (\ref{eqvar}), it will be used in the next section to
formalize the connection between the equivariant model and the
statistical mechanics of three-dimensional random partitions.

\subsection{Linear algebra of noncommutative
  instantons\label{LinalgNC}}

In the previous section we described charge $k$ noncommutative
instantons in six dimensions as elements of the algebra $M_{N\times
  N}(\complex)\otimes\alg$ acting on the free module
$W\otimes\alg\cong\alg^{\oplus N}$ of rank $N$ over the
noncommutative algebra $\alg$. The K-theory of the algebra $\alg$ is
the abelian group $\zed\oplus\zed$, with positive cone
$\nat_0\oplus\nat_0$. Thus every projective module
$\Ecal_{k,N}$ over $\alg$ is labelled by a pair of positive integers
$k,N$, which represent exactly the instanton number and rank
respectively. Explicitly, one has~\cite{Konechny}
\be
\Ecal_{k,N}=\Hcal^{\oplus k}\oplus\alg^{\oplus N}
\label{EcalkNdef}\ee
where as before $\Hcal$ is the Fock space (\ref{Fockspdef}). The key
to obtaining a set of matrix equations
which describes six-dimensional noncommutative instantons is the
existence of a natural isomorphism $\Ecal_{k,N}\cong\alg^{\oplus N}$ of
$\alg$-modules.

A connection
$\nabla:\Ecal_{k,N}\to\Ecal_{k,N}\otimes_\alg\Omega^1_\alg$ induces a
decomposition of the covariant coordinates
$Z^i\in\End_\alg(\Ecal_{k,N})$ with respect to the splitting
(\ref{EcalkNdef}) as
\be
Z^1=\begin{pmatrix}B_1&I_0\\I'&R_1\end{pmatrix} \ , \quad
Z^2=\begin{pmatrix}B_2&J'\\J_0&R_2\end{pmatrix}  \quad \mbox{and}
\quad Z^3=\begin{pmatrix}B_3&K'\\K_0&R_3\end{pmatrix} \ .
\label{Zalphasplit}\ee
Using irreducibility of the Fock module $\Hcal$, we note that the rank
one free module $\alg$ can be decomposed into countably many copies of
$\Hcal$ as $\alg=\bigoplus_{n\in\nat_0}\,P_n\cdot\alg$, where
$P_n=|n\rangle\langle n|$ is an orthogonal system of projectors onto
one-dimensional subspaces. For each $n\in\nat_0$, the mapping
$P_n\cdot f\mapsto\langle n|f$ establishes an isomorphism
$P_n\cdot\alg\cong\Hcal$ of $\alg$-modules. Using this identification,
in (\ref{Zalphasplit}) we regard $B_i\in
M_{k\times k}(\complex)$ as linear operators acting on a
finite-dimensional hermitean vector space $V\cong\complex^k$, while
$R_i\in\End_\alg(W\otimes\alg)$ with $W\cong\complex^N$ the
finite-dimensional Chan--Paton multiplicity space as in the previous
section. The off-diagonal entries in eq.~(\ref{Zalphasplit}) are
operators
\be
I_0,J',K'\in\Hom_\alg(W\otimes\alg,V) \qquad \mbox{and} \qquad
I',J_0,K_0\in\Hom_\alg(V,W\otimes\alg) \ .
\label{offdiagops}\ee
In the following we will use the gauge choice $I'=J'=K'=0$.

Then the first instanton equation of (\ref{adhmform}) yields the sets
of equations
\be
\begin{array}{rlrl}
[B_1,B_2]+I_0\,J_0=0& \qquad \mbox{and} & \qquad [R_1,R_2]-J_0\,I_0=0&
\ , \\[4pt]
[B_1,B_3]+I_0\,K_0=0& \qquad \mbox{and} & \qquad [R_1,R_3]-K_0\,I_0=0&
\ , \\[4pt]
[B_2,B_3]=0& \qquad \mbox{and} & \qquad [R_2,R_3]=0&
\end{array}
\label{insteqBRdiag}\ee
along with
\beq
R_1\,J_0-J_0\,B_1=0 \ , \qquad
R_1\,K_0-K_0\,B_1=0 \qquad \mbox{and} \qquad
I_0\,R_2-B_2\,I_0=0 
\label{insteqBRoffd}\eeq
plus
\beq
R_3\,J_0-J_0\,B_3=R_2\,K_0-K_0\,B_2 \qquad \mbox{and} \qquad
I_0\,R_3-B_3\,I_0=0 \ .
\label{offdiagplus}\eeq
The second instanton equation in (\ref{adhmform}) yields the sets of
equations
\bea
\sum_{i=1}^3\,
\big[B_i\,,\,B_i^\dag\,\big]+I_0\,I_0^\dag-J_0^\dag\,J_0-K_0^\dag\,K_0
&=&3~1_V \ ,
\nonumber\\[4pt]
\sum_{i=1}^3\,
\big[R_i\,,\,R_i^\dag\,\big]-I_0^\dag\,I_0+J_0\,J_0^\dag+K_0\,K_0^\dag
&=&3~1_{W\otimes\alg} \ ,
\nonumber\\[4pt]
I_0\,R_1^\dag-J_0^\dag\,R_2-K_0^\dag\,R_3&=&B_1^\dag\,I_0-B_2\,J_0^\dag-
B_3\,K_0^\dag \ .
\label{2ndinsteqBR}\eea
Let us now decompose the off-diagonal operators in
eq.~(\ref{Zalphasplit}) as
\be
I_0=I\otimes\psi_I~\in~
\Hom_\alg\big(W\otimes\alg^{\oplus N}\,,\,V\big)
\cong\Hom(W,V)\otimes\End_\alg\big(\alg^{\oplus N}\big)
\label{I0decomp}\ee
with $I\in M_{k\times N}(\complex)$ and
\be
J_0=J\otimes\psi_J,K_0=K\otimes\psi_K~\in~
\Hom_\alg\big(V\,,\,W\otimes\alg^{\oplus N}\big)
\cong\Hom(V,W)\otimes\End_\alg\big(\alg^{\oplus N}\big)
\label{J0K0decomp}\ee
with $J,K\in M_{N\times k}(\complex)$, where the elements of $M_{N\times
  N}(\complex)\otimes\alg$ satisfy
$\psi_I\,\psi_J=\psi_I\,\psi_K=1_{W\otimes\alg}$. In this way we
arrive, from (\ref{insteqBRdiag}) and (\ref{2ndinsteqBR}), at the
matrix equations (\ref{ADHMeqs}).

We have thus found that the set of ADHM-type data
\be
(B_i;I,J,K)~\in~\Hom(V,V)^{\oplus 3}~\oplus~
\Hom(W,V)~\oplus~\Hom(V,W)^{\oplus 2}
\label{ADHMdata}\ee
can be used to characterize the noncommutative instantons. However,
this set of data is not likely to be complete, i.e. a solution
to the set of matrix equations (\ref{ADHMeqs}) need not yield a
solution $(R_i;\psi_I,\psi_J,\psi_K)$ to the remaining equations in
(\ref{insteqBRdiag})--(\ref{2ndinsteqBR}). One can also reformulate
the equivariant equations (\ref{eqvar}) of the $\Omega$-background as
matrix equations by writing the adjoint scalar field $\Phi$ in the
block form
\be
\Phi=\begin{pmatrix}\phi&P_0\\P'&\varrho\end{pmatrix}
\label{Phisplitting}\ee
with respect to the splitting of the projective module
(\ref{EcalkNdef}). Then the matrix equations for the $k\times k$
matrix $\phi$ in the gauge $P'=0$ are given by
(\ref{matrixphieqs}) with $P_0=P\otimes\psi_I$. In the next section
these equations will be interpreted as the compensation of the toric
action $\torus^3$ by a gauge transformation in an associated
topological matrix model.

Finally, let us see how to explicitly map solutions of the matrix
equations (\ref{ADHMeqs}) onto noncommutative instanton solutions on
the free module $\alg^{\oplus N}$. The partial isometry equations
(\ref{partialisoeqs}) identify the $k$-dimensional subspace $V:=\ker
U_n$. In particular, the linear operator $U_1^\dag$ has a trivial
kernel, while $U_1$ has a one-dimensional kernel which is spanned by
the vacuum vector $|0,0,0\rangle$.
It follows that as $\alg$-modules there is a natural isomorphism $\ker
U_n\cong\Hcal^{\oplus k}$ and hence one can identify the
projective module (\ref{EcalkNdef}) with
\be
\Ecal_{k,N}\cong\ker(U_n)\oplus\alg^{\oplus N} \ .
\label{EcalkNTkiso}\ee
We may now define an invertible mapping
$\mu:\Ecal_{k,N}\xrightarrow{\approx}\alg^{\oplus N}$ by
\be
\mu(\xi,f)=\xi+U_n^\dag\cdot f \qquad \mbox{and} \qquad
\mu^{-1}(f)=\big(\Pi_n(f)\,,\,U_n\cdot f\big) \ .
\label{muisodef}\ee
Under this isomorphism of projective
$\alg$-modules, the image of an arbitrary covariant coordinate
$Z\in\End_\alg(\Ecal_{k,N})$ represented as
\be
Z=\begin{pmatrix}B&I\\J&R\end{pmatrix}
\label{arbcovcoordZ}\ee
is the element of $\End_\alg(\alg^{\oplus N})\cong M_{N\times
  N}(\complex)\otimes\alg$ given by
\be
\mu(Z)=\Pi_n\,B\,\Pi_n+\Pi_n\,I^\dag\,I\,U_n+U_n^\dag\,J\,J^\dag\,
\Pi_n+U_n^\dag\,R\,U_n \ .
\label{mmuZexpl}\ee

An explicit construction of the data (\ref{ADHMdata}) for $U(3)$ gauge
group may be done by an elementary extension of the construction
of~\cite{Lagraa:2005ww} for the four-dimensional noncommutative $U(2)$
ADHM data. The $U(3)$ noncommutative instanton solutions
of~\cite{Ivanova:2006ek} can be written in the form
$A=\Psi^\dag~\dd\Psi$, where
$\Psi\in\Hom_\alg(W_0\otimes\alg,\Ecal_{k,3})$. These solutions can
again be extended to generic $U(N)$ gauge group following the
technique explained in Section~\ref{Nonabsols}. We
will not explore further details of this construction in this paper.

\subsection{Beilinson spectral sequence}

In the remainder of this section we will show that the matrix
equations (\ref{ADHMeqs}) provide a geometrical interpretation of the
noncommutative instantons satisfying the six-dimensional DUY equations
on $\complex^3$. For this, we will generalize an analogous
construction in four dimensions exhibited by Nakajima
in~\cite{nakajima} (see also~\cite{donaldson}
and~\cite{nakajima2}). Since it is more 
convenient to work with compact spaces, we regard an instanton on
$\complex^3$ as a holomorphic bundle on $\PP^3$ with a trivialization
at infinity. To compactify the instanton moduli space we include all
semistable torsion free sheaves on $\PP^3$ with a trivialization
condition at infinity and the appropriate topological quantum numbers
of an instanton on $\complex^3$. This allows for singularities of the
instanton gauge field on $\complex^3$, with the singularity locus the
support where the torsion free sheaves fail to be holomorphic vector
bundles.

We are thus interested in the framed moduli space of torsion free
sheaves given by
\begin{equation} \label{sheavesonP3}
{\mathcal M}_{N, k} \big(\mathbb{P}^3\big) = \left\{ \begin{array}{c}
\mathcal{E} =\text{torsion
free sheaf on}~ \mathbb{P}^3 \\ {\rm rank}(\mathcal{E})= N \ , \quad
c_1(\mathcal{E}) = 0 \\
\mathrm{ch}_2 (\mathcal{E}) = 0 \ , \quad \mathrm{ch}_3 (\mathcal{E})
= k \\ \mathcal{E}\big|_{\ell^i_{\infty}} \cong\mathcal{O}^{\oplus N}_{\ell^i_{\infty}}
\end{array} \right\} ~\Big/~ {\rm isomorphisms}
\end{equation}
where $\ell^i_{\infty}$, $i=1,2,3$ are three projective lines at infinity which will be specified explicitly below. The moduli space (\ref{sheavesonP3})
corresponds to the counting of D0--D6 bound states on $\complex^3$ in
a suitable $B$-field background, as we demonstrate explicitly in the
next section. Following~\cite{nakajima}, we will use the
Beilinson spectral sequence to parametrize a generic torsion free
sheaf $\mathcal{E}$ and then show that this spectral sequence
degenerates into a complex of sheaves on $\PP^3$ which is related
to solutions of the matrix equations (\ref{ADHMeqs}). The Beilinson
theorem in our case implies that for any torsion free sheaf
$\mathcal{E}$ on $\PP^3$ there is a spectral sequence with $s$-th term
$E_s^{p,q}$ which converges to $\mathcal{E}$ if $p+q=0$ and to zero
otherwise.

The Beilinson spectral sequence can be constructed by starting with
the product space ${\mathbb P}^3 \times {\mathbb P}^3$ and the
canonical projections onto the first and second factors
\begin{equation}
\xymatrix@=15mm{
  &{\mathbb P}^3 \times {\mathbb P}^3 \ar[ld]_{p_1}\ar[rd]^{p_2}& \\
  {\mathbb P}^3 & & {\mathbb P}^3 \ .
}
\end{equation}
Then for any coherent sheaf $\mathcal{E}$ on ${\mathbb P}^3$ one has
the projection formula
\begin{equation} \label{identity}
p_{1\,*} \big(p_2^* \mathcal{E} \otimes \mathcal{O}_{\Delta}\big) =
p_{1\,*} \big(p_2^* \mathcal{E}\big|_{\Delta}\big) = \mathcal{E}
\end{equation}
where $\Delta\cong\PP^3 \subset {\mathbb P}^3 \times {\mathbb P}^3$ is
the diagonal. In the following we will use the notation
\beq
\mathcal{E}_1
\boxtimes\Ecal_2:= p_1^*\Ecal_1\otimes p_2^*\Ecal_2
\eeq
for the exterior product over $\PP^3\times\PP^3$ of two sheaves
$\Ecal_1,\Ecal_2$ on $\PP^3$.

The next step consists in replacing the structure sheaf
$\mathcal{O}_{\Delta}$ 
of the diagonal with an appropriate projective resolution, which we
will take to be given by the Koszul complex. Consider the short exact
sequence which defines the tangent bundle to ${\mathbb P}^3$ by
\begin{equation}
\xymatrix@1{
   0 \ar[r] & \mathcal{O}_{{\mathbb P}^3} \ar[r] &
   \mathcal{O}_{{\mathbb P}^3}(1)^{\oplus 4} \ar[r] & T_{\mathbb{P}^3}
   \ar[r] & 0 \ .
}
\end{equation}
If we tensor this sequence with $\mathcal{O}_{{\mathbb P}^3}(-1)$
we get the short exact sequence
\begin{equation}
\xymatrix@1{
   0 \ar[r] & \mathcal{O}_{{\mathbb P}^3} (-1) \ar[r] &
   \mathcal{O}_{{\mathbb P}^3}^{\oplus 4} \ar[r] & \mathcal{Q} \ar[r]
   & 0 \ ,
}
\end{equation}
which defines the universal quotient bundle $\mathcal{Q} =
T_{\mathbb{P}^3} \otimes \mathcal{O}_{{\mathbb P}^3}(-1)$ and its dual
$\mathcal{Q}^* = \Omega^1_{\mathbb{P}^3} \otimes \mathcal{O}_{{\mathbb
    P}^3}(1)$ where $\Omega_{{\mathbb P}^3}^1$ is the sheaf of
one-forms on $\PP^3$. Then the Koszul complex is given
by~\cite{okonek}
\begin{eqnarray}
&0 ~\longrightarrow~ \bigwedge^3 \big( \mathcal{O}_{{\mathbb P}^3}
  (-1) \boxtimes \mathcal{Q}^* \,\big)~\longrightarrow~
    \bigwedge^2 \big( \mathcal{O}_{{\mathbb P}^3} (-1) \boxtimes
      \mathcal{Q}^* \,\big)~\longrightarrow~
    \mathcal{O}_{{\mathbb P}^3} (-1) \boxtimes \mathcal{Q}^*~
    \longrightarrow~
\nonumber\\ & \qquad\qquad~\longrightarrow~
   \mathcal{O}_{{\mathbb P}^3 \times {\mathbb P}^3}~
   \longrightarrow~\mathcal{O}_{\Delta} ~\longrightarrow~ 0 \ . &
\label{Koszul}\end{eqnarray}

{}From this complex we can construct a spectral sequence by
taking an injective resolution of the hyperdirect image of
eq.~(\ref{identity}), and replacing $\mathcal{O}_{\Delta}$ with its
Koszul resolution. More details can be found
in~\cite[Chapter~2~\S3.1]{okonek} and~\cite[Chapter~2]{nakajima}. By 
introducing 
\begin{equation}
C^p := \mbox{$\bigwedge^{-p}$} \big( \mathcal{O}_{{\mathbb P}^3} (-1)
\boxtimes \mathcal{Q}^* \,\big)
\end{equation}
the double complex can be expressed as the Fourier--Mukai transform
\begin{equation}
{\mbf R}^{\bullet} p_{1 \, *} \big( p_2^* \mathcal{E} \otimes
C^{\bullet} \big) \ ,
\end{equation}
where ${\mbf R}^{\bullet}$ is the right derived functor in the bounded
derived category of coherent sheaves on $\PP^3$.
Note that each term of the Koszul resolution has the form
\begin{equation}
C^p = \mathcal{F}^p_1 \boxtimes \mathcal{F}^p_2 \ .
\end{equation}

The Beilinson theorem then implies that for any coherent sheaf
$\mathcal{E}$ on $\mathbb P^3$ there is a spectral sequence
$E_s^{p,q}$ with $E_1$-term
\begin{equation}
E_1^{p,q} = \mathcal{F}_1^p \otimes H^q \left( \mathbb{P}^3 \, , \,
\mathcal{E} \otimes \mathcal{F}_2^p \right)
\end{equation}
which converges to
\begin{equation}
{E}_\infty^{p,q} = \begin{cases}
  ~\mathcal{E}(-r) \ , \qquad \mathrm{if} ~ p+q=0  \ , \\
  ~0 \ , \qquad \mathrm{otherwise} \end{cases}
\end{equation}
for each fixed integer $r\geq0$, where we denote
$\mathcal{E}(-r):=\mathcal{E}\otimes_{\mathcal{O}_{\PP^3}}
\mathcal{O}_{\PP^3}(-r)$. Explicitly, the first term is given by
\begin{equation} E_1^{p,q} = H^q
\big( {\mathbb P}^3 \,,\, \mathcal{E}(-r) \otimes
  \Omega^{-p}_{\PP^3}(-p) \big) \otimes \mathcal{O}_{\PP^3}(p)
\end{equation}
for $p\leq0$. The $E_1$-term complexes of the spectral sequence can be
summarized in the diagram
\begin{equation}
\begin{xy}
\xymatrix@C=20mm{
  E_1^{-3,3} \ar[r]^{\dd_1} & E_1^{-2,3} \ar[r]^{\dd_1} & E_1^{-1,3}
  \ar[r]^{\dd_1} & E_1^{0,3} \\
  E_1^{-3,2} \ar[r]^{\dd_1} & E_1^{-2,2} \ar[r]^{\dd_1} & E_1^{-1,2}
  \ar[r]^{\dd_1} & E_1^{0,2}\\
  E_1^{-3,1} \ar[r]^{\dd_1} & E_1^{-3,1} \ar[r]^{\dd_1} & E_1^{-1,1}
  \ar[r]^{\dd_1} & E_1^{0,1}\\
  E_1^{-3,0} \ar[r]^{\dd_1} & E_1^{-2,0} \ar[r]^{\dd_1} & E_1^{-1,0}
  \ar[r]^{\dd_1} & E_1^{0,0}\\
}
\save="x"!LD+<-3mm,0pt>;"x"!RD+<5mm,0pt>**\dir{-}?>*\dir{>}\restore
\save="x"!RD+<0pt,-3mm>;"x"!RU+<0pt,2mm>**\dir{-}?>*\dir{>}\restore
\save!CD+<0mm,-4mm>*{p}\restore \save!CR+<+3mm,0mm>*{q}\restore
\end{xy}  \label{spectralsequence}
\end{equation}
where all other entries are zero for dimensional reasons and
the only nonvanishing differential
\begin{equation}
\dd_1 \,:\, E_1^{p,q} ~\longrightarrow~ E_1^{p+1,q}
\end{equation}
is determined by the morphisms in the Koszul complex
(\ref{Koszul}). The double complex (\ref{spectralsequence}) is an
object of the derived category. Our goal in the following is to
reduce it to an object of the stable category of coherent sheaves on
$\PP^3$. Physically, this can be thought of as the stabilization of a
topological B-model brane to a Type~IIA D6-brane wrapping $\PP^3$.

The last ingredient we will need comprises a few short exact sequences
that will play an important role in explicit computations. They are
given by
\begin{eqnarray}
\xymatrix@1{
  0 \ar[r] & \mathcal{O}_{\mathbb P^3} (-1) \ar[r]^{~~\times z_0} &
  \mathcal{O}_{\mathbb P^3}
  \ar[r]^{z_0=0} & \mathcal{O}_{p_{\infty}} \ar[r] & 0 \ ,
} \label{boundary} \\[4pt]
\xymatrix@1{
  0 \ar[r] & \Omega^1_{\mathbb P^3} \ar[r] & \mathcal{O}_{\mathbb P^3}
  (-1)^{\oplus 4}
  \ar[r] & \mathcal{O}_{\mathbb P^3} \ar[r] & 0 \ ,
} \label{O1} \\[4pt]
\xymatrix@1{
  0 \ar[r] & \Omega^2_{\mathbb P^3} \ar[r] & \mathcal{O}_{\mathbb P^3}
  (-2)^{\oplus 6}
  \ar[r] & \Omega^1_{\mathbb P^3} \ar[r] & 0 \ ,
} \label{O1O2} \\[4pt]
\xymatrix@1{
  0 \ar[r] & \Omega^3_{\mathbb P^3}
  \ar[r] & \mathcal{O}_{\mathbb P^3} (-3)^{\oplus 4}
  \ar[r] & \Omega^2_{\mathbb P^3} \ar[r] & 0 \ .
} \label{O2}
\end{eqnarray}
The first sequence defines the plane at infinity $p_{\infty}=
{[0:z_1:z_2:z_3]} \cong \mathbb P^2 $, while the other three sequences
are the Euler sequences for differential forms on $\PP^3$ obtained
via truncation of the Koszul complex.

The final stage consists in imposing suitable boundary conditions on
the sheaf $\mathcal{E}$. The appropriate boundary conditions were
found in~\cite{Iqbal:2003ds} and consist in imposing that the sheaf
$\mathcal{E}$ restricted to the projective plane $\PP^2$ at infinity
corresponds to a four-dimensional instanton, viewed as an asymptote of
the corresponding three-dimensional partition. By this we mean that it
has the same cohomological properties as the torsion free sheaves on
$\PP^2$ studied in~\cite{nakajima}. This condition should be imposed
separately along each of the three complex ``directions'' in
$\complex^3$, i.e. the six-dimensional generalized instantons should
behave in a different way when $z_i \rightarrow \infty$, $i=1,2,3$,
corresponding to generically distinct two-dimensional Young diagrams
(partitions). Since $\complex^3 \cong
\PP^3 / \PP^2$, the three corresponding projective planes at infinity
are given in homogeneous coordinates by
$p_\infty^1={[z_0:0:z_2:z_3]}$, $p_\infty^2={[z_0:z_1:0:z_3]}$ and
$p_\infty^3={[z_0:z_1:z_2:0]}$, and each one contains a line
$\ell_\infty^i$ at infinity defined by $z_0=0$. We then impose the
boundary condition that $\mathcal{E}$ is trivial {\it separately} on
each one of these three lines. With this choice of trivializations one
automatically has $c_1(\Ecal)=0$, and the sheaves
$\Ecal|_{p^i_\infty}$ have quantum numbers 
\beq
c_1\big(\Ecal|_{p^i_\infty}\big)=0 \qquad \mbox{and} \qquad
{\rm ch}_2\big(\Ecal|_{p^i_\infty}\big)=
c_2\big(\Ecal|_{p^i_\infty}\big)=k_i \ .
\label{EcalP2}\eeq
These are torsion free sheaves on $\PP^2$ of rank $N$ trivialized on a
line $\ell_\infty^i$ which represent framed four-dimensional
instantons of charge $k_i$. By T-duality, they correspond to D6--D2
bound states on $\PP^3$ of D2-brane charge $k_i$.

The vacuum moduli space (\ref{sheavesonP3}) is recovered by setting
$k_i=\int_{p_\infty^i}\,{\rm ch}_2(\Ecal)=0$. These are the boundary
conditions appropriate to the gauge theory on $\complex^3$ and will be
the case studied in this section and in Section~\ref{TopMQM}. In
Section~\ref{Nonabpartfn} we will need the more general non-trivial
asymptotics (\ref{EcalP2}). We will now impose this boundary condition
in three steps, dealing with the different terms of the spectral
sequence. A closely related moduli space of instanton sheaves has been rigorously constructed recently by Jardim in~\cite{jardim0}.

\subsection{Homological algebra}

The first step is to determine $E_1^{-3,q}$ and $E_1^{0,q}$. These
terms can be treated simultaneously because $\Omega_{\PP^3}^3 \cong
\mathcal{O}_{\PP^3}(-4)$. Let us tensor the sequence (\ref{boundary})
by $\mathcal{E}(-r)$ to get
\begin{equation}
\xymatrix@1{
  0 \ar[r] & \mathcal{E}(-r-1) \ar[r] & \mathcal{E}(-r)
  \ar[r] & \mathcal{E}(-r)\big|_{p_{\infty}} \ar[r] & 0 \ ,
}
\end{equation}
where we have used the fact that $\mathcal{O}_{\PP^2}$ is a locally
free sheaf to set ${\rm
  Tor}^1(\mathcal{E}(-r)|_{p_\infty},\mathcal{O}_{p_\infty})=0$. Applying
the snake lemma one finds the associated long exact sequence in
cohomology
\begin{equation} \label{LSboundary}
\xymatrix@1{
  0 \ar[r] & H^0 \left( \PP^3 \,,\, \mathcal{E}(-r-1) \right)
  \ar[r] &  H^0 \left( \PP^3 \,,\, \mathcal{E}(-r) \right)
  \ar[r] & H^0 \big( \PP^2 \,,\, \mathcal{E}(-r)\big|_{p_\infty}\,
  \big) \\ \vspace{-5mm}
\ar[r] & H^1 \left( \PP^3 \,,\, \mathcal{E}(-r-1) \right)
  \ar[r] &  H^1 \left( \PP^3 \,,\, \mathcal{E}(-r) \right)
  \ar[r] & H^1 \big( \PP^2 \,,\, \mathcal{E}(-r)\big|_{p_\infty}\,
  \big) \\
\ar[r] & H^2 \big( \PP^3 \,,\, \mathcal{E}(-r-1) \big)
  \ar[r] &  H^2 \left( \PP^3 \,,\, \mathcal{E}(-r) \right)
  \ar[r] & H^2 \big( \PP^2 \,,\, \mathcal{E}(-r)\big|_{p_\infty}\,
  \big) \\
\ar[r] & H^3 \left( \PP^3 \,,\, \mathcal{E}(-r-1) \right)
  \ar[r] &  H^3 \left( \PP^3 \,,\, \mathcal{E}(-r) \right)
  \ar[r] & 0  \ .
}
\end{equation}

Since $\mathcal{E}|_{p_{\infty}}$ is a sheaf on $\PP^2$ which is
trivial when restricted on a projective line $\PP^1 \subset \PP^2$,
we are in the same situation as in~\cite[Chapter~2]{nakajima}. It
follows that
\begin{eqnarray} \label{nakaconditions}
& H^0 \big( \PP^2 \,,\, \mathcal{E}(-r)\big|_{p_\infty}\, \big) = 0
\qquad \mathrm{for} \qquad r \ge 1 \ , \\[4pt]
& H^2 \big( \PP^2 \,,\, \mathcal{E}(-r)\big|_{p_\infty}\, \big) = 0
\qquad \mathrm{for} \qquad r \le 2 \ .
\end{eqnarray}
By imposing this condition on the cohomology long exact sequence
(\ref{LSboundary}) we find
\begin{eqnarray}
& H^3 \left( \PP^3 \,,\, \mathcal{E}(-r-1) \right)= H^3 \left( \PP^3
  \,,\,  \mathcal{E}(-r) \right) \qquad \mathrm{for} \qquad r \le 2  \ ,
\label{H3r20}\\[4pt]
& H^0 \left( \PP^3 \,,\, \mathcal{E}(-r-1) \right) = H^0 \left( \PP^3
\,,\, \mathcal{E}(-r) \right) \qquad \mathrm{for} \qquad r \ge 1 \ .
\end{eqnarray}
By Serre's vanishing theorem~\cite{okonek},
$H^q(\PP^3,\mathcal{E}(m))=0$ is trivial for $m\gg0$ and $q \neq
0$. Combined with Serre duality this implies
\begin{eqnarray} \label{E^0q}
& H^3 \left( \PP^3 \,,\, \mathcal{E}(-r) \right) = 0 \qquad
\mathrm{for} \qquad r \le 3 \ , \\[4pt]
& H^0 \left( \PP^3 \,,\, \mathcal{E}(-r) \right) = 0 \qquad
\mathrm{for} \qquad r \ge 1 \ . \label{E0qH0}
\end{eqnarray}

The second step is to determine $E_1^{-1,q}$. Let us start by
tensoring the exact sequence (\ref{boundary}) with $\mathcal{E}(-r)
\otimes\Omega^1_{\PP^3}(1)$ to get
\begin{equation}
\xymatrix@1{
  0 \ar[r] & \Omega^1_{\PP^3}(1) \otimes \mathcal{E}(-r-1) \ar[r] &
  \Omega^1_{\PP^3}(1) \otimes \mathcal{E}(-r)
  \ar[r] & \left( \Omega^1_{\PP^3}(1) \otimes
  \mathcal{E}(-r) \right)\big|_{p_{\infty}} \ar[r] & 0 \ .
}
\end{equation}
Since $\left( \Omega^1_{\PP^3}(1) \otimes \mathcal{E}(-r)
\right)\big|_{p_{\infty}}$ is a sheaf defined on $p_{\infty} \cong
\PP^2$ which trivializes on a line $\PP^1 \subset \PP^2$, we are again
exactly in the situation of~\cite[Chapter~2]{nakajima} and hence
\begin{eqnarray} \label{conditionsonQ}
& H^0 \big( \PP^2 \,,\, ( \Omega^1_{\PP^3}(1) \otimes
\mathcal{E}(-r) )\big|_{p_{\infty}}\, \big) = 0 \qquad
\mathrm{for} \qquad r \ge 1  \ , \nonumber\\[4pt] & H^2 \big( \PP^2
  \,,\,  (\Omega^1_{\PP^3}(1) \otimes \mathcal{E}(-r) )\big|_{p_{\infty}}
\,\big) = 0 \qquad \mathrm{for} \qquad r \le 1 \ .
\end{eqnarray}
{}From the corresponding long exact sequence in cohomology we have
\begin{equation}
\xymatrix@1{
  H^2 \big( \PP^2 \,,\, ( \Omega^1_{\PP^3}(1) \otimes
  \mathcal{E}(-r))\big|_{p_{\infty}}\, \big)
\ar[r] & H^3 \left( \PP^3 \,,\, \Omega^1_{\PP^3}(1) \otimes
\mathcal{E}(-r-1) \right) &  \\
  \ar[r] &  H^3 \left( \PP^3 \,,\, \Omega^1_{\PP^3}(1) \otimes
    \mathcal{E}(-r) \right)
  \ar[r] & 0  \ .
}
\end{equation}
By imposing the conditions (\ref{conditionsonQ}) and (\ref{H3r20}) we
get
\be
H^3 \left( \PP^3 \,,\, \Omega^1_{\PP^3}(1) \otimes \mathcal{E}(-r)
\right) = 0 \qquad \mathrm{for} \qquad r \le 2 \ .
\label{E^-1q}\ee

We get other conditions by tensoring the Euler sequence for one-forms
(\ref{O1}) with $\mathcal{E}(-r)$ to get
\begin{equation}
\xymatrix@1{
  0 \ar[r] & \mathcal{E}(-r) \otimes \Omega^1_{\PP^3}(1) \ar[r] &
  \mathcal{E}(-r)^{\oplus 4}
  \ar[r] & \mathcal{E}(-r+1) \ar[r] & 0 \ .
}
\end{equation}
The corresponding long exact cohomology sequence is
\begin{equation} \label{LSO1}
\xymatrix@1{
  0 \ar[r] & H^0 \left( \PP^3 \,,\, \mathcal{E}(-r) \otimes
    \Omega^1_{\PP^3}(1) \right)
  \ar[r] &  H^0 \left( \PP^3 \,,\, \mathcal{E}(-r) \right)^{\oplus 4}
  \ar[r] &  H^0 \left( \PP^3 \,,\, \mathcal{E}(-r+1) \right) \\
\ar[r] & H^1 \left( \PP^3 \,,\, \mathcal{E}(-r) \otimes
\Omega^1_{\PP^3}(1) \right)
  \ar[r] &  H^1 \left( \PP^3 \,,\, \mathcal{E}(-r) \right)^{\oplus 4}
  \ar[r] &  H^1 \left( \PP^3 \,,\, \mathcal{E}(-r+1) \right) \\
\ar[r] & H^2 \left( \PP^3 \,,\, \mathcal{E}(-r) \otimes
\Omega^1_{\PP^3}(1) \right)
  \ar[r] &  H^2 \left( \PP^3 \,,\, \mathcal{E}(-r) \right)^{\oplus 4}
  \ar[r] &  H^2 \left( \PP^3 \,,\, \mathcal{E}(-r+1) \right) \\
\ar[r] & H^3 \left( \PP^3 \,,\, \mathcal{E}(-r)  \otimes
\Omega^1_{\PP^3}(1) \right)
  \ar[r] &  H^3 \left( \PP^3 \,,\, \mathcal{E}(-r) \right)^{\oplus 4}
  \ar[r] &  H^3 \left( \PP^3 \,,\, \mathcal{E}(-r+1) \right)
  \ar[r] & 0 \ .
}
\end{equation}
{}From the first line of this sequence, by imposing (\ref{E0qH0}) we
get
\begin{equation}
H^0 \left( \PP^3 \,,\, \Omega^1_{\PP^3}(1) \otimes \mathcal{E}(-r)
\right) = 0 \qquad \mbox{for} \qquad r \ge 1 \ .
\label{H0Om1}\end{equation}

The third and final step is to determine $E_2^{-2,q}$. {}From the Euler
sequence for three-forms (\ref{O2}) we get
\begin{equation}
\xymatrix@1{
  0 \ar[r] & \mathcal{E}(-r-2) \ar[r] &
  \mathcal{E}(-r-1)^{\oplus 4}
  \ar[r] &
  \mathcal{E}(-r) \otimes \Omega^2_{\PP^3}(2)  \ar[r] & 0  \ .
}
\end{equation}
Let us look at the corresponding long exact sequence in cohomology. It
contains, in particular, the line
\begin{equation}
\xymatrix@1{
  0 \ar[r] & H^3 \left( \PP^3 \,,\, \mathcal{E}(-r-2) \right) \ar[r] &
  H^3 \left( \PP^3 \,,\, \mathcal{E}(-r-1) \right)^{\oplus 4}
  \\ \ar[r]  &  H^3 \left( \PP^3 \,,\,
  \mathcal{E}(-r) \otimes \Omega^2_{\PP^3}(2) \right)  \ar[r] & 0 \ .
}
\end{equation}
By imposing the condition (\ref{E^0q}) one gets
\begin{equation}
H^3 \left( \PP^3 \,,\,
  \mathcal{E}(-r) \otimes \Omega^2_{\PP^3}(2) \right) = 0 \qquad
\mbox{for} \qquad r \le 2 \ .
\end{equation}

The only short exact sequence that we haven't yet used is the Euler
sequence for two-forms (\ref{O1O2}) from which it follows that
\begin{equation}
\xymatrix@1{
  0 \ar[r] & \mathcal{E}(-r) \otimes \Omega^2_{\PP ^3} \ar[r] &
  \mathcal{E}(-r-2)^{\oplus 6}
  \ar[r] &
  \mathcal{E}(-r) \otimes \Omega^1_{\PP^3} \ar[r] & 0  \ .
}
\end{equation}
The corresponding long exact cohomology sequence yields
\begin{equation} \label{LSO1O2}
\xymatrix@1{
  0 \ar[r] & H^0 \left( \PP^3 \,,\, \mathcal{E}(-r) \otimes \Omega^2_{\PP
      ^3} \right)
  \ar[r] &  H^0 \left( \PP^3 \,,\, \mathcal{E}(-r-2) \right)^{\oplus 6}
  \ar[r] &  H^0 \left( \PP^3 \,,\, \mathcal{E}(-r) \otimes
    \Omega^1_{\PP^3} \right) \\
\ar[r] & H^1 \left( \PP^3 \,,\, \mathcal{E}(-r) \otimes \Omega^2_{\PP
^3} \right)
  \ar[r] &  H^1 \left( \PP^3 \,,\, \mathcal{E}(-r-2) \right)^{\oplus 6}
  \ar[r] &  H^1 \left( \PP^3 \,,\, \mathcal{E}(-r) \otimes
    \Omega^1_{\PP^3} \right) \\
\ar[r] & H^2 \left( \PP^3 \,,\, \mathcal{E}(-r) \otimes \Omega^2_{\PP
^3} \right)
  \ar[r] &  H^2 \left( \PP^3 \,,\, \mathcal{E}(-r-2) \right)^{\oplus 6}
  \ar[r] &  H^2 \left( \PP^3 \,,\, \mathcal{E}(-r) \otimes
    \Omega^1_{\PP^3} \right) \\
\ar[r] & H^3 \left( \PP^3 \,,\, \mathcal{E}(-r) \otimes \Omega^2_{\PP
^3} \right)
  \ar[r] &  H^3 \left( \PP^3 \,,\, \mathcal{E}(-r-2) \right)^{\oplus 6}
   &   \\ & \ar[r] &  H^3 \left( \PP^3 \,,\, \mathcal{E}(-r)
     \otimes \Omega^1_{\PP^3} \right)
  \ar[r] & 0 \ .
}
\end{equation}
By using (\ref{E0qH0}) and (\ref{H0Om1}) we get
\begin{equation}
H^0 \left( \PP^3 \,,\,
  \mathcal{E}(-r) \otimes \Omega^2_{\PP^3}(2) \right) = 0 \qquad
\mbox{for} \qquad r \ge 1 \ .
\end{equation}

If we collect all the results obtained so far, we see that the
values $r=1,2$ are rather special since then all the cohomology groups
$H^0$ and $H^3$ vanish and the spectral sequence becomes
\begin{equation}
\begin{xy}
\xymatrix@C=20mm{
  0  & 0  & 0  & 0 \\
  E_1^{-3,2} \ar[r]^{\dd_1} & E_1^{-2,2} \ar[r]^{\dd_1} & E_1^{-1,2}
  \ar[r]^{\dd_1} & E_1^{0,2}\\
  E_1^{-3,1} \ar[r]^{\dd_1} & E_1^{-3,1} \ar[r]^{\dd_1} & E_1^{-1,1}
  \ar[r]^{\dd_1} & E_1^{0,1}\\
  0  & 0  & 0  & 0 \\
}
\save="x"!LD+<-3mm,0pt>;"x"!RD+<5mm,0pt>**\dir{-}?>*\dir{>}\restore
\save="x"!RD+<0pt,-3mm>;"x"!RU+<0pt,2mm>**\dir{-}?>*\dir{>}\restore
\save!CD+<0mm,-4mm>*{p}\restore \save!CR+<+3mm,0mm>*{q}\restore
\end{xy}  \label{spectralsequence2}
\end{equation}
A great deal of information can now be obtained with the help of the
Riemann--Roch theorem which computes the Euler character
\begin{equation}
\chi (\mathcal{F}) = \sum_{q=0}^3\, (-1)^q \,\dim_\complex H^q \left(
  \PP^3\,,\, \mathcal{F} \right) = \int_{\PP^3}\,
\mathrm{ch}(\mathcal{F}) \wedge\mathrm{td}\big(\PP^3\big)
\label{RRformula}\end{equation}
for any sheaf $\mathcal{F}$ on $\PP^3$. We are
interested in the cases where $\mathcal{F}$ is $\mathcal{E}(-r)$,
$\mathcal{E}(-r) \otimes \Omega^1_{\PP^3} (1)$ and $\mathcal{E}(-r)
\otimes \Omega^2_{\PP^3} (2)$. Using (\ref{sheavesonP3}) we
parametrize the Chern character of $\Ecal$ as 
\begin{equation}
\mathrm{ch} (\mathcal{E}) = N + k\, \xi^3
\label{ChernE}\end{equation}
where $\xi=c_1(\mathcal{O}_{\PP^3}(1))$ is the hyperplane class which
generates the complex cohomology ring of the projective space
$\PP^3$.

The pertinent values of the Chern and Todd characteristic classes may
be computed from the exact sequences (\ref{boundary})--(\ref{O2}) and
are given by
\begin{eqnarray}
\mathrm{ch}\big(\mathcal{O}_{\PP^3}(-r)\big) &=&
\e^{- r\, \xi} \ , \nonumber\\[4pt]
\mathrm{ch}\big(\Omega^1_{\PP^3}\big) &=&
3-4 \xi + 2 \xi^2 - \mbox{$\frac{2}{3}$}\,\xi^3
\ , \nonumber\\[4pt] \mathrm{ch}\big(\Omega^2_{\PP^3}\big) &=&
3-8 \xi+10 \xi^2 -\mbox{$\frac{22}{3}$}\,
\xi^3 \ , \nonumber\\[4pt]  \mathrm{td}\big(\PP^3\big) &=& 1 + 2 \xi
+\mbox{$\frac{11}{6}$}\, \xi^2 + \xi^3
\ .
\label{ChernTodds}\end{eqnarray}
To apply the Riemann--Roch formula (\ref{RRformula}) we use the
multiplicativity property of the Chern character $\mathrm{ch}(\Ecal_1
\otimes\Ecal_2) = \mathrm{ch}(\Ecal_1)\wedge\mathrm{ch}(\Ecal_2)$ and
the fact that only the top form components $\xi^3$ survive the
integration over $\PP^3$. We will only need to deal with the case
$r=2$ explicitly below, for which the relevant results are
\begin{eqnarray} 
\chi\big(\mathcal{E}(-3)\big) &=& k \ , \nonumber\\[4pt]
\chi\left(\mathcal{E}(-2)
\otimes \Omega^2_{\PP^3}(2)\right) &=& 3 k+N \ , \nonumber\\[4pt]
\chi\left(\mathcal{E}(-2) \otimes \Omega^1_{\PP^3}(1)\right) &=& 3 k \ ,
\nonumber\\[4pt] \chi\big(\mathcal{E}(-2)\big) &=& k \ .
\label{Eulerk=2}\end{eqnarray}

\subsection{Nonlinear Beilinson monad\label{Monad}}

We will now impose an additional condition on the sheaf $\mathcal{E}$
which ensures that the spectral sequence becomes a four-term complex
and that the Euler characteristics computed in eq.~(\ref{Eulerk=2})
above are the stable dimensions (rather than virtual dimensions) of
the vector spaces which appear in the complex. The condition we impose
kills all the $H^1$ cohomology groups and simplifies the spectral
sequence in the case $r=2$. It is given by
\begin{equation} \label{ExtraCondition}
H^1 \left( \PP^3 \,,\, \mathcal{E}(-2) \right) = 0 \ .
\end{equation}
This is the same condition which is placed on the holomorphic bundles
over $\PP^3$ that are used in the twistor construction of framed
instantons in four dimensions~\cite{donaldson}. It is also similar to
the one defining the admissable sheaves of~\cite{jardim}.

If we look at the long exact sequence (\ref{LSboundary}) for $r=2$
with the condition (\ref{nakaconditions}) and impose
(\ref{ExtraCondition}), we immediately see that
\begin{equation}
H^1 \left( \PP^3 \,,\, \mathcal{E}(-3) \right) = 0 \ .
\end{equation}
Moreover, if we consider the sequence (\ref{LSO1}) with $r=2$ and
impose both (\ref{E0qH0}) and (\ref{ExtraCondition}) we find
\begin{equation}
H^1 \left( \PP^3 \,,\, \mathcal{E}(-2) \otimes \Omega^1_{\PP^3} (1)
\right) = 0 \ .
\end{equation}
Finally, let us look at the cohomology sequence (\ref{LSO1O2}) for
$r=0$. By imposing (\ref{ExtraCondition}) as well as (\ref{H0Om1}) one
finally finds
\begin{equation}
H^1 \left( \PP^3 \,,\, \mathcal{E}(-2) \otimes \Omega^2_{\PP^3}
\right) = 0 \ .
\end{equation}

The condition (\ref{ExtraCondition}) thus automatically kills
the $E_1^{p,1}$ line and the spectral sequence reduces to
\begin{equation}
\begin{xy}
\xymatrix@C=20mm{
  0  & 0  & 0  & 0 \\
  E_1^{-3,2} \ar[r]^{\dd_1} & E_1^{-2,2} \ar[r]^{\dd_1} & E_1^{-1,2}
  \ar[r]^{\dd_1} & E_1^{0,2}\\
  0 & 0 & 0 & 0 \\
  0  & 0  & 0  & 0 \\
}
\save="x"!LD+<-3mm,0pt>;"x"!RD+<5mm,0pt>**\dir{-}?>*\dir{>}\restore
\save="x"!RD+<0pt,-3mm>;"x"!RU+<0pt,2mm>**\dir{-}?>*\dir{>}\restore
\save!CD+<0mm,-4mm>*{p}\restore \save!CR+<+3mm,0mm>*{q}\restore
\end{xy}  \label{spectralsequence3}
\end{equation}
This is our candidate for a generalized ADHM-like complex. The
cohomology computations above also serve to show that the spectral
sequence degenerates at the $E_2$-term. If we remember the Beilinson
theorem, then we can write the cohomology of the differential complex
in (\ref{spectralsequence3}) as
\begin{equation}
\begin{xy}
\xymatrix@C=20mm{
  0  & 0  & 0  & 0 \\
  E_{\infty}^{-3,2} =0  & E_{\infty}^{-2,2} = \mathcal{E}(-2) &
  E_{\infty}^{-1,2}=0 & E_{\infty}^{0,2}=0 \\
  0 & 0 & 0 & 0 \\
  0  & 0  & 0  & 0 \\
}
\save="x"!LD+<-3mm,0pt>;"x"!RD+<5mm,0pt>**\dir{-}?>*\dir{>}\restore
\save="x"!RD+<0pt,-3mm>;"x"!RU+<0pt,2mm>**\dir{-}?>*\dir{>}\restore
\save!CD+<0mm,-4mm>*{p}\restore \save!CR+<+3mm,0mm>*{q}\restore
\end{xy}
\label{specseqfinal}\end{equation}

Equivalently, we have the complex
\begin{equation}
\xymatrix@1{
  0 \ar[r] & V \otimes \mathcal{O}_{\PP^3} (-3) \ar[r]^a &
  {B} \otimes \mathcal{O}_{\PP^3} (-2)
  \ar[r]^b &
  {C} \otimes \mathcal{O}_{\PP^3} (-1) \ar[r]^c &
  {D} \otimes \mathcal{O}_{\PP^3} (0) \ar[r] & 0 \ ,
} \label{finalcomplexforP3}
\end{equation}
where we have defined the complex vector spaces
\bea
V&=& H^2\big(
\PP^3 \,,\, \mathcal{E}(-3)\big) \ , \nonumber\\[4pt] B&=&H^2 \big(
\PP^3 \,,\, \mathcal{E}(-2)\otimes\Omega_{\PP^3}^2(2)\big) \ ,
\nonumber\\[4pt]
C&=&H^2 \big(
\PP^3 \,,\, \mathcal{E}(-2)\otimes\Omega_{\PP^3}^1(1)\big) \ ,
\nonumber\\[4pt] D&=&H^2 \big(
\PP^3 \,,\, \mathcal{E}(-2)\big) \ .
\label{complexvecsp}\eea
It is straightforward to show that there is a natural identification
$D=V$. Using the vanishing cohomology groups above, the long
exact sequence (\ref{LSboundary}) truncates for $r=2$ to
\beq
\xymatrix@1{
  0 \ar[r] & H^1 \big( \PP^2 \,,\, \mathcal{E}(-2)\big|_{p_\infty}\,
  \big) \ar[r] & H^2 \big( \PP^3 \,,\, \mathcal{E}(-3) \big)
  \ar[r] &  H^2 \left( \PP^3 \,,\, \mathcal{E}(-2) \right)
  \ar[r] & 0  \ .
}
\label{longtruncate}\eeq
Since the four-dimensional instanton numbers in (\ref{EcalP2}) are
given by~\cite{nakajima} $k_i=\dim_\complex
H^1(\PP^2,\Ecal(-2)|_{p_\infty^i})$, for the vacuum moduli space
(\ref{sheavesonP3}) one has $H^1(\PP^2,\Ecal(-2)|_{p_\infty})=0$ and
hence
\beq
H^2 \big( \PP^3 \,,\, \mathcal{E}(-3) \big)
  =  H^2 \left( \PP^3 \,,\, \mathcal{E}(-2) \right) \ .
\label{H223iso}\eeq
A similar argument provides a natural identification
\beq
C=V\oplus V\oplus V \ .
\label{CVid}\eeq
Indeed, using (\ref{LSboundary}) with $r=1$ identifies
$V=H^2(\PP^3,\Ecal(-1))$, since $H^2(\PP^2,\Ecal(-1)|_{p_\infty})=0$
exactly as above~\cite{nakajima}. Then the exact sequence (\ref{LSO1})
with $r=2$ implies (\ref{CVid}).

By (\ref{specseqfinal}), the cohomology of the complex
(\ref{finalcomplexforP3}) gives the sheaf $\mathcal{E}(-2)$. To get a
complex whose cohomology is simply $\mathcal{E}$ all we have to do is
tensor by $\mathcal{O}_{\PP^3} (2)$ to get the equivalent complex
\begin{equation}
\xymatrix@1{
  0 \ar[r] & V \otimes \mathcal{O}_{\PP^3} (-1) \ar[r]^{~~~a} &
  {B} \otimes \mathcal{O}_{\PP^3}
  \ar[r]^{\!\!\!\!\!\!\!b} &
  {V^{\oplus3}} \otimes \mathcal{O}_{\PP^3} (1) \ar[r]^{~~c} &
  {V} \otimes \mathcal{O}_{\PP^3} (2) \ar[r] & 0  \ .
}
\label{ADHMcomplexfinal}\end{equation}
The crucial point is that the complex vector spaces appearing here are
\emph{finite dimensional}. If we look at eqs. (\ref{Eulerk=2}), then
one has $\dim_\complex V=k$ and $\dim_\complex B=3k+N$. 

Looking back at eq.~(\ref{specseqfinal}), we see that the only
non-trivial cohomology group of the complex (\ref{ADHMcomplexfinal})
is the coherent sheaf
\be
\mathcal{E}=\underline{\ker}(b)\,\big/\,\underline{\rm im}(a) \ .
\label{cohsheafcoh}\ee
In particular, the map $c$ is an epimorphism, $a$ is a monomorphism,
and $\underline{\ker}(c)=\underline{\rm im}(b)$. Using this last
condition, we can truncate (\ref{ADHMcomplexfinal}) to a three-term
complex by replacing ${V^{\oplus3}} \otimes \mathcal{O}_{\PP^3}(1)$
with the locally free kernel sheaf
$\mathcal{K}_c:=\underline{\ker}(c)$, which describes the sheaf
$\mathcal{E}$ as the cohomology of a \emph{nonlinear monad}
\be
\xymatrix@1{
\mathcal{M}^\bullet~:~  0 \ar[r] & V \otimes \mathcal{O}_{\PP^3} (-1)
\ar[r]^{~~~a} &
  {B} \otimes \mathcal{O}_{\PP^3}
  \ar[r]^{~~b} & \mathcal{K}_c\ar[r] & 0
}
\label{ADHMmonad}\ee
with $\dim_\complex V=k$ and $\dim_\complex
B=3k+N$, for which the map $b$ is an epimorphism and $a$ is a
monomorphism with $b\,a=0$. 

Since $\Ecal$ is the only nonvanishing cohomology of the complex
(\ref{ADHMmonad}), we can compute its Chern character through
\beq
{\rm ch}(\Ecal)={\rm ch}(B\otimes\mathcal{O}_{\PP^3})-{\rm ch}
\big(V \otimes \mathcal{O}_{\PP^3} (-1)\big)-{\rm ch}(\mathcal{K}_c) \
.
\label{chEcalKc}\eeq
By comparing with (\ref{ChernE}) and (\ref{ChernTodds}), we see that
the bundle $\mathcal{K}_c$ on $\PP^3$ has non-trival
characteristic classes in all degrees determined entirely by the
instanton number $k$. In particular, one has
\beq
{\rm rank}(\mathcal{K}_c)=2k \qquad \mbox{and} \qquad
c_1(\mathcal{K}_c)=k \ .
\label{rankc1Kc}\eeq
One also finds from (\ref{chEcalKc}) that $\mathcal{K}_c$ cannot be of
the form $K\otimes\mathcal{O}_{\PP^3}(1)$ for some $2k$-dimensional
hermitean vector space $K$, reflecting the nonlinearity
of the monad (\ref{ADHMmonad}).

One can adapt the proof of~\cite[Lemma~4.1.3]{okonek} to show that the
complex (\ref{ADHMcomplexfinal}) determines the sheaf $\Ecal$ uniquely
up to isomorphism. This is equivalent to the requirement that
\beq
{\rm Ext}^q\big(\mathcal{K}_c\,,\,\mathcal{O}_{\PP^3}(-1)\big)=
{\rm Ext}^q(\mathcal{K}_c,\mathcal{O}_{\PP^3})=0
\label{ExtKc0}\eeq
for all $q$, which follows from the fact that the kernel sheaf is
locally free and guarantees that the nonlinear monad
(\ref{ADHMmonad}) is unique up to isomorphism. In the following we
will find it more convenient to work instead with the
equivalent four term complex (\ref{ADHMcomplexfinal}).

\subsection{Barth description of nonlinear monads\label{Barth}}

We will now give a description of the moduli space
(\ref{sheavesonP3}), under the conditions spelled out above, in terms
of linear algebra by generalizing the Barth description of linear
monads~\cite{donaldson}. This will contain as a special
case the matrix equations (\ref{ADHMeqs}) describing the
six-dimensional noncommutative instantons. For this, we regard the
morphism $c$ in (\ref{ADHMcomplexfinal}) as an element of
$(V^*)^{\oplus3}\otimes V\otimes\widetilde{W}$, where
$\widetilde{W}=H^0(\PP^3,\mathcal{O}_{\PP^3}(1))$ is the vector space
spanned by the homogeneous coordinates $z_0,z_1,z_2,z_3$ of
$\PP^3$. The map $c$ can then be written as
\be
c=c_0\,z_0+c_1\,z_1+c_2\,z_2+c_3\,z_3 \ ,
\label{cmapz}\ee
where $c_i:V^{\oplus3}\to V$ are constant linear maps. Similarly,
we can represent the morphisms $a$ and $b$ of the four-term complex
(\ref{ADHMcomplexfinal}) as
\be
a=a_0\,z_0+a_1\,z_1+a_2\,z_2+a_3\,z_3 \qquad \mbox{and} \qquad
b=b_0\,z_0+b_1\,z_1+b_2\,z_2+b_3\,z_3
\label{abmapsz}\ee
where $a_i:V\to B$ and $b_i:B\to V^{\oplus3}$ are constant linear
maps. The monadic condition $b\,a=0$ then implies that
\beq
b_i\,a_i=0 \qquad \mbox{and} \qquad b_i\,a_j+b_j\,a_i=0
\label{monadiccondab}\eeq
for each $i,j=0,1,2,3$ with $i<j$. On the other hand, the condition
$\underline{\ker}(c)=\underline{\rm im}(b)$ implies that
\beq
\ker(c_i)={\rm im}(b_i) \qquad \mbox{and} \qquad c_i\,b_j+c_j\,b_i=0
\ .
\label{cohtrivcondcb}\eeq

Restricting the complex (\ref{ADHMcomplexfinal}) to the line at
infinity $\ell_\infty=[z_0:z_1:0:0]\cong\PP^1\subset\PP^3$ we get
\be
\xymatrix@1{
0 \ar[r] & V \otimes \mathcal{O}_{\PP^3} (-1)\big|_{\ell_\infty}
\ar[r]^{~~a_\infty} &
  {B} \otimes \mathcal{O}_{\PP^3}\big|_{\ell_\infty}
  \ar[r]^{\!\!\!\!\!b_\infty} &
  {V^{\oplus3}} \otimes \mathcal{O}_{\PP^3} (1)\big|_{\ell_\infty}
  \ar[r]^{c_\infty} & 
  {V} \otimes \mathcal{O}_{\PP^3} (2)\big|_{\ell_\infty} \ar[r] & 0
}
\label{monadinfty}\ee
where $a_\infty=a_0\,z_0+a_1\,z_1$, $b_\infty=b_0\,z_0+b_1\,z_1$ and
$c_\infty=c_0\,z_0+c_1\,z_1$. The kernel sheaf
$\mathcal{K}_b=\underline{\ker}(b)$ is locally free, and one has the
exact sequence of sheaves 
\be
\xymatrix@1{
0 \ar[r] & V \otimes \mathcal{O}_{\PP^3} (-1)\big|_{\ell_\infty}
\ar[r]^{~~~~~~a_\infty} &
  \mathcal{K}_b\big|_{\ell_\infty}
  \ar[r] &
  \mathcal{E}\big|_{\ell_\infty} \ar[r] & 0
}
\label{exactseqkerb}\ee
where the third arrow comes from the projection onto
(\ref{cohsheafcoh}). Using the associated long exact sequence in
cohomology, along with $H^q(\PP^1,\mathcal{O}_{\PP^1}(-1))=0$,
$q=1,2$ and
$\mathcal{E}|_{\ell_\infty}\cong\mathcal{O}_{\ell_\infty}^{\oplus N}$,
we conclude that $H^1(\PP^1,\mathcal{K}_b|_{\ell_\infty})=0$ and
$H^0(\PP^1,\mathcal{K}_b|_{\ell_\infty})\cong
H^0(\PP^1,\mathcal{E}|_{\ell_\infty})\cong\mathcal{E}_p$ where
$\mathcal{E}_p$ is the fibre of $\mathcal{E}$ at some point
$p\in\ell_\infty$. We set
\be
W=H^0\big(\PP^1\,,\,\mathcal{K}_b|_{\ell_\infty}\big) \ .
\label{Wdef}\ee
This is a complex vector space of dimension ${\rm
  rank}(\mathcal{E})=N$ and a choice of basis for $W$ corresponds to a
choice of trivialization for $\mathcal{E}|_{\ell_\infty}$. More
general framing data, for which $\mathcal{E}|_{\ell_\infty}$ can
contain a sum of non-trivial line bundles, are considered
in~\cite{Diaconescu}.

Similarly, from the exact sequence
\be
\xymatrix@1{
0 \ar[r] & \mathcal{K}_b\big|_{\ell_\infty}
\ar[r] &
  B \otimes \mathcal{O}_{\PP^3}\big|_{\ell_\infty}
  \ar[r]^{\!\!\!\!\!\!\!\!b_\infty} &
   V^{\oplus3} \otimes \mathcal{O}_{\PP^3}(1)\big|_{\ell_\infty}
   \ar[r]^{c_\infty} & 
  {V} \otimes \mathcal{O}_{\PP^3} (2)\big|_{\ell_\infty}\ar[r]
 & 0 \ ,
}
\label{kerbexactseq}\ee
along with $H^0(\PP^1,\mathcal{O}_{\PP^1})=\complex$ and
$H^1(\PP^1,\mathcal{K}_b|_{\ell_\infty})=0$, we obtain
\be
\xymatrix@1{
0 \ar[r] & W \ar[r] & B
\ar[r]^{\!\!\!\!\!\!\!\!\!\!\!\!\!\!\!\!\!\!\!\!\!\!\!\!\!\!\!\!\!\!
  \!\!\!\!b_\infty} &
   V^{\oplus3} \otimes H^0\big(\PP^1\,,\,\mathcal{O}_{\PP^1}(1)\big)
   \ar[r]^{~~c_\infty} & 
  {V} \otimes H^0\big(\PP^1\,,\,\mathcal{O}_{\PP^1} (2)\big)\ar[r] & 0
  \ . 
}
\label{WBKexactseq}\ee
Using the identifications $H^0(\PP^1,\mathcal{O}_{\PP^1}(1))=\complex 
z_0\oplus\complex z_1$ and
$H^0(\PP^1,\mathcal{O}_{\PP^1}(2))=\complex z_0^2\oplus\complex
z_0\,z_1\oplus\complex z_1^2$, we can rewrite (\ref{WBKexactseq}) as
\be
\xymatrix@1{
0 ~\ar[r] &~ W~ \ar[r] &~ B~ \ar[r]^{\!\!\!\!\!\!{{b_0}\choose{b_1}}} &
  ~ V^{\oplus6}~ \ar[r]^{{\gamma_{0,1}}} &
  ~ V^{\oplus3}~\ar[r] &~ 0 \ ,
}
\label{WBKrewrite}\ee
where generally we define
\beq
\gamma_{i,j}:=\begin{pmatrix}c_i&0\\ c_j&c_i\\ 0&c_j \end{pmatrix}
\,:\, V^{\oplus3}\oplus V^{\oplus3}~\longrightarrow~V\oplus V\oplus V
\label{gammaijc}\eeq
for $0\leq i<j\leq3$. It follows that $W=\ker(b_0)\cap\ker(b_1)$ and
that $\gamma_{0,1}^\dag:(V^*)^{\oplus3}\to(V^*)^{\oplus6}$ is
injective.

Let us now apply the same argument to the dual complex
\be
\xymatrix@1{
0 \ar[r] & V^*\otimes\mathcal{O}_{\PP^3}(-2)\big|_{\ell_\infty}
\ar[r]^{\!\!\!\!c_\infty^\dag} &
(V^*)^{\oplus3} \otimes \mathcal{O}_{\PP^3} (-1)\big|_{\ell_\infty}
\ar[r]^{~~~~~~b^\dag_\infty} &
  {B}^* \otimes \mathcal{O}_{\PP^3}\big|_{\ell_\infty}
  \ar[r]^{\!\!\!\!\!\!a^\dag_\infty} &
  V^* \otimes \mathcal{O}_{\PP^3} (1)\big|_{\ell_\infty} \ar[r] & 0
}
\label{dualmonadrestr}\ee
whose cohomologies before restriction to $\ell_\infty$ are
$\underline{\mathcal{H}om}(\Ecal,\mathcal{O}_{\PP^3})$ and
$\underline{\Ecal xt}^1(\Ecal,\mathcal{O}_{\PP^3})$, where we have
used ${\rm
  Ext}^1(\mathcal{O}_{\PP^3}(2),\mathcal{O}_{\PP^3})=0$. After 
tensoring with $\mathcal{O}_{\PP^1}(1)$, this yields the exact
sequence
\be
\xymatrix@1{
0 ~\ar[r] & ~H^0\big(\PP^1\,,\,\underline{\ker}(c_\infty^\dag)\big)~
 \ar[r] & ~(V^*)^{\oplus3}~
 \ar[r]^{\!\!{{b_0}\choose{b_1}}^\dag} & 
   ~B^* \oplus B^*~
   \ar[r]^{{\alpha_{0,1}^\dag}} & ~ (V^*)^{\oplus3}~ \ar[r] &~ 0
}
\label{BveeVexactseq}\ee
which implies that the linear map
\beq
\alpha_{0,1}:=\begin{pmatrix} a_0& a_1& 0\\ 0&a_0&a_1
\end{pmatrix}\,:\,V\oplus V\oplus V~ \longrightarrow~ B\oplus B
\label{alpha01a}\eeq
is injective. Using the exact sequence (\ref{WBKrewrite}) along with
$\dim_\complex B=3\dim_\complex V+\dim_\complex W$, we can thus
identify
\be
B=V\oplus V\oplus V\oplus W \ .
\label{BVWident}\ee
Furthermore, since
\be
\ker(b_0)\,\big/\,{\rm im}(a_0)=\mathcal{E}_{[1:0:0:0]}=W=\ker(b_0)
\cap\ker(b_1)
\label{kerb1ima1}\ee
one has ${\rm im}(a_0)\cap\ker(b_1)=0$. It follows that
$b_0\,a_1=-b_1\,a_0:V\to V^{\oplus3}$ is injective.

Using the monadic conditions $b_0\,a_1+b_1\,a_0=0$ and
$b_0\,a_0=0=b_1\,a_1$, we can choose bases for the vector spaces $V$,
$W$, $B$ and $C=V^{\oplus3}$ such that the injection $b_0\,a_1$ is
given by
\be
b_0\,a_1=\begin{pmatrix}0_{k\times k}\\0_{k\times k}\\
1_{k\times k}\end{pmatrix}
\label{b1a2expl}\ee
along with
\bea
a_0=\begin{pmatrix}0_{k\times k}\\ 1_{k\times k}\\ 0_{k\times k}\\
0_{N\times k} \end{pmatrix} \qquad & \mbox{and} & \qquad
b_0=  \begin{pmatrix}1_{k\times k}&0_{k\times k}&0_{k\times k}&
0_{k\times N}\\ 0_{k\times k}&0_{k\times k}&0_{k\times k}&
0_{k\times N}\\ 0_{k\times k}&0_{k\times k}&1_{k\times k}&
0_{k\times N}\end{pmatrix} \ , \nonumber\\[4pt]
a_1=\begin{pmatrix}0_{k\times k}\\ 0_{k\times k}\\ 1_{k\times k}\\
0_{N\times k} \end{pmatrix} \qquad & \mbox{and} & \qquad
b_1=  \begin{pmatrix}0_{k\times k}&0_{k\times k}&0_{k\times k}&
0_{k\times N}\\ -1_{k\times k}&0_{k\times k}&0_{k\times k}&
0_{k\times N}\\ 0_{k\times k}&-1_{k\times k}&0_{k\times k}&
0_{k\times N}\end{pmatrix} \ .
\label{a12b12expl}\eea
Using the conditions $\ker(c_0)={\rm im}(b_0)$, $\ker(c_1)={\rm
  im}(b_1)$ and $c_0\,b_1+c_1\,b_0=0$ we can then write the maps
\beq
c_0=\begin{pmatrix}0_{k\times k}&1_{k\times k}&0_{k\times k}
\end{pmatrix} \qquad \mbox{and} \qquad
c_1=\begin{pmatrix}1_{k\times k}&0_{k\times k}&0_{k\times k}
\end{pmatrix} \ .
\label{c0c1expl}\eeq
Using the equations $b_2\,a_0+b_0\,a_2=0=b_2\,a_1+b_1\,a_2$ and
$b_3\,a_0+b_0\,a_3=0=b_3\,a_1+b_1\,a_3$, together with
$c_2\,b_0+c_0\,b_2=0=c_2\,b_1+c_1\,b_2$ and
$c_3\,b_0+c_0\,b_3=0=c_3\,b_1+c_1\,b_3$, we can parametrize the
remaining linear maps in the form
\bea
a_2=\begin{pmatrix} B_1'\\ B_1\\ B_2\\ J \end{pmatrix}
\qquad & \mbox{and} & \qquad
b_2= \begin{pmatrix} C_1 & -B_1' & 0_{k\times k} & 0_{k\times N} \\
C_2 & 0_{k\times k}&B_1'&0_{k\times N} \\
C_3&-B_2&B_1&I \end{pmatrix} \ , \nonumber\\[4pt]
a_3=\begin{pmatrix} B_3\\ B_2'\\ B_3'\\ K \end{pmatrix}
\qquad & \mbox{and} & \qquad
b_3=\begin{pmatrix} C_1' & -B_3 & 0_{k\times k} & 0_{k\times N} \\
C_2' & 0_{k\times k}&B_3&0_{k\times N} \\
C_3'&-B_3'&B_2'&I' \end{pmatrix}
\label{a34b34expl}\eea
with $B_i,B_i',C_i,C_i' \in M_{k\times k}(\complex)$, $J,K\in M_{N\times
  k}(\complex)$ and $I,I'\in M_{k\times N}(\complex)$, along with
\beq
c_2=\begin{pmatrix}-C_2& C_1&-B_1' \end{pmatrix} \qquad \mbox{and}
\qquad c_3=\begin{pmatrix} -C_2'&C_1'&-B_3\end{pmatrix} \ .
\label{c2c3expl}\eeq

It remains to satisfy the remaining equations in (\ref{monadiccondab})
and (\ref{cohtrivcondcb}). {}From the last three equations
$b_2\,a_2=0=b_3\,a_3$ and $b_2\,a_3+b_3\,a_2=0$ of the monadic
condition we get the respective sets of matrix relations
\be
\begin{array}{rlrrl}
B_1'\,B_1&=~C_1\,B_1' \qquad &\mbox{and} &\qquad
B_3\,B_2'~=&C_1'\,B_3 \ , \\[4pt]
B_1'\,B_2&=~-C_2\,B_1' \qquad &\mbox{and} &\qquad
B_3\,B_3'~=&-C_2'\,B_3 \ , \\[4pt]
[B_1,B_2]+I\,J&=~-C_3\,B_1' \qquad &\mbox{and} &\qquad
[B_2',B_3']+I'\,K~=&-C_3'\,B_3
\end{array}
\label{ab2233rem}\ee
along with
\bea
B_3\,B_1+B_1'\,B_2'&=&C_1\,B_3+C_1'\,B_1' \ , \nonumber\\[4pt]
B_3\,B_2+B_1'\,B_3'&=&-C_2\,B_3-C_2'\,B_1' \ , \nonumber\\[4pt]
[B_2,B_2']+[B_3',B_1]-I\,K-I'\,J&=&C_3\,B_3+C_3'\,B_1' \ .
\label{ab2332rm}\eea
On the other hand, from the remaining conditions $c_2\,b_2=0=c_3\,b_3$
we get the respective additional matrix equations
\be
\begin{array}{rlrrl}
[C_1,C_2]&=~B_1'\,C_3 \qquad &\mbox{and} &\qquad
[C_1',C_2']~=&B_3\,C_3' \ , \\[4pt]
B_1'\,B_2&=~-C_2\,B_1' \qquad &\mbox{and} &\qquad
B_3\,B_3'~=&-C_2'\,B_3 \ , \\[4pt]
B_1'\,I&=~0 \qquad &\mbox{and} &\qquad
B_3\,I'~=&0 \ ,
\end{array}
\label{bc2233rem}\ee
while from $c_2\,b_3+c_3\,b_2=0$ we find
\bea
[C_1,C_2']+[C_1',C_2]&=&B_3\,C_3+B_1'\,C_3' \ , \nonumber\\[4pt]
B_3\,B_2+B_1'\,B_3'&=&-C_2\,B_3+C_2'\,B_1' \ , \nonumber\\[4pt]
B_3\,B_1+B_1'\,B_2'&=&C_1\,B_3+C_1'\,B_1' \ , \nonumber\\[4pt]
B_3\,I&=&-B_1'\,I' \ .
\label{bc2332rem}\eea

To investigate the solution space of the somewhat complicated system
of matrix equations (\ref{ab2233rem})--(\ref{bc2332rem}), we should
recall that the boundary conditions appopriate to the six-dimensional
generalized instantons on $\complex^3$ require trivializations on
\emph{three} independent lines $\PP^1\subset\PP^3$. We will choose the
remaining two lines to be given in homogeneous coordinates by
$[z_0:0:z_2:0]$ and $[0:z_1:z_2:0]$. We then restrict the complex
(\ref{ADHMcomplexfinal}) to these lines exactly as before. By an
identical analysis to that given above one arrives at analogous
constraints on the kernels and ranges of the linear maps involved,
which then imposes additional restrictions on the matrices satisfying
eqs.~(\ref{ab2233rem})--(\ref{bc2332rem}) above.

For instance, the injectivity of the linear maps $b_0\,a_2$ and
$b_1\,a_2$ lead respectively to the kernel constraints
\beq
\ker(B_1')\cap\ker(B_2)=0 \qquad \mbox{and} \qquad
\ker(B_1')\cap\ker(B_1)=0 \ ,
\label{kerB1B2B30}\eeq
which automatically guarantees all other vanishing conditions on the
maps $a_i$ and $b_i$. Likewise, injectivity of $\gamma_{0,2}^\dag$
and $\gamma_{1,2}^\dag$ lead respectively to the range constraints
\beq
{\rm im}(B_1')\cap{\rm im}(C_2)=0 \qquad \mbox{and} \qquad
{\rm im}(B_1')\cap{\rm im}(C_1)=0 \ .
\label{imB1C1C20}\eeq
Finally, the framing data determine isomorphisms between the
Chan--Paton space (\ref{Wdef}) and the two subspaces of
vectors:
\begin{itemize}
\item $(v,w)\in\big(\ker(B_1')\cap\ker(C_2)\big)\oplus W$ for which
  $B_2(v)=I(w)$; and
\item $(v,w)\in V\oplus W$ for which $B_1(v)=-I(w)$.
\end{itemize}
We interpret these extra constraints as stability conditions on the
linear maps satisfying the matrix equations
(\ref{ab2233rem})--(\ref{bc2332rem}). The issue of stability will be
discussed further in Section~\ref{Stability} below and in
Section~\ref{TopMQM}.

To illustrate the use of these extra conditions, suppose that a vector
$v\in V$ satisfies the equation $B_1'\,B_1(v)=C_1\,B_1'(v)$. Then by
(\ref{imB1C1C20}) one has $B_1(v)=B_1'(v)=0$, and hence $v=0$ by
(\ref{kerB1B2B30}). This implies that the linear map
$B_1'\,B_1-C_1\,B_1':V\to V$ is bijective, which contradicts the first
equation of (\ref{ab2233rem}). Thus we set $B_1'=0$. Then
(\ref{ab2233rem}) implies the first equation of the set
(\ref{ADHMeqs}). It is straightforward to find that the next two
equations of (\ref{ADHMeqs}) follow similarly, after some
identifications amongst the various matrices. As in
Section~\ref{LinalgNC}, there are generically more moduli and
equations than needed to identify a torsion free sheaf on $\complex^3$
with the set of matrix equations~(\ref{ADHMeqs}). The role of these
extra moduli will be explained in Section~\ref{QuiverMQM}.

\subsection{Stability\label{Stability}}

There is a natural free action of the group $GL(k,\complex)$ on the
data $(B_i,B_i',C_i,C_i',J,K,I,I')$ of Section~\ref{Barth} above, of
the form in (\ref{Ukactionmatrixeqs}) with $g\in GL(k,\complex)$. As
in the case of linear monads~\cite{okonek}, one can show that any
isomorphism of a complex (\ref{ADHMcomplexfinal}) which preserves the
trivializations on the lines $\ell^i_\infty$ and the choice of bases
for the vector spaces $V$, $W$, $B$ and $C$ made above has this
form. We can restrict the complex (\ref{ADHMcomplexfinal}) to
$\complex^3=\PP^3\setminus\PP^2$ by setting $z_0=1$ and replacing
$\mathcal{O}_{\PP^3}(-r)$ by $\mathcal{O}_{\complex^3}$
everywhere. Then on the fibre at $z=(z_1,z_2,z_3)\in\complex^3$, the
morphisms $a$, $b$ and $c$ induce homomorphisms of vector spaces
\be
\xymatrix@1{
V \ar[r]^{\!\!\!\!\!\!\!\!\!\!\!\!\sigma_z} &
{\begin{matrix} V\oplus V\oplus V
   \\ \oplus \\ W \end{matrix}} \ar[r]^{\tau_z} & V\oplus V \oplus
V \ar[r]^{~~~~~~\eta_z} & V \ .
}
\label{ADHMcomplex}\ee

Following the analog statement for linear monads on
$\PP^d$~\cite{jardim}, one can show that whenever $N\geq3$ the
cohomology sheaf (\ref{cohsheafcoh}) is nondegenerate, and hence that
the cohomology of a generic complex (\ref{ADHMcomplexfinal}) is a
torsion free sheaf on $\PP^3$ with the stated properties. Then from
(\ref{kerB1B2B30}), along with suitable identifications of the various
matrices, it follows that the localized maps $a_z\in\Hom(V,B)$ are
injective for all  $z\in\PP^3$. On the other hand, surjectivity of
$c_z\in\Hom(C,V)$ and exactness $\ker(c_z)={\rm im}(b_z)$ for all
$z\in\PP^3$, together with (\ref{imB1C1C20}) and appropriate matrix
identifications, presumably implies a certain
algebraic stability condition analogous to the four-dimensional
case~\cite{nakajima}. Using stability along with the fact that $B_2$
and $B_3$ commute, one can repeat the proof
of~\cite[Proposition~2.7]{nakajima} step by step to show that in the
abelian case the maps
\be
J=K=0 \qquad \mbox{for} \quad N=1
\label{JK0N1}\ee
and the cohomology sheaf (\ref{cohsheafcoh}) is
isomorphic to the ideal $\mathcal{I}$ given in the description of the
Hilbert scheme ${\sf Hilb}^k(\complex^3)$ in terms of instantons in
noncommutative gauge theory on $\complex^3$ presented in the
previous section. The algebraic stability condition also follows from 
the last matrix equation in (\ref{ADHMeqs}) which encodes the
noncommutative deformation. We will see this explicitly in the next
section wherein we shall make the instanton complex
(\ref{ADHMcomplex}) equivariant with respect to the toric action
following~\cite[Chapter 5]{nakajima} and~\cite{Bruzzo:2002xf}, and
utilize the powerful localization techniques of a suitably defined
topological matrix model. This is tantamount to introducing the matrix
equations (\ref{matrixphieqs}) of the $\Omega$-background.

\bigskip

\section{Topological matrix quantum mechanics\label{TopMQM}}

In this section we will carry on with the case $X =
\complex^3$ to avoid complications coming from a non-trivial
topology of the ambient space. In this context the gauge theory
computes bound states of D6 and D0 branes with a suitable
$B$-field turned on. As in the four-dimensional case there are
two ways of doing the instanton counting computation. One way is to
write down the noncommutative gauge theory directly, solve for the
critical points and compute the fluctuation determinants around each
critical point, as we did in Section~\ref{NCGT}. The other way is to
consider the effective field theory of a gas of $k$ D0-branes coupled
to the D6-branes. This is equivalent to a topological matrix quantum
mechanics on the resolved moduli space. For instance, we may think of
this field theory as arising from quantization of the collective
coordinates around a fixed point of the original gauge theory. In the
static limit, relevant for considerations involving the BPS ground
states, this gives rise to an ADHM-like formalism which provides a
dynamical realization of the matrix equations describing the torsion
free sheaves of the previous section. This will nicely tie the
computation in terms of noncommutative instantons into one involving
the abelian category of holomorphic D-branes (coherent sheaves).

The set of fields and equations of motion involved in the matrix
quantum mechanics can be nicely interpreted as a representation of a
quiver with relations, an oriented graph consisting of nodes and
arrows together with linear combinations of paths in the graph. To
provide a representation of the quiver diagram in the category of
complex vector spaces, one associates a set of vector spaces to the
nodes and linear maps (``fields'') to the arrows which respect the
relations (``constraints''). In the large
volume limit one can represent a D-brane configuration, viewed as an
object in the abelian category of coherent sheaves, as a
representation of a quiver. This ties in nicely with the varieties
naturally associated to quivers that were introduced by
Nakajima to study ordinary instanton moduli spaces.\footnote{The
original motivation was to understand moduli spaces of
four-dimensional instantons on singular varieties, but the formalism
can be extended to smooth spaces as well and was indeed used
in~\cite{nakajima}.} One may regard the quiver construction
presented in the following as a generalization of this formalism
to study moduli spaces of solutions of the DUY equations.
Physically we are left with a supersymmetric quantum mechanics
that lives on the moduli space of the quiver variety, whose
supersymmetric ground states correspond to BPS bound states in the
original D-brane picture. The quiver quantum mechanics that
describes the BPS states of the D6--D0 system with an appropriate
$B$-field turned on was introduced in~\cite{jafferis}. The connection
between the supersymmetric quantum mechanics and the noncommutative
gauge theory description of the D6--D0 system was suggested originally
in~\cite{CIMM}--\cite{Witten:2000mf}, where it was also shown that this system is
non-supersymmetric in the absence of a $B$-field. This means that the
noncommutative deformation is crucial for obtaining stable instanton
solutions and that a description involving torsion free sheaves is
unavoidable in this context.

Ultimately the quiver approach and the noncommutative gauge theory
give equivalent descriptions of the same physical problem, but it is
important to bear in mind that in the noncommutative computation
one is expanding the gauge theory path integral around critical
points and computing the quantum fluctuations around the classical
solutions. On the other hand, the matrix quantum mechanics already
contains the fluctuation factors. By fixing the number of D0-branes to
be $k$ (and likewise the rank of the matrices), we are automatically
working in what from the gauge theory perspective would be the
charge $k$ instanton sector. The quiver matrix model computes directly
the equivariant volume of this moduli space. Finally, one takes back
this information to the gauge theory. In the gauge theory, by
expanding around any critical point one is left with a matrix
model on the instanton moduli space. This is what the quiver matrix
model computes.

In this section we will begin by briefly reviewing the typical
computation of ratios of fluctuation determinants in cohomological
field theory, following the treatment of~\cite{Moore:1998et}, and then
apply this formalism to the quiver matrix quantum mechanics. To
compute the quiver partition function we will also give a
classification of the critical points of the matrix model. The path
integral localizes onto fixed points of the toric action and each
fixed point is given by a solution of the equations
(\ref{ADHMeqs}). The equivariant character of the complex
(\ref{ADHMcomplex}) at a fixed point represents the linearized
contribution of the fixed point to the partition function. Finally,
the computation will be completed by using the localization
formula. We will obtain exact agreement with the results coming from
noncommutative gauge theory, reproducing in particular the partition
function (\ref{genfnnonabinvs}).

\subsection{Cohomological field theory formalism\label{CohFT}}

Let us start with a set of equations $\vec{\mathcal{E}}= \vec0$. In
our applications these will either be the ADHM-like equations of the
previous section or the six-dimensional DUY equations of
Section~\ref{NCGT}. These equations will be functions of some complex
fields which we denote collectively by $X_i$ and are assumed to
transform in the adjoint representation of some $U(k)$ gauge symmetry
group. To construct a supersymmetric field theory one needs to
supplement these fields with superpartners to form multiplets $(X_i ,
\Psi_i)$ with BRST transformations
\begin{equation} Q X_i = \Psi_i \, \qquad \mbox{and}
\qquad \, Q \Psi_i = [\phi , X_i] \ .
\end{equation}
The $X_i$ are coordinates on the field space, which
after imposing the constraints and gauge invariance becomes the
moduli space, and $\Psi_i$ are their differentials.

To these fields we add the Fermi multiplet of antighosts and auxiliary
fields $(\vec{\chi} , \vec{H}\,)$ associated with the equations
$\vec{\mathcal{E}} = \vec0$. Schematically, the bosonic part of the
action contains a term
\begin{equation}
S_{\rm bos}=
\ii\Tr\,\vec{\mathcal{E}} \cdot \vec{H} + \Tr\,\vec{H}^2 \ ,
\end{equation}
which on-shell gives $\vec{H} = -\frac\ii2\,\vec{\mathcal{E}}$ and
$S_{\rm bos}=\frac14\,\Tr\,\vec{\mathcal{E}}\,^2$. Because of this, the
multiplet $(\vec{\chi} , \vec{H}\,)$ has the same quantum numbers as
the equations $\vec{\mathcal{E}}$. Finally, one adds the gauge
multiplet $(\phi ,\overline{\phi} ,  \eta)$ necessary to close the
BRST algebra, where $\phi$ is the generator of gauge transformations
which is related to the Higgs field $\Phi$ of the noncommutative gauge
theory via the decomposition (\ref{Phisplitting}). Their BRST
transformations are given by
\be
Q\phi=0 \ , \quad Q\overline{\phi}=\eta \quad \mbox{and} \quad
Q\eta=\big[\phi\,,\,\overline{\phi}~\big] \ .
\label{gaugeBRST}\ee
We will split the set of equations $\vec{\mathcal{E}}$ into two sets
$(\vec{\mathcal{E}}_c = \vec0 ,\mathcal{E}_r = 0)$ of complex and real
equations, respectively, which play a role analogous to the F-term and
D-term conditions. The latter condition can be regarded as a stability
condition.

We will also want to work equivariantly with respect to some
toric action $\mathbb{T}^d$. Let us choose the rotations $X_i
\rightarrow X_i ~\e^{- \ii \epsilon_i}$ for some parameters
$\epsilon_i$ that generate the toric action. The BRST transformations
are then modified to
\begin{equation}
Q X_i = \Psi_i \, \qquad \mbox{and}
\qquad \, Q \Psi_i = [\phi , X_i] - \epsilon_i\,X_i \ ,
\end{equation}
where as usual the transformation of the fermions reflects the
infinitesimal transformation of $X_i$ under the symmetry group $U(k)
\times\mathbb{T}^d$. Let us consider now the Fermi multiplet
$(\vec{\chi}_c , \chi_r , \vec{H}_c , H_r)$ associated with the
equations $(\vec{\mathcal{E}}_c , \mathcal{E}_r)$. The equations
$\vec{\mathcal{E}}_c$ transform as an adjoint field under
$U(k)$ and as $\e^{\ii \epsilon_c}$ under the toric action, where
$\epsilon_c$ denotes some linear combination of the toric
parameters $\epsilon_i$ that may be different for each equation.
The second equation $\mathcal{E}_r$ again transforms in the adjoint
representation of $U(k)$ but is invariant under the toric action.
These conditions determine the BRST transformation rules for the Fermi
multiplets to be
\be \label{transfchi}
\begin{array}{rllrl}
Q \chi_{c,i} &=& H_{c,i} & \quad \mbox{and} \qquad Q H_{c,i} &= \ [\phi ,
\chi_{c,i}] - \epsilon_{c,i} \,\chi_{c,i} \ , \\[4pt]
Q \chi_r &=& H_r & \quad \mbox{and} \qquad Q H_r &= \ [\phi , \chi_r] \ ,
\end{array}
\ee
where the label $i$ runs over the set of equations of
$\vec{\mathcal{E}}_c$. Finally, the gauge multiplet transforms again
as in (\ref{gaugeBRST}).

The action of the cohomological gauge theory can be represented as
\begin{equation}
S = Q \Tr\left( \eta\, \big[\phi \,,\, \overline{\phi}~\big] -
\vec{\chi} \cdot \vec{\mathcal{E}} + g\, \vec{\chi} \cdot \vec{H} +
\Psi_i\,\big[X_i \,,\, \overline{\phi}~\big] \right)
\end{equation}
and the path integral localizes onto the solutions of the equations
$\vec{\mathcal{E}}=\vec0$. Note that the path
integral is independent of the coupling constant $g$, as usual in
cohomological gauge theories. Let us now explicitly evaluate the path
integral. The first step is to use the $U(k)$ gauge invariance to
diagonalize the gauge generator $\phi$. This produces a Vandermonde
determinant $\mathrm{det} \left(\mathrm{ad}\, \phi \right)$ in the
path integral measure. Then one would like to integrate out the
fields $\vec{\chi}$, which appear quadratically in the
action. However, this is not immediately possible. By looking at the
BRST transformations (\ref{transfchi}) we see that the mass matrix of
$\chi_r$, $\Tr\,\vec\chi_c\cdot[\phi , \vec\chi_c]$, can have zero
modes (while this is not the case for $\vec\chi_c$ if $\epsilon_{c,i}$
are generic). To cure this problem we add the term
\begin{equation}
t_1 \,Q \Tr\,\chi_r \,\overline{\phi}
\end{equation}
to the action and take the limits $t_1 \rightarrow \infty$ and $g
\rightarrow \infty$ with $g\ll t_1$. This is a legitimate procedure
since the quantum field theory is topological and independent of
$t_1$. But in this way the resulting action has no kinetic term for
the fermions, and so we add it by hand via the term
\begin{equation}
t_2 \,Q \Tr\big(X_i\, \Psi_i^{\dagger} - X_i^\dag\,\Psi_i\big) \ .
\end{equation}
Again this is legitimate since we are adding a BRST-exact term to
the action and so the path integral will not change.

We can now proceed in three steps:
\begin{itemize}
\item Take the limit $t_1 \rightarrow \infty$. The relevant part of
  the action is
\be
t_1\,\Tr \big(H_r \,\overline{\phi} + \chi_r\, \eta\big)
\ee
and these fields can be trivially integrated out. This means
that in the following expressions we can neglect them since their
contributions to the path integral are suppressed in the $t_1
\rightarrow \infty$ limit.
\item Now take the limit $g \rightarrow \infty$. The relevant part
  of the action is quadratic in $\vec{\chi}_c$ (and in
  $\vec{H}_c$). We can integrate them out and the result is a factor
\begin{equation}
\mathrm{det} \left( \mathrm{ad}\, \phi - \epsilon_{c,i} \right)
\end{equation}
for each of the fields $\chi_{c,i}$. Note that we get a
determinant and not its square root since $\chi_{c,i}$ is a
complex field. The determinant is in the numerator since the field
$\chi_{c,i}$ obeys Fermi statistics.
\item Finally, we take the limit $t_2 \rightarrow \infty$ and
  integrate out the fields $X_i$ (and $\Psi_i$) which now appear
  quadratically in the action. We obtain a factor
\begin{equation}
\frac{1}{\mathrm{det} \left( \mathrm{ad}\, \phi - \epsilon_i\right)
}
\end{equation}
for each of the fields. Again we get determinants since $X_i$
are complex fields, but now in the denominator since they are
bosonic.
\end{itemize}
We have dropped various normalization factors coming from the gaussian
integrals that involve the couplings $g$, $t_1$ and $t_2$, which as
expected cancel between the bosonic and fermionic integrations, and
some ratio of the toric parameters $\epsilon_i$ which depends on the
choice of equivariant action.
It is important to remember that we still have the Cartan
subalgebra integral over the diagonal field $\phi$ left to do.

When the equations are the ADHM-like or the DUY equations the
resulting integral represents the instanton fluctuation factor in the
charge $k$ instanton sector. In this approach $k$ is fixed and is the
rank of the matrices $X_i$. The fluctuation determinant has the
following structure. In the numerator there appear the Vandermonde
determinant and a determinant due to the equations
$\vec{\mathcal{E}}_c = \vec0$. In the denominator the factors are
related to the fields $X_i$ and reflect their quantum numbers. The
determinant is of the form (constraint)/(fields) and this
structure generalizes to any topological matrix quantum mechanics
\cite{Nekrasov:2004vw}. We will make extensive use of this in the
following.\footnote{The Vandermonde determinant is not really a
  constraint but in a certain sense is related to the equation
  $\mathcal{E}_r = 0$. Recall that it arises when one uses gauge
  invariance to diagonalize $\phi$.}

\subsection{Quiver matrix quantum mechanics\label{QuiverMQM}}

We are now ready to systematically construct the topological
quiver quantum mechanics of~\cite{jafferis} and compute its path
integral. We introduce two vector spaces $V$ and $W$ of complex
dimensions $\dim_\complex V = k$ and $\dim_\complex W = N$. In the
noncommutative gauge theory the space $V$ is an ``internal''
finite-dimensional subspace of the Fock module $\Hcal$, represented
geometrically by sheaf cohomology groups in (\ref{complexvecsp}),
while $W$ is an ``external'' Chan--Paton space determined
geometrically by framing data through the sheaf cohomology group
(\ref{Wdef}). In the quiver description they represent the spaces
where the fields of the matrix quantum mechanics take values. In the
D-brane picture $V$ is spanned by the gas of $k$ D0-branes, while $W$
represents the $N$ (spectator) D6-branes. In this description we fix
the topological sector and restrict attention to instantons of charge
$k$. As in the previous sections we will keep the number $N$ of
D6-branes arbitrary for formal considerations, but concrete
computations will require an explicit choice of gauge symmetry
breaking pattern.

The fields of the quiver are given by
\begin{eqnarray}
X_i &=& (B_1 , B_2 , B_3 , \varphi , I) \ , \nonumber\\[4pt]
\Psi_i &=& (\psi_1 ,\psi_2 , \psi_3 , \zeta , \rho) \ .
\end{eqnarray}
The matrices $B_i$ arise from 0--0 strings and represent the
position of the coincident D0-branes inside the D6-branes. One
may regard the fields $(B_1 , B_2 , B_3 , \varphi)$ as arising from
the reduction to zero dimensions of the six-dimensional Yang--Mills
multiplet $(Z_1,Z_2,Z_3,\rho)$. On the other hand, the field $I$
describes open strings stretching from the D6-branes to the D0-branes
and is characteristic of the quiver formalism. It characterizes the
size and orientation of the D0-branes inside the D6-branes, and is
required to make the system supersymmetric. Thus the bosonic fields
are defined as linear maps
\bea
(B_1 , B_2 , B_3 , \varphi) &\in&\Hom (V,V) \ , \nonumber\\[4pt]
I&\in&\Hom(W,V) \ .
\label{bosnaive}\end{eqnarray}
The matrices $B_i$ and $I$ originate in the noncommutative gauge
theory through the decompositions of the covariant coordinates
(\ref{Zalphasplit}) and of the Higgs field (\ref{Phisplitting}).
As we will localize below onto the maximal torus of the $U(N)$
gauge group, we can neglect the remaining fields $J$ and $K$ using
(\ref{JK0N1}).

The fields $B_i$ and $\varphi$ all lie in the adjoint
representation of $U(k)$ where $k$ is the number of D0-branes (or the
instanton number). {}From the dimensional reduction, we can identify
their transformation properties under the toric $\mathbb{T}^3$ action
as
\begin{eqnarray}
B_i ~&\longmapsto~&B_i ~\e^{- \ii \epsilon_i} \ , \nonumber\\[4pt]
\varphi~&\longmapsto~&\varphi ~\e^{- \ii (\epsilon_1 + \epsilon_2 +
\epsilon_3)} \ .
\end{eqnarray}
We will frequently use the notation $\epsilon = \epsilon_1 +
\epsilon_2 + \epsilon_3$ (with $\epsilon=0$ when we wish to make the
Calabi--Yau condition (\ref{epsconstr}) explicit). On the other hand,
$I$ is a $U(k) \times U(N)$ bifundamental field where $N$ is the
number of D6-branes (or the rank of the six-dimensional gauge
theory). Under the full symmetry group $U(k)\times U(N)\times\torus^3$
it transforms as
\begin{equation}
I ~\longmapsto~ g_{U(k)}\, I\, g_{U(N)}^{\dagger} ~\e^{- \ii
\varepsilon} \ .
\end{equation}
The transformation of $I$ under the toric action $\mathbb{T}^3$ is not
fixed by any constraint. In the following we will argue that
$\varepsilon = 0$ (although the precise value of the toric parameter
$\varepsilon$ is not important for the evaluation of the path
integral).

\begin{figure}[hbt]
\bigskip
\begin{center}
\epsfxsize=2 in\epsfbox{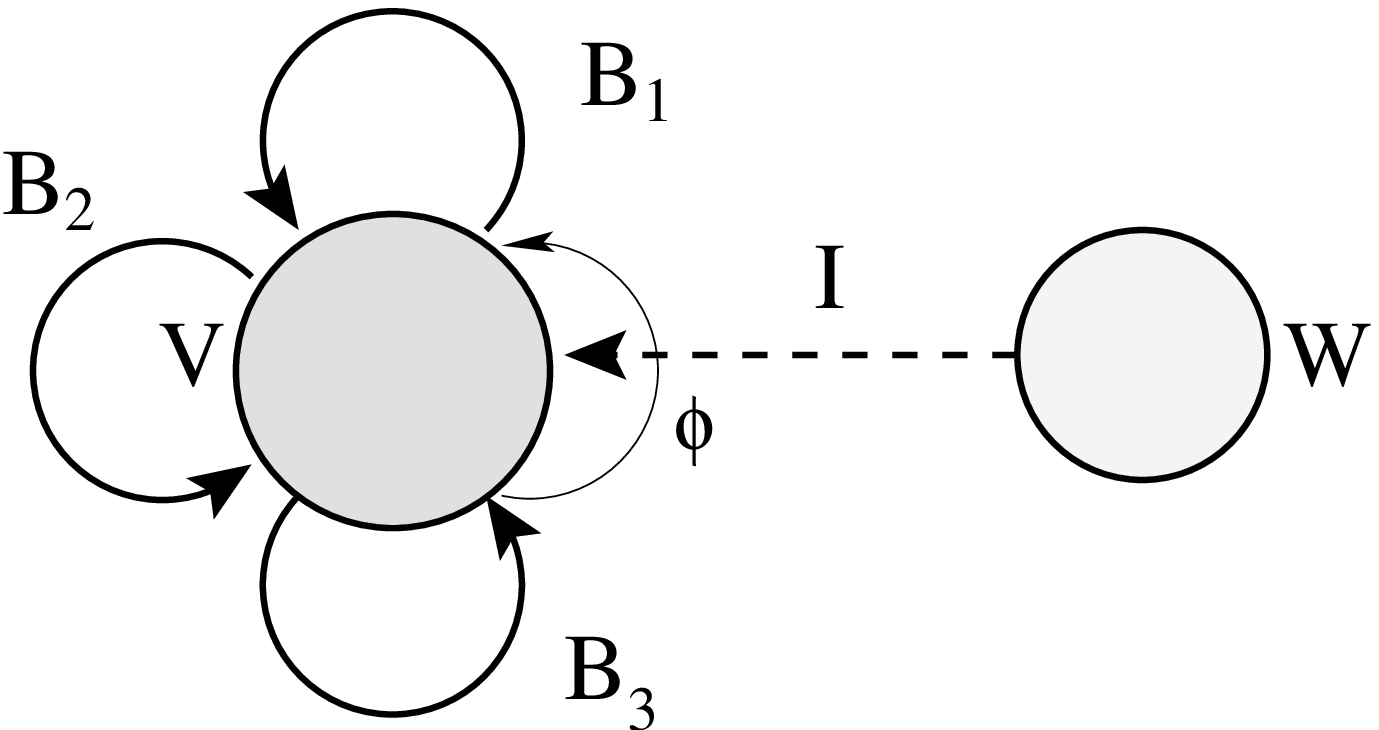}
\end{center}
\caption{Quiver description of the Hilbert scheme
$\textsf{Hilb}^k (\complex^3)$.} \label{quiver}
\end{figure}

The corresponding BRST transformations read
\be
\begin{array}{rllrl}
Q B_i &=& \psi_i & \quad \mbox{and} \qquad Q \psi_i& = \ [\phi , B_i] -
\epsilon_i\,
B_i \ , \\[4pt] Q \varphi& =& \zeta & \quad \mbox{and} \qquad
Q \zeta &= \ [\phi ,\varphi] - \epsilon \,\varphi \ , \\[4pt]
Q I &=& \rho & \quad \mbox{and} \qquad Q \rho& = \
\phi \,I - I \,\mbf a - \varepsilon \,I \ ,
\end{array}
\label{BRSTmatrices}\ee
where $\mbf a = \mathrm{diag} (a_1 , \dots , a_N)$ is a background
field which parametrizes an element of the Cartan subalgebra
$\mathfrak{u}(1)^{\oplus N}$. In the noncommutative gauge theory on the
D6-branes, $\mbf a$ plays the role of the vev of the Higgs field
$\mu(\Phi)$, defined by a mapping analogous to (\ref{mmuZexpl}). In
the present approach the fields $\mbf a$ and $\phi$ parametrize
distinct D6 and D0 brane gauge transformations. {}From
(\ref{BRSTmatrices}) it follows that they are only related to each
other on-shell at the BRST fixed points. Figure~\ref{quiver} depicts
the relevant fields that will enter into the quiver description of the
Hilbert scheme below.

For the bosonic fields we will consider the equations of motion
\begin{eqnarray}
\mathcal{E}_i \,&:&\, [B_i , B_j] + \epsilon_{i j k} \,\big[B_k^{\dagger}
\,,\, \varphi\big] = 0 \ , \nonumber\\[4pt]
\mathcal{E}_r \,&:&\, \sum_{i=1}^{3}\, \big[B_i \,,\,
B_i^{\dagger}\,\big] + \big[\varphi \,,\, \varphi^{\dagger}\,\big] + I\,
I^{\dagger} = r \ , \nonumber\\[4pt] \mathcal{E}_I \,&:&\,
I^{\dagger} \,\varphi = 0 \ .
\label{quiverdefeqs}\end{eqnarray}
The equations $\mathcal{E}_i$ and $\mathcal{E}_I$ are relations for the
quiver depicted in Fig.~\ref{quiver}, while the equation
$\mathcal{E}_r$ is a cyclic vector for the representation of the
quiver in the moduli space of coherent sheaves. The Fayet--Iliopoulos
parameter $r>0$ is determined by the
noncommutative deformation of the original gauge theory and it
determines the mass of the D0--D6 fields in terms of the asymptotic
$B$-field required to preserve supersymmetry in the D6--D0 bound
states.

The extra field $\varphi$ and equation $\Ecal_I$ can in fact
be seen to arise from the extra moduli and matrix equations that we
found in (\ref{ab2233rem})--(\ref{bc2332rem}), e.g. by identifying
$B_1'=\varphi^\dag$. Its appearence there is natural since the
projective space $\PP^3$ is not Calabi--Yau. They can also be seen to
arise from the noncommutative instanton equations (\ref{adhmform}) by
decomposing the field $\rho$ similarly to (\ref{Phisplitting}) with
$\varphi\in{\rm End}_\alg(\Hcal^{\oplus k})$. For the quiver matrix
model appropriate to the dynamics on $\complex^3$, however, one should
set $\varphi=0$ and arrive at the system of matrix equations
(\ref{ADHMeqs}). These extra moduli can also play the role of the
extra fields required when considers non-trivial asymptotic boundary
conditions, such as those of (\ref{EcalP2}), as we do in the next
section.

We now have to add the Fermi multiplets $(\vec{\chi} ,
\vec{H}\,)$, which contain the antighost and auxiliary fields
$\vec{\chi}= (\chi_1 , \chi_2 , \chi_3 , \chi_r , \xi)$ and $\vec{H} =
(H_1 , H_2 , H_3 , H_r , h)$. As we stressed earlier, the auxiliary
fields are determined by the equations $\vec{\mathcal{E}}$ on-shell
and so must carry the same quantum numbers. This implies that the
antighosts are defined as maps
\begin{eqnarray}
(\chi_1 , \chi_2 , \chi_3 , \chi_r) &\in& \Hom (V,V) \ ,
\nonumber\\[4pt] \xi &\in&\Hom (V,W) \ ,
\label{fermnaive}\end{eqnarray}
since $\xi$ corresponds to the equation $\mathcal{E}_I$ which maps a
vector in $V$ to a vector in $W$.

Let us take a closer look at the defining equations
(\ref{quiverdefeqs}). As in Section~\ref{CohFT} above, the complex
equations $\mathcal{E}_i$ live in the adjoint representation of $U(k)$
but transform under the toric action with a factor
$\e^{ - \ii (\epsilon - \epsilon_i)}$. The real equation
$\mathcal{E}_r$ lives in the adjoint representation of $U(k)$ and is
invariant under the toric action. Finally, the equation
$\mathcal{E}_I$ transforms under $U(k)\times U(N)\times\mathbb{T}^3$
as
\begin{equation}
I^{\dagger}\, \varphi ~\longmapsto~ \e^{ \ii (\varepsilon -
\epsilon)}\, g_{U(N)} \,I^{\dagger}\, g_{U(k)}^{\dagger} \,g_{U(k)}\,
\varphi\, g_{U(k)}^{\dagger} = \e^{ \ii (\varepsilon - \epsilon)}\,
g_{U(N)}\, I^{\dagger}\, \varphi\, g_{U(k)}^{\dagger} \ .
\end{equation}
We now have all the ingredients necessary to write down the BRST
transformations for the remaining fields as
\be
\begin{array}{rllrl}
Q \chi_i &=& H_i & \quad \mbox{and} \qquad Q H_i &= \ [\phi , \chi_i] -
(\epsilon - \epsilon_i)\, \chi_i \ , \\[4pt]
Q \chi_r &=& H_r & \quad \mbox{and} \qquad Q H_r &= \ [\phi , \chi_r] \ ,
\\[4pt] Q \xi &=& h & \quad \mbox{and} \qquad Q h &= \ \mbf a \,
\xi - \xi\, \phi + (\varepsilon -\epsilon)\, \xi \ ,
\end{array}
\ee
to which we add the gauge multiplet to close the algebra
(\ref{gaugeBRST}).

The action that corresponds to this system of
fields and equations is given by
\begin{eqnarray}
S &=&  Q \Tr\Big(\chi_i^{\dagger}\,(H_i -
\mathcal{E}_i) + \chi_r \,(H_r - \mathcal{E}_r) + \xi^{\dagger}\, (h -
\mathcal{E}_I) + \psi_i \,\big[~\overline \phi \,,\,
B_i^{\dagger}\big] \cr && \qquad\quad +\,
\zeta \,\big[~\overline{\phi} \,,\, \varphi^{\dagger}\big] + \rho\,
 \overline{\phi}\,I^{\dagger} + \eta \,\big[\phi \,,\,
 \overline{\phi}~
\big] + \mbox{h.c.} \Big) \ .
\end{eqnarray}
This action is topological and the path integral can be treated as
we did in the Section~\ref{CohFT} above. The critical points are
determined by the zeroes of the BRST charge. We are interested in
the class of minima where $\varphi$ vanishes (and the fermions are
set to zero). The fixed point equations are then
\begin{eqnarray} \label{fixedpoint}
\left( B_i \right)_{ab} \,\left( \phi_a - \phi_b - \epsilon_i
\right) &=& 0 \ , \nonumber\\[4pt]
I_{a , l}\, \left( \phi_a - a_l - \varepsilon\right) &=& 0
\end{eqnarray}
where we have diagonalized both $\phi$ (producing a Vandermonde
determinant $\mathrm{det} (\mathrm{ad}\, \phi)$ in the path integral
measure) and $\mbf a$ by $U(k)$ and $U(N)$ gauge transformations,
respectively. We will give a more precise classification of the fixed
points in terms of three-dimensional partitions in
Section~\ref{3Dparts} below.

Regardless of what the structure of the fixed point set is, we can
write down directly the fluctuation determinants with the general
rules outlined in Section~\ref{CohFT} above. The fields give a
contribution in the denominator determined by their quantum numbers,
while the constraints similarly appear in the numerator. The main
difference from the noncommutative gauge theory is that now we have an
additional field and an additional constraint. Putting everything
together we get the partition function
\begin{eqnarray}
Z &=& \oint~\prod_{i=1}^k \,\mathrm{d}\phi_i~
\frac{\mathrm{det}\left( \mathrm{ad\, \phi} \right)\,
\mathrm{det}\left( \mathrm{ad\, \phi} + \epsilon_1 + \epsilon_2
\right)\, \mathrm{det}\left( \mathrm{ad\, \phi} + \epsilon_1 +
\epsilon_3 \right)\, \mathrm{det}\left( \mathrm{ad\, \phi} +
\epsilon_2 + \epsilon_3 \right)}{\mathrm{det}\left( \mathrm{ad\,
\phi} + \epsilon \right)\, \mathrm{det}\left( \mathrm{ad\, \phi} +
\epsilon_1 \right)\, \mathrm{det}\left( \mathrm{ad\, \phi} +
\epsilon_2 \right)\, \mathrm{det}\left( \mathrm{ad\, \phi} +
\epsilon_3 \right)}
 \cr && \qquad\qquad\qquad\quad\times ~
\frac{\mathrm{det}\big( - \phi\otimes1_W + 1_V\otimes\mbf a +
(\varepsilon - \epsilon)\big)}
{\mathrm{det}\left( \phi\otimes1_W -1_V\otimes\mbf a - \varepsilon
  \right)}
\label{quivercontint}\end{eqnarray}
where we have again dropped several factors including the volume of the
gauge group. As in the noncommutative gauge theory, the ratio of
determinants formally cancels up to a sign when $\epsilon=0$. Thus the
integration in (\ref{quivercontint}) is ill-defined as a Lebesgue
integral and must be defined via an appropriate contour integration.

At this point one should proceed to evaluate the integral of $\phi$
over the Cartan subalgebra $\mathfrak{u}(1)^{\oplus k}$. This is the main
difference from the noncommutative gauge theory, even though the
ratios of determinants formally look alike. In the noncommutative
gauge theory the ratio of determinants is a function of only the
equivariant parameters, since there we compute the determinants by
taking products of eigenvalues of operators over the Hilbert space
$\Hcal$. Here things are different, as we are really dealing with a
finite-dimensional $k \times k$ matrix model, and the ratio is a
function of the eigenvalues of $\phi$. The integral over the maximal
torus of the group $U(k)$ still has to be performed and it requires
an appropriate prescription to pick an integration contour which
encircles the relevant poles. Note that the fixed points of the
toric action appear as poles in the denominator. The evaluation of
(\ref{quivercontint}) by residues is in fact equivalent to the use of
the localization formula applied to the equivariant Euler
characteristic of the complex (\ref{ADHMcomplex}), a fact that will be
exploited in the following. 

It is amusing to note that the analogous problem was solved in four
dimensions by lifting the theory to five dimensions (with the fifth
dimension compactified), where a natural prescription for the
contour integral can be found~\cite{Nekrasov:2004vw,Losev:2003py}. The
rationale behind this procedure is that an instanton counting problem
in four dimensions can be lifted to a soliton counting problem in five
dimensions. It is tempting to speculate that the ratio of determinants
in (\ref{quivercontint}) (and in more general instances) may have a
clearer origin in the context of topological M-theory on the product
of a Calabi--Yau threefold with $\mathbb{S}^1$.

\subsection{Fixed points and three-dimensional partitions\label{3Dparts}}

We will now clarify how the matrix quantum mechanics
equations (\ref{fixedpoint}) can be solved in terms of plane
partitions. We will focus on the abelian case where the
parametrization of the Hilbert scheme allows for an explicit
classification of the fixed points. In this framework we can show
explicitly how the melting crystal picture is encoded in the
matrix model formalism and thus how the gravitational quantum foam
emerges naturally from the gauge theory variables. The following
analysis is a generalization of the arguments of~\cite{nakajima}.
We will go through the argument in two steps. First we will
recover the Hilbert scheme of points by the matrix quantum
mechanics equations of motion. Then we will show how to construct
a three-dimensional partition given a fixed point in the Hilbert
scheme and viceversa.

Consider the Hilbert scheme of $k$ points in $\mathbb{C}^3$ given by
\begin{equation}
{\sf Hilb}^k\big(\mathbb{C}^3\big)= \big\{\text{ideals } \mathcal{I}
\subset \mathbb{C}[B_1 , B_2 
, B_3] ~\big|~\dim_\complex \mathbb{C}[B_1
, B_2 , B_3]/\mathcal{I} = k \big\} \ .
\end{equation}
Then we claim that
\begin{equation}
{\sf Hilb}^k\big(\mathbb{C}^3\big)\cong \left\{ (B_1 , B_2 , B_3 , I)
  ~ \begin{array}{| c} ~[B_1 , B_2] = [B_1  ,B_3] = [B_2 , B_3] = 0 \\
\text{Stability: there is no proper} \\ \text{subspace $S \subset
\mathbb{C}^k$ such that} \\ \text{$B_i (S) \subset S$ and
$\mathrm{im}(I) \subset S$ }
\end{array} \right\} ~ \Big/~
GL(k,\mathbb{C})
\end{equation}
where $B_i \in \End(\mathbb{C}^k)$ and $I \in \Hom(\mathbb{C} ,
\mathbb{C}^k)$. In the quiver language one has $\mathbb{C} = W$ and
$\mathbb{C}^k = V$. The action of the gauge group $GL(k,\mathbb{C})$
is given in (\ref{Ukactionmatrixeqs}).

The correspondence can be seen as follows. Suppose we are given an
ideal $\mathcal{I} \in {\sf Hilb}^k(\complex^3)$. Then we can define
$V = \mathbb{C}[B_1 , B_2 , B_3] / \mathcal{I}$, and $B_i \in
\End(V)$ to be given as multiplication by $B_i \mod \mathcal{I}$ and $I
\in \Hom(\mathbb{C} , V)$ by $I(1) = 1 \mod \mathcal{I}$. Then all
the $B_i$ commute since they are realized as multiplication and the
stability condition holds since products of the
$B_i$'s times 1 span the whole of the polynomial ring $\mathbb{C}[B_1
, B_2 , B_3]$. Conversely, suppose we are given a quadruple of maps
$(B_1 , B_2 , B_3 , I)$. We introduce the map
\begin{eqnarray}
\mu \,:\, \mathbb{C}[B_1 , B_2 , B_3] &\longrightarrow& \mathbb{C}^k 
\cr f &\longmapsto& f(B_1 , B_2 , B_3) I(1)
\end{eqnarray}
which is well-defined since the $B_i$'s commute. Consider now the
subspace $\mathrm{im}(\mu) \subset \mathbb{C}^k$. This subspace is
$B_i$-invariant since it is the subspace spanned by the $B_i$
themselves and $\mathrm{im} (I) \subset \mathrm{im} (\mu)$. Then the
stability condition implies $\mathrm{im} (\mu) = \mathbb{C}^k$. This
means that the map $\mu$ is surjective. Then $\mathcal{I} := \ker \mu$
is an ideal with $\dim_\complex \mathbb{C}[B_1 , B_2 , B_3] /
\mathcal{I} = k$. Explicitly, one has
\begin{equation}
\mathcal{I} = \big\{ f(z)\in
\mathcal{O}_{\complex^3}\cong\mathbb{C}[B_1 , B_2 , B_3] ~\big|~ f(B_1
, B_2 , B_3) I(1) = 0 \big\} \ .
\label{idealabelian}\end{equation}

This ideal is isomorphic to the rank one cohomology sheaf
(\ref{cohsheafcoh}). By restricting the complex
(\ref{ADHMcomplexfinal}) with  $N=1$ to
$\complex^3=\PP^3\setminus\PP^2$ as before, the image of
(\ref{cohsheafcoh}) in $\mathcal{O}_{\complex^3}$ induced by the
localized maps $a_z$ and $b_z$ with $J=K=0$, and suitable matrix
identifications in Section~\ref{Barth}, is precisely the ideal
(\ref{idealabelian}). The proof
parallels~\cite[Proposition~2.7]{nakajima}. As an explicit example,
the charge~2 abelian instanton moduli space is
$\mathcal{M}_{1,2}(\complex^3)\cong
\complex^3\times\mathcal{O}_{\PP^2}(-1)$, where the first factor is
the space of $2\times 2$ matrices $(B_1,B_2,B_3)$ which parametrize
the center of mass of the instantons, while the second factor is the
resolution of the relative position singularity at the origin which
gives the size and orientation of the instanton configuration.

We have thus constructed an explicit correspondence between elements of
the Hilbert scheme and commuting matrices (with a stability
condition) which correspond to the quiver quantum mechanics. We will
now consider a fixed point and show that it can be parametrized by a
three-dimensional partition. We consider a torus $\mathbb{T}^3$
acting on $\mathbb{C}^3$ with generators $(t_1,t_2,t_3)$. This action
lifts to the Hilbert scheme. A fixed point given by $(B_1 , B_2 , B_3
, I)$ is characterized by the condition that an equivariant rotation
is equivalent to a gauge transformation
\begin{eqnarray} \label{condfixedpt}
t_1 \,B_1 &=& g\, B_1\, g^{-1} \ , \nonumber\\[4pt]
t_2\, B_2 &=& g\, B_2\, g^{-1} \ , \nonumber\\[4pt] t_3\, B_3
&=& g\, B_3 \,g^{-1} \ , \nonumber\\[4pt] I &=& g\, I
\end{eqnarray}
with $g\in GL(k,\complex)$. We use the weight decomposition
\beq
V = \bigoplus_{i,j,k\in\zed}\, V(i-1,j-1,k-1)
\label{Vweightdecomp}\eeq
with
\begin{equation} \label{weightdecomp}
V(i-1,j-1,k-1) = \big\{v \in V ~\big|~ g^{-1}\, v = t_1^{i-1}\,
t_2^{j-1}\, t_3^{k-1} \,v \big\} \ .
\end{equation}
The notation has been chosen such that $V(0,0,0)$ denotes the
subspace spanned by gauge-invariant vectors.

Consider a generic triple of integers $(i,j,k)$. According
to the definition (\ref{weightdecomp}) the only non-vanishing
components of the maps $(B_1,B_2,B_3,I)$ with respect to the splitting
(\ref{Vweightdecomp}) are given by
\begin{eqnarray}
B_1 \,:\, V(i,j,k) &\longrightarrow& V(i-1 , j , k) \ ,
\nonumber\\[4pt] B_2 \,:\, V(i,j,k)
&\longrightarrow& V(i , j-1 , k) \ , \nonumber\\[4pt] B_3 \,:\,
V(i,j,k) &\longrightarrow& V(i , j , k-1) \ , \nonumber\\[4pt] I \,:\,
W &\longrightarrow& V(0,0,0) \ .
\label{mapsnonzero}\end{eqnarray}
This can be seen as follows. By the stability condition the vector
space $V$ is spanned by elements of the form $B_1^p\, B_2^q\, B_3^m\,
I(1)$ with $p,q,m \in \nat_{0}$. In particular, the subspace
$V(0,0,0)$ is spanned by $I(1)$ and is one-dimensional. Consider now
acting on the vector $I(1)$ that generates this space with, say,
$B_1$. This gives the vector $B_1 I(1)$. Now we see that
\begin{equation}
g^{-1}\, \big(B_1 I(1)\big) = g^{-1}\, B_1\, g\, g^{-1}\, I(1) =
t_1^{-1}\, B_1\, I(1) 
\end{equation}
due to (\ref{condfixedpt}). This means that $B_1 I(1) \in
V(-1,0,0)$. The more general cases in (\ref{mapsnonzero}) are treated
similarly.

To make contact with the fixed point equation we parametrize the gauge
group by the $k\times k$ matrix $\phi$ and diagonalize it at the fixed
points. We also write $t_i = \e^{- \ii \epsilon_i}$. The defining
equation in (\ref{weightdecomp}) can be satisfied by picking the $k$
eigenvalues of $\phi$ and setting them equal to
\begin{equation}
\phi_{(i,j,k);l} = a_l+\epsilon_1\, (i-1) + \epsilon_2\, (j-1) +
\epsilon_3\, (k-1) \ .
\end{equation}
Using (\ref{mapsnonzero}), this reproduces the expected result from
(\ref{fixedpoint}). Then this 
weight decomposition of the vector space $V$ is equivalent to the
classification of the fixed points as done in the previous
sections. At each fixed point the eigenvalues of $\phi$ are determined
by the weights of the toric action. Following~\cite{nakajima} we will
now see how the definition (\ref{weightdecomp}) implies that the allowed
values of $\phi$ (i.e. the allowed non-trivial spaces
$V(i-1,j-1,k-1)$) are in correspondence with plane partitions.

We note that $V(i,j,k) = 0$ if one of $(i,j,k)$ is strictly
positive. Only negative or zero values are allowed. If we generalize this
reasoning to vectors of the form $B_i^p I(1)$ we conclude that
\begin{eqnarray}
\dim_\complex V(i,0,0) &\ge& \dim_\complex V(i-1,0,0) \ ,
\nonumber\\[4pt] \dim_\complex V(0,j,0) 
&\ge& \dim_\complex V(0,j-1,0) \ , \nonumber\\[4pt] \dim_\complex
V(0,0,k) &\ge& \dim_\complex V(0,0,k-1) \ ,
\end{eqnarray}
and that these dimensions can only be either zero or one. Intuitively,
we are ``constructing'' the exterior boxes of the three-dimensional
partition and each box is represented by a non-trivial vector
space. Now we will go to the boxes in the interior by using the
commutativity relations $[B_i , B_j]= 0$. We will
proceed by induction.

The commutativity relation $ [B_1 , B_2] = 0 $ ensures that the
configurations
\begin{eqnarray}
\xymatrix@1{
   V(i,j-1,k) \cong \complex \ar[r]^{B_1} & V(i-1,j-1,k) \neq 0 \cr
   V(i,j,k) \cong \complex \ar[u]^{B_2} \ar[r] & 0 \ar[u]
}
\end{eqnarray}
and
\begin{eqnarray}
\xymatrix@1{
   0 \ar[r] & V(i-1,j-1,k) \neq 0 \cr
   V(i,j,k) \cong \complex \ar[u] \ar[r]_{B_1} & V(i-1 , j , k) \cong
   \complex \ar[u]_{B_2}
}
\end{eqnarray}
are impossible, where the notation $V(i,j,k) \cong \complex$ stands
for the {\it assumption} that these spaces are one-dimensional. On the
other hand, the diagram
\begin{eqnarray}
\xymatrix@1{
   V(i,j-1,k) \cong \complex \ar[r]^{B_1} & V(i-1,j-1,k) \cr
   V(i,j,k) \cong \complex \ar[u]^{B_2} \ar[r]_{B_1} & V(i-1 , j , k)
   \cong \complex
   \ar[u]_{B_2}
}
\end{eqnarray}
implies that $V(i-1 , j-1 , k)$ has dimension 0 or 1. We can
think of $B_1$ and $B_2$ as the ``directions'' of the base of the
three-dimensional partitions. Then these conditions tell us that the
base is a usual Young tableau oriented as in~\cite{nakajima}.

Now we have to put boxes on top of this ``base'' Young tableau. What
are the allowed configurations? The equation $[B_1 , B_3] = 0$ implies
that the two diagrams 
\begin{eqnarray}
\xymatrix@1{
   V(i,j,k-1) \cong \complex \ar[r]^{B_1} & V(i-1,j,k-1) \neq 0 \cr
   V(i,j,k) \cong \complex \ar[u]^{B_3} \ar[r] & 0 \ar[u]
}
\end{eqnarray}
and
\begin{eqnarray}
\xymatrix@1{
   0 \ar[r] & V(i-1,j,k-1) \neq 0 \cr
   V(i,j,k) \cong \complex \ar[u] \ar[r]_{B_1} & V(i-1 , j , k) \cong
   \complex
   \ar[u]_{B_3}
}
\end{eqnarray}
are forbidden. Again the commutativity of
\begin{eqnarray}
\xymatrix@1{
   V(i,j,k-1) \cong \complex \ar[r]^{B_1} & V(i-1,j,k-1) \cr
   V(i,j,k) \cong \complex \ar[u]^{B_3} \ar[r]_{B_1} & V(i-1 , j , k)
   \cong \complex
   \ar[u]_{B_3}
}
\end{eqnarray}
implies that $\dim_\complex V(i-1 , j , k-1)$ is either 1 or 0. Finally, the
equation $[B_2 , B_3 ] = 0$ forbids the two configurations
\begin{eqnarray}
\xymatrix@1{
   V(i,j,k-1) \cong \complex \ar[r]^{B_2} & V(i,j-1,k-1) \neq 0 \cr
   V(i,j,k) \cong \complex \ar[u]^{B_3} \ar[r] & 0 \ar[u]
}
\end{eqnarray}
and
\begin{eqnarray}
\xymatrix@1{
   0 \ar[r] & V(i,j-1,k-1) \neq 0 \cr
   V(i,j,k) \cong \complex \ar[u] \ar[r]_{B_2} & V(i , j-1 , k) \cong
   \complex \ ,
   \ar[u]_{B_3}
}
\end{eqnarray}
while the commutativity of
\begin{eqnarray}
\xymatrix@1{
   V(i,j,k-1) \cong \complex \ar[r]^{B_2} & V(i,j-1,k-1) \cr
   V(i,j,k) \cong \complex \ar[u]^{B_3} \ar[r]_{B_2} & V(i , j-1 , k)
   \cong \complex
   \ar[u]_{B_3}
}
\end{eqnarray}
implies that $\dim_\complex V(i , j-1 , k-1)$ is either 1 or 0.

By induction this implies that each vector space $V(i,j,k)$ in
(\ref{Vweightdecomp}) is either 
one-dimensional or zero-dimensional. Moreover, the allowed
configurations correspond to three-dimensional partitions
where we put each box in the position of a one-dimensional vector
space. For example, we can construct the partition starting from a
two-dimensional Young tableau lying on the plane $(B_1 , B_2)$. On top
of each position $(i,j)$ corresponding to a box of the tableau we put
$\pi_{i,j}$ boxes. The diagrams above corresponding to the equations
$[B_1 , B_3] = [B_2 , B_3] = 0$ tell us that we can pile the boxes
only in the ``correct'' way, so that
\begin{equation}
\pi_{i,j} \ge \pi_{i+r , j+s} \qquad \mathrm{ for } \qquad r,s \ge
0
\end{equation}
with $k=\sum_{i,j\leq0}\,\pi_{i,j}$. Pictorially, the diagram
\begin{eqnarray}
\xymatrix@1{
    & & \scriptstyle V(0,0,-2) \ar[r] & \scriptstyle V(0,-1,-2) & &\cr
    & & \scriptstyle V(0,0,-1) \ar[dl] \ar[u] \ar[r] &\scriptstyle
    V(0,-1,-1) \ar[u] & &\cr
    &\scriptstyle V(-1,0,-1) &\scriptstyle V(0,0,0) \ar[u]^{B_3}
    \ar[dl]^{B_1} \ar[r]^{B_2} &
    \scriptstyle V(0,-1,0) \ar[dl] \ar[r] \ar[u] &\scriptstyle
    V(0,-2,0) \ar[dl] \ar[r] &
    \scriptstyle V(0,-3,0) \cr
    &\scriptstyle V(-1,0,0) \ar[dl] \ar[u] \ar[r] &\scriptstyle
    V(-1,-1,0) \ar[r] & \scriptstyle V(-1,-2,0)
    & &\cr
    \scriptstyle V(-2,0,0) & & & & &\cr
 }
\end{eqnarray}
is a simple plane partition. Note that one has to specify from the
beginning which vector spaces are non-trivial in order to get a
partition.

This construction easily generalizes to the nonabelian theory in
the broken phase $U(1)^N$. In this situation we can assume that
the relevant moduli space splits into a direct sum of sectors
labeled by the Higgs vevs $a_l$, each one essentially identical
to the Hilbert scheme of points. Consequently, the classification
above of the fixed points can be easily generalized by simply adding a
label $a_l$ (colour) to each vector space, $V_{a_l}(i,j,k)$. This
agrees with the classification suggested in~\cite{Sduality}, while
in~\cite{jafferis} the equations of the topological matrix model have
been interpreted in terms of skew partitions with $N$ ``corners'' on
the interior boundary. In the string picture this corresponds to a
situation in which the $N$ D6-branes are well separated in the
transverse space. The D0-brane bound states with each of the D6-branes
give $N$ copies of the melting crystal configuration.

Note that in this framework the role played by $I$ (representing the
0--6 open strings) is to simply label the colour of the
three-dimensional partition. In our approach we use the $N$
non-trivial components $I(1)_{(0,0,0);l}$ labelled by the Higgs vevs
$a_l$ to build $N$ individual three-dimensional partitions. Notice
also that the $k$ non-vanishing components of $B_i$ and $I$ at a fixed
point, corresponding to the number of boxes of the associated
three-dimensional Young diagram, are fixed completely by the matrix
equations (\ref{ADHMeqs}). This follows from the fact that the
matrices in the last equation of (\ref{ADHMeqs}) (the D-term
constraint) are all $k\times k$ and the diagonal components yield
exactly $k$ constraints. Thus the fixed points are indeed isolated.

Geometrically, these coloured partitions can be understood as fixed
points in the framed moduli space (\ref{sheavesonP3}) as
follows. There is a natural action of the torus $\torus^3\times
U(1)^N$ on $\mathcal{M}_{N,k}(\PP^3)$ induced by the $\torus^3$-action
on the toric manifold $\PP^3$ and the action of the maximal 
torus $U(1)^N$ on $W\otimes\mathcal{O}_{\ell^i_{\infty}}$, $i=1,2,3$, where
$W=\bigoplus_l\,W_l$ decomposes into irreducible representations
$W_l\cong\complex$ of $U(1)$. Then the $\torus^3\times U(1)^N$ fixed
points on $\mathcal{M}_{N,k}(\PP^3)$ are the coherent, torus invariant
sheaves $\mathcal{E}_{\vec\pi}=\mathcal{I}_{a_1}\oplus\cdots
\oplus\mathcal{I}_{a_N}$ with pointlike support
$Z=Z_{a_1}\sqcup\cdots\sqcup Z_{a_N}$ at the origin
of $\complex^3$, where $\mathcal{I}_{a_l}$ is a $\torus^3$-invariant
ideal sheaf supported on the $\torus^3$-fixed zero-dimensional
subscheme $Z_{a_l}\subset\complex^3=\PP^3\setminus p^i_\infty$ with the framing
$\mathcal{I}_{a_l}|_{\ell^i_\infty}\cong
W_l\otimes\mathcal{O}_{\ell^i_{\infty}}$. In particular, the weight
decomposition (\ref{Vweightdecomp}) coincides with
$H^0(\mathcal{O}_Z)$. 

\subsection{Quiver variety for Donaldson--Thomas
  data} \label{sec:DTdata}

At this stage we have provided a classification of the critical
points in the abelian theory that can be easily generalized to the
$U(1)^N$ picture. To complete the localization program we have to now
compute the quantum fluctuation factor around each critical point. This
can be done explicitly in our ADHM-type formalism since it provides a
direct parametrization of the compactified instanton moduli space.
The heuristic structure of ``fields'' and ``constraints'' that we
have exploited above to write down the ratio of determinants in
(\ref{quivercontint}) can be made precise in terms of an equivariant
index that counts BPS states. Geometrically, we will provide a
description of the (virtual) tangent space around each fixed point
(the ``fields''), and on top of it we build the normal bundle (the
``constraints''). This finally gives the last missing piece of
information needed to use the localization formula. We will assume
that the gauge symmetry is broken down to the maximal torus $U(1)^N$
and recover the abelian theory as a particular case. 

Recall that in the quiver description above the two vector spaces $V$
and $W$ with $\dim_\complex V = k$ and $\dim_\complex W = N$ represent
respectively the 
gas of $k$ D0-branes and the $N$ D6-branes. Naively, the bosonic
fields are elements (\ref{bosnaive}), while the fermionic fields
associated with the equations of motion live in (\ref{fermnaive}).
However, this is not really what we need since we have to compute
contributions coming from the fixed points and each fixed point is
characterized by the fact that the equivariant transformation
mixes with the linear transformations of the vector spaces $V$ and
$W$. To make this apparent we introduce a three-dimensional
$\mathbb{T}^3$-module $Q$ that acts on $V$. At a fixed point
$f\in\mathcal{M}$ we have to supplement the conditions
(\ref{bosnaive}) and (\ref{fermnaive}) with the information that
a gauge transformation is equivalent to an equivariant rotation.
Let us take the three-torus to be $\mathbb{T}^3 = (t_1 = \e^{\ii
\epsilon_1} , t_2 = \e^{\ii \epsilon_2} , t_3 = \e^{\ii
\epsilon_3})$, and introduce the following notation. 
$T_i$ is the one-dimensional module generated by $t_i$, $T_i\, T_j$ is
generated by $t_i\, t_j$ and $T_1\, T_2\, T_3$ by $t_1\, t_2\, t_3$,
and similarly for the dual modules $T_i^*:=T_i^{-1}$. We write $E_{l}$
for the module over $\torus^3 \times U(1)^N$ generated by $e_l=\e^{\ii
  a_l}$. For brevity we omit tensor product symbols between the
$T$-modules.

With this notation, at the fixed points of the $\torus^3 \times
U(1)^N$ action, which correspond from above to coloured partitions
$\vec\pi=(\pi_1,\dots,\pi_N)$, we decompose the vector spaces
\begin{eqnarray}
V_{\vec\pi} &=& \sum_{l=1}^N \,e_l~ \sum_{(i,j,k)\in \pi_l}\,
t_1^{i-1} \,t_2^{j-1}\,t_3^{k-1} \ , \nonumber\\[4pt] W_{\vec\pi} &=&
\sum_{l=1}^N\,e_l
\label{decompos}\end{eqnarray}
as $\torus^3 \times U(1)^N$ representations viewed as polynomials in
$t_1$, $t_2$, $t_3$ and $e_l$, $l=1,\dots,N$. For each $l$, the
sum over boxes of $\pi_l$ in $V_{\vec\pi}$ is the trace of the
$\torus^3$-action (i.e. the $\torus^3$-character) on
$\complex[B_1,B_2,B_3]/\mathcal{I}_{a_l}$.
Taking into account the action
of the torus on the vector space $V$ we can write
\begin{eqnarray}
B_1 &\in& \Hom(V_{\vec\pi},V_{\vec\pi}) \otimes T_1^{-1} \ ,
\nonumber\\[4pt] B_2 &\in& 
\Hom(V_{\vec\pi},V_{\vec\pi}) \otimes T_2^{-1} \ , \nonumber\\[4pt]
B_3 &\in& 
\Hom(V_{\vec\pi},V_{\vec\pi}) \otimes T_3^{-1} \ , \nonumber\\[4pt]
\varphi &\in& 
\Hom(V_{\vec\pi},V_{\vec\pi}) \otimes (T_1\, T_2 \,
T_3)^{-1} \ , \nonumber\\[4pt] I &\in& \Hom(W_{\vec\pi},V_{\vec\pi}) \
.
\end{eqnarray}
It is important to stress that these decompositions only hold at
the fixed points and not generically on the instanton moduli space. We
have chosen $I$ as before to be invariant under the toric action. This
amounts to setting $\varepsilon=0$, which we will see later on is
consistent with the known abelian case. Similarly for the constraints
(again at the fixed points) 
\begin{eqnarray}
\chi_1 &\in& \Hom (V_{\vec\pi},V_{\vec\pi}) \otimes T_1^{-1}\, T_2^{-1} \ ,
\nonumber\\[4pt] \chi_2 
&\in& \Hom (V_{\vec\pi},V_{\vec\pi}) \otimes T_1^{-1} \,T_3^{-1} \ ,
\nonumber\\[4pt]  \chi_3 &\in& 
\Hom (V_{\vec\pi},V_{\vec\pi}) \otimes T_2^{-1}\, T_3^{-1} \ ,
\nonumber\\[4pt] \chi_r 
&\in& \Hom (V_{\vec\pi},V_{\vec\pi}) \ , \nonumber\\[4pt] \xi &\in&
\Hom (V_{\vec\pi},W_{\vec\pi}) 
\otimes T_1^{-1}\, T_2^{-1} \,T_3^{-1} \ .
\end{eqnarray}

We will call an element
\beq
(B_1,B_2,B_3,\varphi,I)~\in~\big(Q\otimes{\rm Hom}(V,V)\big)~\oplus~
\big(\,\mbox{$\bigwedge^3Q$}\otimes{\rm Hom}(V,V)\big)~\oplus~
{\rm Hom}(W,V)
\label{DTdata}\eeq
a Donaldson--Thomas datum,  
where 
\begin{eqnarray}
Q &=& T_1^{-1} + T_2^{-1} + T_3^{-1} \ , \nonumber\\[4pt]
\mbox{$\bigwedge^2 Q$} &=& 
T_1^{-1}\, T_2^{-1} + T_1^{-1}\, T_3^{-1} + T_2^{-1}\,
T_3^{-1} \ , \nonumber\\[4pt] 
\mbox{$\bigwedge^3 Q$} &=& \det Q \ = \ T_1^{-1}\, T_2^{-1}\, T_3^{-1} \ .
\end{eqnarray}
Recall that there is a natural $GL(k,\complex)$ action on this
data. If we impose the stability condition on (\ref{DTdata}), then this
group action is free. Then we may define the geometric invariant
theory quotient of the subspace of (\ref{DTdata}) given by
$\mu_{\rm c}^{-1}(0)/\!/\,GL(k,\complex)$, where $\mu_{\rm c}=(\Ecal_i,\Ecal_I)$ is
a complex moment map. This is the \emph{quiver variety} for
Donaldson--Thomas data. This can also presumably be defined by
relaxing stability and taking instead a hyper-K\"ahler quotient of the
data (\ref{DTdata}) given by $\mu_{\rm c}^{-1}(0)\cap\mu_{\rm r}^{-1}(r)/\!/\,U(k)$,
where $\mu_{\rm r}=\Ecal_r$ is a real moment map. However, like the other
moduli spaces considered in this paper, we are not aware of any scheme
(or stack) construction on this set. Our calculations below suggest that there is an isomorphism $\mathcal{M}_{k,N}\cong\mu_{\rm c}^{-1}(0)\cap\mu_{\rm r}^{-1}(r)/\!/\,U(k)$ with the moduli space of framed, stable coherent sheaves (\ref{sheavesonP3}).

\subsection{Localization formula and character} \label{sec:loc}

Let $(B_1 , B_2 , B_3 , \varphi, I)$ be a Donaldon--Thomas datum
corresponding to the fixed point $\vec\pi$. Let us study the local
geometry of the instanton moduli space around this fixed
point. Consider the complex
\begin{equation} \label{adhmdefcomplex}
\xymatrix@1{
  \Hom(V_{\vec\pi} , V_{\vec\pi})
   \quad\ar[r]^{\!\!\!\!\!\!\!\!\!\!\!\!\!\!\!\!\sigma} &\quad
   {\begin{matrix} \Hom(V_{\vec\pi} , V_{\vec\pi}) \otimes Q 
   \\ \oplus \\
   \Hom(W_{\vec\pi} , V_{\vec\pi}) \\ \oplus  \\ \Hom(V_{\vec\pi} ,
   V_{\vec\pi}) \otimes \bigwedge^3 
   Q \end{matrix}}\quad \ar[r]^{\tau} & \quad
   {\begin{matrix} \Hom(V_{\vec\pi} , V_{\vec\pi}) \otimes \bigwedge^2
       Q \\ \oplus \\ 
       \Hom(V_{\vec\pi},W_{\vec\pi}) \otimes \bigwedge^3 Q \ .
   \end{matrix}}
}
\end{equation}
The map $\sigma$ is an infinitesimal (complex) gauge
transformation
\begin{eqnarray}
\sigma (\phi) = \left( \begin{matrix} \phi\, B_1 - B_1\,
    \phi \\ 
\phi\, B_2 - B_2\, \phi \\ \phi\, B_3 - B_3\,
\phi \\ \phi\, I - I \,\mbf a \\ \phi\, \varphi - \varphi\,
\phi \end{matrix} \right) \ ,
\end{eqnarray}
while the map $\tau$ is the differential of the equations that
define the moduli space $[B_i, B_j] = 0$ given by
\begin{eqnarray}
\tau\left( \begin{matrix} Y_1 \\ Y_2 \\ Y_3 \\ s \\ Y_4
\end{matrix} \right) = \left( \begin{matrix} [B_1 , Y_2] + [Y_1 ,
B_2] \\ [B_1 , Y_3] + [Y_1 , B_3] \\ [B_2 , Y_3] + [Y_2 , B_3] \\
s^{\dagger}\, \varphi + I^{\dagger}\, Y_4
\end{matrix} \right) \ ,
\end{eqnarray}
where one can think of $Y_i$ as $\delta B_i$ and so on.

The complex (\ref{adhmdefcomplex}) is the matrix quantum mechanics
analog of the instanton deformation complex (\ref{defcomplex}). In a
similar way, its first cohomology is a local model of the Zariski
tangent space to the moduli space, while its second cohomology
parametrizes obstructions. This is exactly the information we need to
integrate (\ref{eulerobstr}). We conjecture, along the lines implied in Section~\ref{sec:DTdata} above, that this integral reproduces the localization of a virtual cycle on the geometric moduli space (\ref{sheavesonP3}).

We want to apply the
Duistermaat--Heckman localization formula (\ref{DHformula}), or better
its supersymmetric generalizations~\cite{Bruzzo:2002xf}, to the
integral (\ref{eulerobstr}). This involves the ratio of the top Chern
class of the obstruction bundle over the weights coming from the
tangent space. For what concerns the computation of the Chern classes
we can decompose the tangent and normal bundles over the moduli
space as Whitney sums of line bundles by using the splitting principle
as $T \mathcal{M} =\bigoplus_i \,\mathcal{L}_i$ and $\mathcal{N} =
\bigoplus_i\,\mathcal{Q}_i$. Accordingly, the equivariant Chern
polynomials are given by
\begin{eqnarray}
c(T \mathcal{M}) &=& \prod_{i=1}^n\, \big( c_1 (\mathcal{L}_i) + w_i [T
\mathcal{M}] \big)\ , \nonumber\\[4pt] c( \mathcal{N}) &=& \prod_{i=1}^n
\,\big( c_1 (\mathcal{Q}_i) + w_i [\mathcal{N}] \big) \ ,
\end{eqnarray}
where the local weights $w_i$ in general depend on the toric parameters
$(\epsilon_1 , \epsilon_2 , \epsilon_3 , \mbf a)$. The same
information is contained in the equivariant Chern characters
\begin{eqnarray}
\mathrm{ch}(T \mathcal{M}) &=& \sum_{i=1}^n \,\e^{c_1 (\mathcal{L}_i) +
w_i [T \mathcal{M}]} \ , \nonumber\\[4pt] \mathrm{ch}( \mathcal{N})
&=& \sum_{i=1}^n \,\e^{c_1 (\mathcal{Q}_i) + w_i [\mathcal{N}]} \ .
\end{eqnarray}

The Chern classes of the line bundles do not contribute to the
localization formula since they have to be evaluated at a point
(and the critical points are isolated). Then we can use directly
the above expansions to extract the relevant weights to be used in
the localization formula. In practise this is accomplished via the
transform~\cite{sw,Bruzzo:2002xf}
\begin{equation} \label{conversion} \sum_{i=1}^n\,
n_i~\e^{w_i} ~\longmapsto ~\prod_{i=1}^n\,w_i^{n_i} \ .
\end{equation}
This means that all the relevant data that enter in the
localization formula are already contained in the equivariant
index of the complex (\ref{adhmdefcomplex}). To be precise
the index computes the virtual sum $H^1\ominus H^0\ominus H^2$ of
cohomology groups. We assume that $H^0$ vanishes, which is
equivalent to restricting attention to irreducible connections. Using
(\ref{conversion}) we see then that the equivariant index computes 
exactly the inverse of the ratio of the weights that enter in the
localization formula. The equivariant index is given in terms of
the characters of the representation evaluated at the fixed point
as
\begin{eqnarray} \label{character}
\chi_{\vec\pi} \big(\mathbb{C}^3\big)^{[k]} &=&  V^*_{\vec\pi}
\otimes V_{\vec\pi} 
\otimes \big(T_1^{-1} + T_2^{-1} + T_3^{-1} + T_1^{-1}\, T_2^{-1}\,
T_3^{-1}\big) +  W^*_{\vec\pi} \otimes V_{\vec\pi} \cr & & -
 V^*_{\vec\pi} 
\otimes V_{\vec\pi} \otimes \big(1 + T_1^{-1}\, T_2^{-1} + T_1^{-1}\, T_3^{-1} +
T_2^{-1}\, T_3^{-1}\big) -  V^*_{\vec\pi} \otimes W_{\vec\pi}
\otimes T_1^{-1}\, 
T_2^{-1}\, T_3^{-1} \nonumber\\[4pt] &=&  W^*_{\vec\pi}
\otimes V_{\vec\pi} - 
\frac{ V^*_{\vec\pi} \otimes W_{\vec\pi}}{t_1\, t_2\, t_3} +
 V^*_{\vec\pi} 
\otimes V_{\vec\pi} ~\frac{(1-t_1)\, (1-t_2)\, (1-t_3)}{t_1\, t_2\,
  t_3} \ . 
\end{eqnarray}

In the abelian case, $W \cong\mathbb{C} $ and we can formally
set $W=1$. Then the character (\ref{character}) reproduces exactly
the vertex character computed in~\cite{MNOP}. It is not clear what is the
precise relation of the above construction with the more geometric
approach of~\cite{MNOP}, since in the gauge theory picture there
are several matter fields which do not figure into the computations
of~\cite{MNOP}. Nevertheless, it is reassuring that we can reproduce 
the result of~\cite{MNOP} without resorting to the evaluation of virtual
fundamental classes but only with arguments well rooted in our
physical intuition. It would be very interesting to understand in the
language of~\cite{MNOP} what the meaning is of the vector space $W$. A
final remark about the equivariant transformation of $I$, i.e. the
choice $\varepsilon = 0$. This can be now justified since it is
the choice that reproduces the abelian character.

{}From this discussion we can immediately write down the partition
function
\begin{equation}
\mathcal{Z}_{\rm DT}^{U(1)^N}\big(\complex^3\big)
 = \sum_{\vec\pi}\, \mathcal{Z}_{\vec\pi} ~\e^{\ii \vartheta\, |\vec\pi|}
 \ ,
\end{equation}
where $\mathcal{Z}_{\vec\pi}$ is what the matrix integral
(\ref{quivercontint}) computes and
is given by (\ref{character}) through the rule
(\ref{conversion}). However, its explicit form is not very
illuminating. To obtain a more manageable form let us simplify the
character a bit. We begin by looking at the abelian theory with
$N=1$. Then
\begin{equation}
\chi_{\pi, {\rm ab}} \big(\mathbb{C}^3\big)^{[k]} =  V_{\pi} -
\frac{ V^*_{\pi} }{t_1\, t_2\, t_3} +
 V^*_{\pi} \otimes V_{\pi}~ 
\frac{(1-t_1)\, (1-t_2)\, (1-t_3)}{t_1\, t_2\, t_3}
\end{equation}
where at a fixed point
\begin{equation}
V_{\pi} = \sum_{ (i,j,k)\in {\pi}}\, t_1^{i-1}\,
t_2^{j-1}\, t_3^{k-1} \ .
\end{equation}
One can easily see that
\begin{equation}
\chi_{\pi , {\rm ab}} \big(\mathbb{C}^3\big)^{[k]} = {\sf T}_{\pi}^+ +
{\sf T}_{\pi}^-
\end{equation}
where
\begin{eqnarray}
{\sf T}_{\pi}^+ &=& V_{\pi} - V_{\pi} \otimes  V^*_{\pi}~
\frac{(1-t_1)\,(1-t_2)}{t_1 \,t_2} \ , \nonumber\\[4pt]
{\sf T}_{\pi}^- &=& - \frac{ 
 V^*_{\pi}}{t_1\, t_2 \,t_3} + V_{\pi}\otimes
 V^*_{\pi}~\frac{(1-t_1)\,(1-t_2)}{t_1 \,t_2 \,t_3} \ .
\label{calTab}\end{eqnarray}
With the dual operation $t\mapsto t^*= t^{-1}$, this
splitting of the character has the remarkable property 
\begin{equation} \label{splitproperty}
\big({\sf T}_{\pi}^+\big)^* \,\big|_{t_1 \,t_2 \,t_3 = 1} = -
{\sf T}_{\pi}^- \big|_{t_1 \,t_2 \,t_3 = 1} \ .
\end{equation}
Note that this property is true only when one imposes the Calabi--Yau 
condition $t_1 \,t_2\, t_3 = 1$.

What is remarkable about the property (\ref{splitproperty}) is that
due to (\ref{conversion}) the contribution to the full fluctuation
determinant is a minus sign to some power and this power can be
computed to be ${\sf T}_{\pi}^+ (t_1 = t_2 = t_3 =
1)$. Explicitly, one has
\begin{eqnarray}
\chi_{\pi,{\rm ab}} \big(\complex^3\big)^{[k]} \big|_{t_1 \,t_2\,
  t_3=1} &=& {\sf T}^+_{\pi} 
\big|_{t_1\, t_2\, t_3=1} + {\sf T}^-_{\pi} \big|_{t_1\, t_2\,
  t_3=1} \nonumber\\[4pt] &=& 
{\sf T}^+_{\pi} \big|_{t_1\, t_2\, t_3=1} -
\big({\sf T}^+_{\pi}\big)^*\,
\big|_{t_1\, t_2\, t_3=1} \nonumber\\[4pt] &=& \sum_{i=1}^n\, n_i~
\e^{w_i[\epsilon_1 , 
\epsilon_2 , \epsilon_3]} - \sum_{i=1}^n\, n_i~ \e^{-w_i[\epsilon_1 ,
\epsilon_2 , \epsilon_3]} \ .
\end{eqnarray}
By using the transform (\ref{conversion}) we can write the
contribution to the partition function as
\begin{equation}
\mathcal{Z}_\pi=\prod_{i=1}^n\, \frac{w_i [\epsilon_1 , \epsilon_2 ,
\epsilon_3]^{n_i}}{\big(-w_i [\epsilon_1 , \epsilon_2 ,
\epsilon_3]\big)^{n_i}} = \prod_{i=1}^n\, (-1)^{n_i} = (-1)^{\sum_i\, n_i} \
, 
\end{equation}
and the sum over the multiplicities $n_i$ can be obtained by
taking ${\sf T}^+_{\pi} \big|_{t_1\, t_2\, t_3=1}$ and setting the
weights $w_i$ to zero, i.e. setting $\epsilon_i$ to zero or $t_i$ to
one. It is easy to see that ${\sf T}_{\pi}^+ (t_1 =
t_2 = t_3 = 1) = |\pi|$ since the second term of
${\sf T}^+_{\pi}$ in (\ref{calTab}) vanishes at $t_i=1$. Thus
we conclude that the partition function in this case is given by
\begin{equation}
\mathcal{Z}_{\rm DT}^{U(1)}\big(\complex^3\big) = \sum_{\pi}\,
(-1)^{|\pi|}~ \e^{\ii \vartheta\, |\pi|} \ ,
\end{equation}
which reproduces the MacMahon function (\ref{DTMq}) with the usual
redefinition $- \e^{\ii \vartheta} = \e^{- g_s} = q$.

These arguments carry over to the $U(1)^N$ theory with only minor
modifications. Now we have to consider the full character
(\ref{character}). The splitting
\begin{eqnarray} \label{nabsplit}
{\sf T}_{\vec\pi}^+ &=& V_{\vec\pi}\otimes  W^*_{\vec\pi}
- V_{\vec\pi}\otimes  V^*_{\vec\pi}~ 
\frac{(1-t_1)\,(1-t_2)}{t_1\, t_2} \ , \nonumber\\[4pt]
{\sf T}_{\vec\pi}^- &=& - 
\frac{ V^*_{\vec\pi}\otimes W_{\vec\pi}}{t_1\, t_2\, t_3} +
V_{\vec\pi}\otimes  V^*_{\vec\pi}~ 
\frac{(1-t_1)\,(1-t_2)}{t_1\, t_2\, t_3}
\label{calTnonab}\end{eqnarray}
is helpful in simplifying the computation. At the fixed points
$\vec\pi$ the vector spaces $V$ and $W$ decompose as in
(\ref{decompos}). The property (\ref{splitproperty}) still holds but 
now the dual involution is
defined as $(\epsilon_1 , \epsilon_2 , \epsilon_3 , \mbf a)
\mapsto (-\epsilon_1 , -\epsilon_2 , -\epsilon_3 , -\mbf a)$.
By a similar argument as above we need only evaluate
${\sf T}_{\vec\pi}^+$ at $(\epsilon_1 , \epsilon_2 , \epsilon_3 ,
\mbf a) = (0,0,0,\mbf0)$. The second term in (\ref{calTnonab}) again
drops out and the first one gives
\begin{equation}
\sum_{l=1}^N ~\sum_{l'=1}^N ~\sum_{(i,j,k)\in\pi_{l'}}\,1 = N\,
\sum_{l'=1}^N\, |\pi_{l'}| \ .
\end{equation}

To complete the computation of the partition function, the only
missing ingredient now is the instanton
action. Analogously to~\cite{nakajima2,Losev:2003py}, we write the
universal sheaf $\mathcal{E}$ on the moduli space
$\mathcal{M}_{N,k}(\PP^3)$ as
\beq
\mathcal{E}=W~\oplus~ V\otimes\big(S^-\ominus S^+\big)
\label{Espinordef}\eeq
where $S^\pm$ are the positive/negative chirality spinor bundles over
$\PP^3$, localized at a point of the fibre of
$\complex^3=\PP^3/\PP^2$. At a 
critical point $\vec\pi$ we regard (\ref{Espinordef}) as a virtual
$\torus^3\times U(1)^N$ representation. By using the correspondence
between spinors and differential forms given by twisting the spinor
bundles to get $S^{\pm}\cong\Omega_{\PP^3}^{{\rm even}/{\rm
  odd},0}$~\cite{Blau:1997pp}, we can derive the Chern character
\bea
\mathrm{ch}(\mathcal{E}_{\vec\pi}) &=&
W_{\vec\pi}+(t_1+t_2+t_3+t_1\,t_2\,t_3-1-t_1\,t_2-t_2\,t_3-t_1\,t_3)~
V_{\vec\pi} \nonumber\\[4pt] &=&
 W_{\vec\pi} - (1-t_1) \,(1-t_2) \,(1-t_3)~ V_{\vec\pi} \ .
\label{Chernmatrix}\eea
In the quiver formalism we work in an instanton sector with fixed
charge $k=\dim_\complex V_{\vec\pi}$ (so that the instanton action is
proportional to $k$), and this corresponds to the total number of
boxes of the partition by the same arguments used in the
classification of the fixed points. Each subspace in the weight
decomposition of the vector space $V_{\vec\pi}$ is one-dimensional and
corresponds to a box in the partition $\vec\pi$. The generating
function for the nonabelian invariants is thus
\be
\mathcal{Z}_{\rm DT}^{U(1)^N}\big(\complex^3\big) =\sum_{\vec\pi}\,
(-1)^{N\, |\vec\pi|} ~\e^{\ii \vartheta \,|\vec\pi|} \ ,
\ee
which coincides with eq.~(\ref{genfnnonabinvs}).

\subsection{Comparison with the noncommutative gauge theory}

The results described in this section have a clear interpretation in
terms of the noncommutative gauge theory of Section \ref{NCGT} that
was spelled out in detail in Section \ref{Matrixsheaves}. We can
formally take this a step further and stress some close similarities
between the two approaches. Recall that for the classification of the
fixed points in the noncommutative gauge theory one simply needs to
construct the Hilbert space on which the noncommutative algebra
$\mathcal{A}$ is represented. In particular, the ratio of fluctuation
determinants has the same structure, the main difference being that in
the noncommutative gauge theory we are dealing with the determinants
of operators acting on a separable Hilbert space which gives directly
the result, while in the matrix model after the computation of the
determinants there is still the integral over the Cartan subalgebra to
do. The evaluation of the contour integral (\ref{quivercontint}) was
finally sidestepped by constructing an explicit local model for the
instanton moduli space. The computation of the associated index is
equivalent to a direct use of the localization formula.

We can build a formal dictionary to go back and forth between the two
approaches. This could prove very helpful in extending our general
formalism to other setups. The equivariant Chern character
(\ref{Chernmatrix}) can be derived precisely in the noncommutative
gauge theory. If we use the notation
\begin{equation}
\chi_{\mathcal{I}}(t):=\mathrm{ch}_{\vec\pi}(t) = W_{\vec\pi} -
\big(1-\e^{t \,\epsilon_1}\big)\, \big(1-\e^{t\,
  \epsilon_2}\big) \,\big(1-\e^{t \,\epsilon_3}\big)~ V_{\vec\pi}
\end{equation}
along with the redefinitions $t_i=\e^{t\,\epsilon_i}$ and
$e_l=\e^{t\,a_l}$, then the integrand of (\ref{ratioindex})
formally reproduces the character (\ref{character}) up to the
perturbative contribution $W_{\vec\pi}\otimes W_{\vec\pi}^*$ and an
irrelevant overall sign. The sign mismatch was explained above in
terms of the alternating sign in the definition of the index. The role
of the exponentiation and of the integral over $t$ is to reproduce the
transform (\ref{conversion}). Altogether, we can take this as a rule
to compute the equivariant index from the ratio of fluctuation
determinants as computed in the noncommutative gauge theory. This is
perhaps not surprising as field theoretically both approaches are just
two different ways to handle the ill-defined localization in the
original cohomological gauge theory. An application of this formalism
will be presented in the next section where we compute the partition
function of the $U(1)^N$ model on a generic toric Calabi--Yau
manifold.

\bigskip

\section{Partition function on a toric
  manifold\label{Nonabpartfn}}

In the previous sections we have derived a precise dictionary between
the noncommutative gauge theory and an auxiliary matrix quantum
mechanics. We will now apply our formalism in a controlled
setup, switching between the two approaches when convenient. In
particular, we will write down the partition function of the
$U(1)^N$ gauge theory on an arbitrary toric Calabi--Yau
threefold, extending the gauge theory prescription
of~\cite{Iqbal:2003ds} for handling the instanton counting on a
generic toric manifold and the geometric results of~\cite{MNOP} to
compute the quantum fluctuation determinants. Although our final result
does not provide any new  geometrical invariants of threefolds, as it
does not capture the full nonabelian structure of the
Donaldson--Thomas invariants, it does compute the number of BPS bound
states of branes in this specific regime of the theory.

\subsection{Instanton action\label{InstactionX}}

Let $X$ be a nonsingular toric threefold with K\"ahler two-form $k_0$
and Newton polyhedron $\Delta(X)$, the image of $X$ under the moment
map associated to the toric action on $X$. The vertices $f$ of
$\Delta(X)$ correspond to the fixed points of the $\torus^3$-action on
$X$. For each $f$ there is a $\torus^3$-invariant open $\complex^3$
chart centred at the fixed point. On each patch we can choose
coordinates corresponding to the directions $(t_1, t_2, t_3)$, where
$(t_1, t_2, t_3)= (\e^{\ii\epsilon_1} , \e^{\ii \epsilon_2}, \e^{\ii
  \epsilon_3})$ are the 
generators of the toric action. The edges of the polyhedron
$\Delta(X)$ correspond to the $\torus^3$-invariant lines of $X$. They
represent generic projective lines $\PP^1$ which join two fixed points
$f_1$ and $f_2$ in $\Delta(X)$. Instantons of $U(N)$ noncommutative
gauge theory on each $\complex^3$ patch in the Coulomb phase
correspond to sums of monomial ideals
$\mathcal{I}_f=\mathcal{I}_{a_{1,f}}\oplus\cdots
\oplus\mathcal{I}_{a_{N,f}}$ in $\mathcal{O}_{\complex^3}$ associated
to coloured three-dimensional partitions
$\vec\pi_f=(\pi_{1,f},\dots,\pi_{N,f})$. Such collections of ideals 
correspond globally to $\torus^3\times U(1)^N$-invariant torsion free
sheaves $\mathcal{E}$ 
of rank $N$ on $X$ with associated subscheme $Z$ supported on the fixed
points in $\Delta(X)$ and the lines connecting them, together with a
framing $\mathcal{E}_\infty\cong W\otimes\mathcal{O}_X$.

We need to compute $\Tr_{\mathcal{H}_{\mathcal{I}_f}}\big( \e^{t \,\Phi}\big)$,
  where $\Phi$ is a nonabelian Higgs field and the Hilbert space
  $\mathcal{H}_{\mathcal{I}_f}$ corresponds to a three-dimensional
  partition $\vec\pi_f$ with fixed asymptotic behaviour at
  infinity. One can write down directly the Chern character at a fixed
  point $f\in\Delta(X)$ corresponding to the
generalized instanton configuration with fixed asymptotics as
\begin{eqnarray}
\chi_{\mathcal{I}_f} (t) &=&  \sum_{l=1}^N\, e_{l,f}\,
\Big( 1- (1-t_1) \,
(1-t_2)\, \sum_{(i,j)\in\lambda_{3,l,f}} \,t_1^{i-1} \,t_2^{j-1} -
(1-t_1)\, (1-t_3) \,
\sum_{(i,k)\in\lambda_{2,l,f}}\, t_1^{i-1}\, t_3^{k-1} \nonumber\\
&&\qquad\qquad-\,(1-t_2) \,(1-t_3) \,
\sum_{(j,k)\in\lambda_{1,l,f}}\, t_2^{j-1}\, t_3^{k-1} \nonumber\\ &&
\qquad\qquad -\,(1-t_1)\, (1-t_2) \,
(1-t_3)\, \sum_{(i,j,k)\in\pi_{l,f}}\, t_1^{i-1}\, t_2^{j-1}\,
t_3^{k-1} \Big) \ .
\label{nabcharacterX}\end{eqnarray}
The first set of terms represent the vacuum contribution which is fixed
asymptotically by the $3N$ two-dimensional Young tableaux
$(\lambda_{1,l,f},\lambda_{2,l,f},\lambda_{3,l,f})$, the asymptotics
of $\pi_{l,f}$ in the coordinate directions labelling the
corresponding edges emanating from the vertex $f$. We are covering the
toric manifold $X$ with $\complex^3$ patches and solving the
noncommutative gauge theory in each patch. The asymptotic boundary
conditions are necessary to glue the patches together. The ordinary
two-dimensional partition which is the asymptotic condition in the
$i$-th direction $t_i$ is denoted by $\lambda_{i,l,f}$, with the index
$l=1,\dots,N$ reminding us from which sector of the Hilbert space
$\mathcal{H}_{\mathcal{I}_f}$ they come from. The last term
corresponds to the three-dimensional partitions $\pi_{l,f}$ which 
should now be understood as properly renormalized for $n\gg0$
with volume 
\begin{equation}
|\pi_{l,f}| = \Big(\,\sum_{\scriptstyle  (i,j,k)\in \pi_{l,f}  \atop
\scriptstyle i,j,k \le n} \, 1 \,\Big) - (n+1)\, \big(
|\lambda_{1,l,f}| + |\lambda_{2,l,f}| + |\lambda_{3,l,f}| \big) \ .
\end{equation}

In the following we will impose the Calabi--Yau condition
(\ref{epsconstr}) and define $x=
\frac{\epsilon_2}{\epsilon_1}$. The equivariant parameters are
defined differently in each $\complex^3$ patch and in such a way
that they match when gluing the patches together as in the topological
vertex gluing rules. The contribution of the third Chern character to
the instanton action, i.e. the coefficient
$\chi_{\mathcal{I}_f}^{(3)}$ of $t^3$ in the small $t$ expansion of
(\ref{nabcharacterX}), gives 
\bea\label{nabinstaction}
&& \frac{\ii}{48 \pi^3} \,\int_X\, \Tr F_A \wedge F_A \wedge F_A ~=~
\sum_{f\in\Delta(X)}~ \sum_{l=1}^N\, \Big( - \frac{a_{l,f}^3}{6
  \epsilon_{1,f}\, x_f \, 
(1+x_f)} + \sum_{(i,j,k)\in\pi_{l,f}} \, 1 \nonumber\\ &&
\qquad\quad+\,\sum_{(i,j)\in\lambda_{3,l,f}} \,\left( 
-\frac{1}{2} + j + \frac{i-j}{1+x_f} +
\frac{a_{l,f}}{\epsilon_{1,f}\, (1 + x_f)} \right) \\ &&
\qquad\quad+\, 
\sum_{(i,k)\in\lambda_{2,l,f}} \,\left( -\frac{1}{2} + k +
  \frac{k-i}{x_f} - 
\frac{a_{l,f}}{\epsilon_{1,f}\, x_f } \right) +
\sum_{(j,k)\in\lambda_{1,l,f}} \,\left( -\frac{1}{2} + k - j \,x_f + k 
  \,x_f - \frac{a_{l,f}}{\epsilon_{1,f} } \right) \Big) \nonumber
\eea
where we have written down explicitly the sum over fixed points $f$
associated with the vertices of the toric diagram $\Delta(X)$. Each 
vertex contribution has a proper factor associated with the vertex
itself, but it also comes with edge factors associated with the
asymptotic two-dimensional partitions. To properly treat the edge
factors one has to consider both contributions associated to an edge
coming from the two adjacent vertices. This will be done explicitly in
the following.

The first contribution in (\ref{nabinstaction}) is independent of
the partitions $\vec\pi_f$, and as such can be factored out as a
``perturbative'' contribution as before. The second contribution gives 
a factor of $|\vec\pi_f|$ associated to each vertex of the
toric diagram and generalizes the instanton action. The
remaining three terms come from the contribution of the
asymptotics and have to be combined with the analogous contributions
coming from other fixed points. For example, let us choose a fixed
point $f$ and consider the term where we sum over boxes of
$\lambda_{3,l,f}$. The vertex we are considering is joined to another
vertex along an edge and the partitions on the edge have the structure
given by $\lambda_{3,l,f}$. The edge represents a $\torus^3$-invariant
rational curve $\PP^1$ with 
normal bundle $\mathcal{O}_{\PP^1}(-m_1) \oplus \mathcal{O}_{\PP^1}(-m_2)$
determining the local geometry of $\Delta(X)$ near the edge. The
Calabi--Yau condition implies
\beq
m_1+ m_2 = 2 \ .
\label{CYmconstr}\eeq
The two contributions we have to consider are exactly the same but
each one is expressed in terms 
of the equivariant parameters that are associated with the local
coordinates in each $\complex^3$ patch. The relation between the
equivariant parameters is then given by the transition function
between the two coordinate charts as~\cite{MNOP}
\begin{eqnarray}
\epsilon_{1 , f_2} &=& \epsilon_{1 , f_1} + m_1 \,\epsilon_{3 , f_1} \
, \nonumber\\[4pt] \epsilon_{2 , f_2} &=& \epsilon_{2 , f_1} + m_2\,
\epsilon_{3 , 
f_1} \ , \nonumber\\[4pt] \epsilon_{3 , f_2} &=& - \epsilon_{3 , f_1}
\ , \nonumber\\[4pt] x_{f_2}
&=& \frac{\epsilon_{2 ,f_2}}{\epsilon_{1 , f_2}} \ = \ \frac{x_{f_1} -
m_2 - m_2 \,x_{f_1}}{1 - m_1 - m_1\, x_{f_1}} \ ,
\label{transfns}\end{eqnarray}
where $(f_{1},f_2)$ labels the two fixed points joined by the edge.

After a bit of algebra we find
\begin{eqnarray} 
&& \sum_{l=1}^N~ \sum_{(i,j)\in\lambda_{3,l,f_1}}\, \left(  -\frac{1}{2} +
  j + 
\frac{i-j}{1+x_{f_1}} + \frac{a_{l,f_1}}{\epsilon_{1,f_1}\, (1 +
x_{f_1})} - \frac{1}{2} + j + \frac{i-j}{1+x_{f_2}} +
\frac{a_{l,f_2}}{\epsilon_{1,f_2}\, (1 + x_{f_2})} \right) \nonumber\\
&& \qquad ~=~
\sum_{l=1}^N ~\sum_{(i,j)\in\lambda_{3,l,f_1}} \,\left( -1 + 2 j + i \,m_1
  - j\, m_1 
\right) \nonumber\\[4pt] && \qquad ~=~ \sum_{l=1}^N
~\sum_{(i,j)\in\lambda_{3,l,f_1}}\, \big( m_1 
\,(i-1) + m_2\, (j-1) + 1 \big)
\end{eqnarray}
where in the last equality we have used (\ref{CYmconstr}) to express
the edge contribution in a universal form. The Higgs vevs cancel since
we require $a_{l , f_1} = a_{l , f_2}$ as part of the gluing
conditions,\footnote{Each $a_{l,f}$ gives the asymptotic boundary
  condition on the instanton labelled by $f$, and we can glue together
  two instantons at infinity if and only if they have the same value
  of $a_{l,f}$.} and the edge contribution along the ``direction''
$t_3$ is the sum of $N$ terms which all have the same form as in the 
abelian $N=1$ theory. After performing similar computations along the 
other two ``directions'' $t_2$ and $t_1$, the ${\rm ch}_3$ term of the
instanton action gives
\begin{equation}
{\sf I}(\vec\pi_f) = \sum_{f\in\Delta(X)}~ \sum_{l=1}^N \,|\pi_{l,f}|
+ \sum_{e \in 
\Delta(X)} ~\sum_{l=1}^N ~\sum_{(i,j) \in \lambda_{l,e}}\, \big(
m_{1,e} \,(i-1) + m_{2,e} \,(j-1) + 1 \big) \ ,
\label{Ipif}\end{equation}
where the sum over $f$ runs through the vertices of the toric diagram 
$\Delta(X)$ while the sum over $e$ runs through the edges.

To complete the evaluation of the instanton action one needs to also
consider the second Chern character, i.e. the coefficient
$\chi^{(2)}_{\mathcal{I}_f}$ of $t^2$ in the small $t$ expansion of
(\ref{nabcharacterX}). One finds 
\begin{eqnarray}
&& -\frac{1}{8 \pi^2}\, \int_X\, k_0 \wedge \Tr F_A \wedge F_A ~=~
\sum_{f\in\Delta(X)} \,
\frac{H_f \,\chi^{(2)}_{\mathcal{I}_f}}{\epsilon_{1,f}\,
    \epsilon_{2,f}\, \epsilon_{3,f}} \\[4pt]
&& \qquad\qquad ~=~\sum_{f\in\Delta(X)}~ \sum_{l=1}^N \,\bigg(
  \frac{a_{l,f}^2\, H_f}{2 \epsilon_{1 , f}\, 
\epsilon_{2 , f} \,\epsilon_{3 , f}} - \sum_{(i,j)\in\lambda_{3,l}}\,
\frac{H_f}{\epsilon_{3 , f}} - \sum_{(i,k)\in\lambda_{2,l}}\,
\frac{H_f}{\epsilon_{2 , f}} - \sum_{(j,k)\in\lambda_{1,l}}\,
\frac{H_f}{\epsilon_{1 , f}} \bigg) \nonumber
\end{eqnarray}
where $H_f$ is the value at the fixed point $f$ of the hamiltonian $H$
associated with the vector field $\Omega$ that generates the equivariant
rotations, i.e., $\dd H=\imath_\Omega k_0$. The term depending on the
Higgs vevs can again be dropped and we can analyse the edge
contributions as before with the result
\begin{equation}
-\sum_{e\in \Delta(X)}~ \sum_{l=1}^N \,t_e\, |\lambda_{l,e}|
\end{equation}
where $t_e = \frac{H_{f_1} - H_{f_2}}{\epsilon_{e}}$, for each pair
of fixed points $(f_1 , f_2)$ connected by the edge $e$, is the
K\"ahler parameter of the line $\PP^1$ associated to $e$. Altogether,
the instanton weight is given by 
\begin{equation}
\e^{\ii \vartheta\,{\sf I}(\vec\pi_f)}~ \e^{-\sum_{e \in \Delta(X)}~
\sum_{l=1}^N\, t_e\, |\lambda_{l,e}|} \ .
\end{equation}

\subsection{Fluctuation determinants}

The next step in the evaluation of the partition function is to
determine the ratio of quantum fluctuation determinants. According to
our rule of the previous section, we can write down the ratio in the
noncommutative gauge theory and read off from the integrand the
equivariant Euler characteristic of the complex (\ref{adhmdefcomplex})
using the techniques of~\cite{MNOP}. The ratio at a fixed point
$f\in\Delta(X)$ comes out to be
\begin{equation}
\frac{\chi_{\mathcal{I}_f}(t)\,
  \chi_{\mathcal{I}_f}(-t)}{(1-t_1)\, (1-t_2)\, (1-t_3)} 
\end{equation}
as in (\ref{ratioindex}), with $\chi_{\mathcal{I}_f}(t)$ given
by (\ref{nabcharacterX}). The character at
the fixed point $f$ decomposes as
\begin{eqnarray}
\chi_f(X) &=& \Big( W_f- (1-t_1)\, (1-t_2)~V_{12,f} -  (1-t_1)\,
(1-t_3)~V_{13,f} 
\nonumber\\ && \qquad - \,(1-t_2)\, (1-t_3)~V_{23,f} - (1-t_1)\,
(1-t_2)\, (1-t_3) ~V_f\Big) 
\nonumber\\ && \otimes\, \Big(  W^*_f -
\frac{(1-t_1)\,(1-t_2)}{t_1\, t_2} 
~ V^*_{12,f} - \frac{(1-t_1)\,(1-t_3)}{t_1 \,t_3}~
 V^*_{13,f} \label{chifX}
\\ &&\qquad-\, \frac{(1-t_2)\,(1-t_3)}{t_2\, t_3}~
 V^*_{23,f} + \frac{(1-t_1)\, (1-t_2)\, (1-t_3)}{t_1 \,t_2\, t_3}~
 V^*_f \Big) ~\frac{1}{(1-t_1)\, (1-t_2)\,(1-t_3)}\nonumber
\end{eqnarray}
where the $\torus^3\times U(1)^N$ modules are given by
\begin{eqnarray}
W_f &=& \sum_{l=1}^N \,e_{l,f} \ , \nonumber\\[4pt] V_f &=&
\sum_{l=1}^N\, 
e_{l,f}~\sum_{(i,j,k)\in\pi_{l,f}}\, t_1^{i-1}\, t_2^{j-1}\, t_3^{k-1}
\ , \nonumber\\[4pt] V_{\alpha\beta,f} &=& \sum_{l=1}^N
\,e_{l,f}~\sum_{(i,j) \in 
\lambda_{\gamma,l,f}}\, t_{\alpha}^{i-1}\, t_{\beta}^{j-1}
\end{eqnarray}
with $(\alpha , \beta , \gamma)$ a cyclic permutation of $(1,2,3)$.

The computation of the ratio of determinants from (\ref{chifX}) is now
just a tedious but completely straightforward algebraic exercise. By
using splittings of modules with properties analogous to those of
Section~\ref{sec:loc}, one easily shows that all contributions
independent of the Chan--Paton space $W_f$ vanish at
$(\epsilon_1,\epsilon_2,\epsilon_3,\mbf a)=(0,0,0,\mbf0)$. After
dropping the perturbative contribution $W_f\otimes W^*_f$, one thus
finds that the only non-vanishing contributions which survive at
$(\epsilon_1,\epsilon_2,\epsilon_3,\mbf a)=(0,0,0,\mbf0)$ are given by
\bea
&& \frac{W_f\otimes  V^*_f}{t_1 \,t_2 \,t_3} - V_f\otimes
 W^*_f - \frac{(1-t_1)\,(1-t_2)\,(1-t_3)}{t_1 \,t_2\, t_3}~
V_f\otimes  V^*_f \label{chifXCY}\\ &&
\qquad\qquad+\,\frac{1}{1-t_3}~\left( -V_{12,f} 
  \otimes W^*_f - \frac{W_f\otimes 
 V^*_{12,f}}{t_1\, t_2} + \frac{(1-t_1)\,(1-t_2)}{t_1\, t_2}~
V_{12,f}\otimes  V^*_{12,f} \right) \nonumber\\ &&
\qquad\qquad +\, \frac{1}{1-t_2}~ \left( 
-V_{13,f}\otimes  W^*_f - \frac{W_f\otimes
   V^*_{13,f}}{t_1\, t_3} + 
\frac{(1-t_1)\,(1-t_3)}{t_1 \,t_3}~ V_{13,f}\otimes
 V^*_{13,f} \right) 
\nonumber\\ && \qquad\qquad+\, \frac{1}{1-t_1}~ \left(
  -V_{23,f}\otimes  W^*_f - \frac{W_f\otimes 
 V^*_{23,f}}{t_2\, t_3} + \frac{(1-t_2)\,(1-t_3)}{t_2\, t_3}~
V_{23,f} \otimes V^*_{23,f} \right) \ . \nonumber
\eea
After applying the transformation (\ref{conversion}), each of the
four lines in (\ref{chifXCY}) gives a minus sign to some power. 

The first line of (\ref{chifXCY}) coincides with the character
(\ref{character}) computed in Section~\ref{sec:loc} and gives a factor
$(-1)^{N \,\sum_{l=1}^N\, |\pi_{l,f}| }$ for each fixed point
$f\in\Delta(X)$. The remaining terms are a nonabelian generalization
of the edge character computed in~\cite{MNOP} (reproduced formally by
setting $W$ to $1$), and can be understood from the point of view of
the matrix quantum mechanics as follows. By T-duality, the D6--D2
system corresponds to a four-dimensional instanton problem over each
rational curve of $X$. These asymptotics are each described by an ADHM
quiver~\cite{nakajima}, with associated vector spaces $W_f$ of
dimension $N$ and $V_{\alpha\beta,f}$ of dimension
$k_{\gamma,f}=|\vec\lambda_{\gamma,f}|$. The full quiver is the
modification of the D6--D0 quiver of Fig.~\ref{quiver} obtained by
inserting the vector spaces $V_{\alpha\beta,f}$ plus all additional
open string fields~\cite{jafferis}. This modifies the complex
(\ref{adhmdefcomplex}) by including terms from the four-dimensional
ADHM deformation complex~\cite{nakajima,Bruzzo:2002xf}. The additional
contributions in (\ref{chifXCY}) then arise from the usual
characters in four-dimensions. They may be computed in the present
case by carefully matching the edge contributions with the partner
terms coming from other vertices of the toric diagram $\Delta(X)$ as
in Section~\ref{InstactionX} above. 

For example, consider the contribution ${\sf L}_{23,f}$ from the last
line of (\ref{chifXCY}), which is oriented in the $t_1$
``direction''. To this term we have to add its partner term coming
from the vertex which shares the same edge $e$, and which has the same
two-dimensional partition structure of $V_{23,f}$ as its asymptotic
behaviour. Using (\ref{transfns}) with $t_1\leftrightarrow t_3$ the
full edge contribution is then
\beq
{\sf E}_{23,e}={\sf L}_{23,f}(t_1,t_2,t_3,e_{l,f})+{\sf
  L}_{23,f}\big(t_1^{-1}\,,\,t_2\,t_1^{m_1}\,,\,t_3\,t_1^{m_2}\,,\,
e_{l,f}\big) \ . 
\label{E23f}\eeq
We use the splitting ${\sf L}_{23,f}={\sf L}_{23,f}^++{\sf
  L}_{23,f}^-$ with
\beq
{\sf L}_{23,f}^+=\frac1{1-t_1}~\left(-V_{23,f}\otimes
   W^*_f+\frac{1-t_2}{t_2}~  
V_{23,f}\otimes V^*_{23,f}\right)
\label{L23fsplit}\eeq
and $\big({\sf L}^+_{23,f}\big)^*\,\big|_{t_1\,t_2\,t_3=1}=-{\sf
  L}_{23,f}^-\big|_{t_1\,t_2\,t_3=1}$. Together with the condition
(\ref{CYmconstr}), after some algebra one finds that the corresponding 
splitting ${\sf E}_{23,e}^+$ in the limit
$(\epsilon_1,\epsilon_2,\epsilon_3,\mbf a)=(0,0,0,\mbf0)$ is given by
\begin{eqnarray}
&& \sum_{l,l'=1}^N~ \sum_{(j,k)\in\lambda_{1,l,f}}\, (j \,m_1 + k
\,m_2 -1) + \sum_{l,l'=1}^N ~\sum_{(j,k)\in\lambda_{1,l,f}}~
\sum_{(j',k'\,)\in\lambda_{1,l',f}}\, m_1 \\  
&& \qquad~=~ N\, \sum_{l=1}^N ~\sum_{(j,k)\in\lambda_{1,l,f}}\, \big(
m_1\, (j-1) + m_2\, 
(k-1) + 1 \big) + \sum_{l,l'=1}^N\, |\lambda_{1,l,f}|\, |
\lambda_{1,l',f}\, |\, m_1 \ . \nonumber
\end{eqnarray}
In this way one finds that the final result for the ratio of
fluctuation determinants can be expressed as $(-1)^{{\sf
    J}(\vec\pi_f)}$, where  
\bea
{\sf J}(\vec\pi_f) &=& \sum_{f\in\Delta(X)}\, N~ \sum_{l=1}^N
\,|\pi_{l,f}| + \sum_{e \in 
\Delta(X)}\, N ~\sum_{l=1}^N~\sum_{(i,j) \in \lambda_{l,e}}\,
\big( m_{1,e}\, (i-1) + m_{2,e}\, (j-1) + 1 \big) \nonumber\\ && +\,
\sum_{e \in \Delta(X)} ~\sum_{l,l'=1}^N\, |\lambda_{l,e}|\,
|\lambda_{l',e}|\,m_{1,e} 
\eea
generalizes the abelian $N=1$ result of~\cite{MNOP,Iqbal:2003ds}.

\subsection{Partition function}

We can now collect all the ingredients and write down the
Donaldson--Thomas partition function on any toric Calabi--Yau manifold 
$X$ in the $U(1)^N$ phase of the six-dimensional topological
Yang--Mills theory on $X$. One finds
\begin{equation}
\mathcal{Z}_{\rm DT}^{U(1)^N}(X) = \sum_{\vec\pi_f} \,(-1)^{{\sf
    J}(\vec\pi_f)} ~\e^{\ii 
\vartheta \,{\sf I}(\vec\pi_f)}~ \e^{- \sum_{e \in \Delta(X)}
~\sum_{l=1}^N\, |\lambda_{l,e}| \,t_e} \ .
\end{equation}
After some rewriting we have
\begin{equation}
\mathcal{Z}_{\rm DT}^{U(1)^N}(X) = \sum_{\vec\pi_f}\, q^{{\sf I}(\vec\pi_f)}~
(-1)^{(N+1)\, {\sf I}(\vec\pi_f)}~ \prod_{e \in \Delta(X)}\,
(-1)^{\sum_{l,l'=1}^N\, 
|\lambda_{l,e}|\, |\lambda_{l',e}|\, m_{1,e}} ~\e^{- \sum_{l=1}^N\,
|\lambda_{l,e}|\, t_e} \ ,
\label{ZDTU1NX}\end{equation}
where the $1$ in the $N+1$ factor arises from the minus sign in the
definition $q= - \e^{\ii \vartheta}$ and ${\sf I}(\vec\pi_f)$ is the
${\rm ch}_3$ contribution to the instanton action given by
(\ref{Ipif}). The $m_1$-dependent signs in (\ref{ZDTU1NX}) are
naturally interpreted as framings of the corresponding edges, as in
the topological vertex formalism.

As a simple example, let us consider the resolved conifold
$X=X_{\rm con}$, the total space of the rank two holomorphic bundle
$\mathcal{O}_{\PP^1}(-1)\oplus \mathcal{O}_{\PP^1}(-1) \rightarrow
\PP^1$ (viewed as the normal bundle to the local Calabi--Yau curve
$\PP^1$). In this case one has
\begin{eqnarray}
\mathcal{Z}_{\rm DT}^{U(1)^N}(X_{\rm con}) &=&  \sum_{\scriptstyle
  {\pi_{1 , f_1} , 
\dots, \pi_{N , f_1}} \atop \scriptstyle {\pi_{1 , f_2} , \dots,
\pi_{N , f_2}} } \,q^{\sum_{l=1}^N\, (|\pi_{l,f_1}|+|\pi_{l,f_2}|)
+ \sum_{l=1}^N ~\sum_{(i,j) \in \lambda_{l}} \,(i+j+1) } \nonumber\\
&& \qquad\qquad\quad \times \,
(-1)^{{(N+1)\, \left( \sum_{l=1}^N \,(|\pi_{l,f_1}|+|\pi_{l,f_2}|) +
\sum_{l=1}^N ~\sum_{(i,j) \in \lambda_l} \,(i+j+1) \right) }}
\nonumber\\ && \qquad\qquad\quad
\times \,(-1)^{\sum_{l,l'=1}^N \,|\lambda_l|\, |\lambda_{l'}| }~
\e^{-\sum_{l=1}^N\, t\, |\lambda_l|} \ .
\end{eqnarray}
This formula gives
\bea
\mathcal{Z}_{\rm DT}^{U(1)^N}(X_{\rm con}) &=& \Big(\, \sum_{\pi_f}\,
\left( (-1)^{N+1}~ q 
\right)^{|\pi_f|+\sum_{(i,j)\in\lambda} \,(i+j+1)}~ (-1)^{|\lambda|}
~\e^{- t\, |\lambda|}\, \Big)^N \nonumber\\[4pt] &=&
\prod_{n=1}^\infty\,\frac{\big(1-(-1)^{N\,n}\,q^n~\e^{-t}\big)^{N\,n}}
{\big(1-(-1)^{N\,n}\,q^n\big)^{N\,n}} \ .
\eea
Up to an alternating sign, this is just the $N$-th power of the
abelian result which coincides with the topological string amplitude
computed in the melting crystal reformulation of the topological
vertex~\cite{Okounkov:2003sp,Iqbal:2003ds}. The alternating sign will
be explained in the next section.

\bigskip

\section{Summary and applications\label{Appl}}

In this final section we will discuss some properties of the $U(1)^N$
partition functions (\ref{ZDTU1NX}). We will also discuss some open
questions and relations to other models. In particular, we summarize
our results and describe how they could be applied to other settings.

\subsection{The OSV conjecture}

Let us consider for simplicity the threefold $X=\complex^3$. The
partition function (\ref{genfnnonabinvs}) can be rewritten in the form
\be \mathcal{Z}_{\rm DT}^{U(1)^N}\big(\complex^3\big) 
= \sum_{\pi_1, \ldots,
 \pi_N  } \,\e^{(N \ii \pi - g_s)\,
\sum_{l=1}^N\, |\pi_l|} = M ({\tilde q})^N \ , \label{ZDTrewrite}\ee
where $M(\tilde q)$ is the MacMahon function (\ref{MacMfn}) of
$\tilde{q}=\e^{-{\tilde{g}_s}} $ and \be {\tilde g_s} = g_s - N
\ii \pi \ . \label{renstring}\ee 
The partition function is just a power of the
abelian $N=1$ result. This is reasonable since the torus fixed points
of the moduli space (\ref{sheavesonP3}) essentially broke up
accordingly (so that no new non-trivial instantons are present). On
the other hand, the Donaldson--Thomas string coupling constant $g_s$
is modified to (\ref{renstring}) as well. As we now explain, this
modification is natural from the point of view of the OSV
conjecture~\cite{Ooguri:2004zv}. 

Consider Type~IIA string theory on $X\times\real^4$, where
D6, D2 and D0 branes wrap holomorphic cycles of the
Calabi--Yau threefold $X$. The mixed ensemble partition function for
BPS bound states with fixed chemical potentials $\phi_2^a$ and
$\phi_0$ for the D2 and D0 brane charges, and magnetic charge $p^0$ of
the D6-branes, is denoted $Z_{\rm BH}(p^0; \phi^a_2, \phi_0)$. It
factorizes in the limit of large charge $p^0$ and small string
coupling $G_s$ as~\cite{Ooguri:2004zv}
\beq
Z_{\rm BH}\big(p^0\,;\, \phi_2^a, \phi_0\big) = \big|Z_{\rm top}(G_s,
t^a)\big|^2 \ , 
\label{OSV}\eeq
where $Z_{\rm top}(G_s, t^a)$ is the A-model topological string
partition function on $X$ evaluated at the attractor
point~\cite{Ferrara:1995ih,Strominger:1996kf} of the moduli space
given by
\beq
G_s =\frac{4\pi\ii}{X^0}=
{4 \pi \ii \over p^0 + \ii {\phi_0 \over \pi}}
\qquad \mbox{and} \qquad t^a=-\,{2 \phi^a_2 \over p^0 + \ii {\phi_0
    \over \pi}} 
\label{attractor}\eeq
with $t^a$ the K\"ahler parameters of the two-cycles wrapped by the
D2-branes. 

If we suppose now that the six-dimensional cohomological gauge theory
on $X$ computes the topological A-model partition function, then the
$U(N)$ generalization means including multiple D6-branes from the
point of view of the OSV formula (\ref{OSV}). However, the Calabi--Yau
crystal description is valid in the strong coupling
regime~\cite{Okounkov:2003sp} and so we expect the relation 
\beq
g_s = \frac1{G_s} \ .
\eeq
Rewriting the OSV formula (\ref{OSV}) in the S-dual string couplings
${\tilde g_s}$ and $g_s$ with and without the D6-branes, one finds
\beq
\tilde{g}_s = g_s - {\ii p^0 \over 4 \pi} \ .
\eeq
Taking into account the relation $p^0 = \frac{N}{4\pi}$,\footnote{The
  $4\pi$ factor is a matter of convention.} we see that the string
coupling in the Donaldson--Thomas description changes in the way
expected from the OSV conjecture.\footnote{The dependence on $N$ does
  not spoil the S-duality conjecture. In fact, the instanton measure
  does not depend on the Higgs vevs. Nevertheless, it would be
  interesting to check this prediction against a computation in some
  Chern--Simons theory.}

According to the generic prediction (\ref{OSV}), the general formula
(\ref{ZDTU1NX}) should factorize in the large $N$ limit. It would be
interesting to further check the OSV factorization for our
$\complex^3$ example. Factorization including D6-branes has only been
checked for the single example $X=K3
\times\torus^2$~\cite{kappeli}. Furthermore, the conformal field
theory derivation of the OSV formula of~\cite{Beasley:2006us} does not
readily generalize for multiple D6-branes. Given that the string
coupling in our case lives in the strong coupling regime, it would be
interesting to see if and how the OSV factorization works here. A
refined version of the OSV formula was found in~\cite{Denef:2007vg}
using the $U(1)$ Donaldson--Thomas theory for a D6 brane-antibrane
pair on compact Calabi--Yau manifolds. It was also shown there that in
certain limits of the background one can identify the abelian
Donaldson--Thomas partition function with the BPS index for stable
D6--D2--D0 bound states with unit D6-brane charge.

\subsection{Enumerative invariants}

Let us study the small $q$ expansion of the partition function
(\ref{ZDTrewrite}), with the aim of understanding its role in
enumerative geometry. For example, for $U(2)$ gauge group one gets 
\begin{eqnarray}
\mathcal{Z}_{\rm DT}^{U(1)^2}\big(\complex^3\big) &=& \sum_{\pi_1 ,
  \pi_2}\, (-1)^{3 
(|\pi_1|+|\pi_2|)}~ q^{|\pi_1|+|\pi_2|} \nonumber\\[4pt] &=&
\sum_{\pi_1 , 
\pi_2}\,(-1)^{|\pi_1|+|\pi_2|} ~q^{|\pi_1|+|\pi_2|} \nonumber\\[4pt]
&=&\Big(\, \sum_{\pi}\, (- q)^{|\pi|}\, \Big)^2 \\[4pt] &=&
\Big(~\prod_{n=1}^{\infty}\, \big( 1 - (-q)^n \big)^{-n}\, \Big)^2
\nonumber\\[4pt] 
&=&  1-2 q+7 q^2-18 q^3+47 q^4-110 q^5+258 q^6-568 q^7+1237
q^8+O\left(q^9\right) \ . \nonumber
\end{eqnarray}
Similarly, for the $U(N)$ gauge theory one finds
\begin{eqnarray}
\mathcal{Z}_{\rm DT}^{U(1)^N}\big(\complex^3\big) &=& \sum_{\pi_1 ,
  \dots , \pi_N}\, 
(-1)^{(N+1)\, (|\pi_1|+ \cdots + |\pi_N|)}~ q^{|\pi_1| + \cdots +
|\pi_N|} \nonumber\\[4pt] &=& \Big(\, \sum_{\pi}\, (-1)^{(N+1)\, |\pi| 
}~ q^{|\pi|}\, 
\Big)^N \nonumber\\[4pt] &=& \Big(~ \prod_{n=1}^{\infty}\, \big( 1 - 
((-1)^{N+1}\, q)^n \big)^{-n}\, \Big)^N \label{smallqUN}\\[4pt] &=&
1-(-1)^N\, N~ 
q+\mbox{$\left(\frac{1}{2}\, (-1)^{2 N}\, N^2+\frac{5}{2}\, (-1)^{2
      N}\, N\right) 
~q^2$} \nonumber\\ && +\,\mbox{$ \left(-\frac{1}{6}\, (-1)^{3 N}\,
  N^3-\frac{5}{2}\, (-1)^{3 
N}\, N^2-\frac{10}{3}\, (-1)^{3 N}\, N\right)~ q^3$} \nonumber\\ &&
+\,\mbox{$\left(\frac{1}{24}\, (-1)^{4 N}\, N^4+\frac{5}{4}\, (-1)^{4
      N}\, 
N^3+\frac{155}{24}\, (-1)^{4 N}\, N^2+\frac{21}{4}\, (-1)^{4 N}\,
N\right)$}~q^4+O\left(q^5\right) \ . \nonumber
\end{eqnarray}
The numerical invariants computed by the small $q$-expansion
(\ref{smallqUN}) are all integer-valued.

As the nonabelian $U(1)^N$ partition function on $\complex^3$ is given
by the $N$-th power of the MacMahon function, up to sign factors, the
numerical values of the Donaldson--Thomas invariants are the
same. But the continuation of the MacMahon function in
(\ref{ZDTrewrite}) to complex values of the string coupling suggests a 
non-trivial interpretation in terms of Gromov--Witten
theory. By using the correspondence between the Donaldson--Thomas and
Gromov--Witten partition functions~\cite{MNOP}, one sees that the
Gromov--Witten invariants change non-trivially. This change can be
encoded in the topological string
coupling constant associated with the D6-brane charge. In terms of the
usual Gromov--Witten theory, this may arise through geometric
transition via a connection with Chern--Simons gauge theory with
complexified coupling constant, as arises for complex gauge groups.
It would be very interesting to understand the meaning of the new
parameter $N$ in terms of the underlying closed topological string
theory.

However, we should reiterate that our computations are only valid in
the Coulomb phase, where the gauge group is completely broken to its
maximal torus $U(1)^N$. Furthermore, we considered only the target
space $X=\complex^3$ properly, wherein the instanton moduli space
could be characterized by a stable framed representation of the quiver
with relations of Fig.~\ref{quiver} in the category of complex vector
spaces. The precise characterization of the full moduli space and the
construction of the full nonabelian Donaldson--Thomas theory is a much
more difficult task. For a generic threefold $X$ it should involve a
stable twisted representation of the quiver with relations depicted in
Fig.~\ref{quiver} (including ADHM quivers over the rational curves) in
the abelian category of coherent sheaves of
$\mathcal{O}_X$-modules~\cite{King}, with \emph{non-trivial}
framing. A recent proposal for such a construction in the case that
$X$ is a local curve can be found in~\cite{Diaconescu}. Understanding
of the moduli space of the full nonabelian Donaldson--Thomas theory,
exploring its combinatorial nature in terms of random partitions, and
searching for new geometric invariants are all interesting and challenging
tasks, which are however beyond the scope of the present work. In this
paper we have only probed a small corner of the full moduli space, and
developed some potentially useful starting techniques in this
direction.

\subsection{Relations with other models}

It is instructive to compare our results with those
obtained on other backgrounds, some of which can be related to ours
through duality transformations. It was argued
in~\cite{Sduality,Kapustin:2004jm} that the fact that perturbative
A-model topological string amplitudes capture non-perturbative
information about D-brane bound states can be understood as a
consequence of S-duality, when embedding the topological string theory
into the physical Type~IIB superstring theory. This S-duality was used
in~\cite{Dijkgraaf:2006um} to relate the abelian Donaldson--Thomas
invariants on a Type~IIA compactification with the Gopakumar--Vafa BPS
invariants of M2-branes in M-theory, establishing another point of
view on the relationship between four and five dimensional black holes
of~\cite{Gaiotto:2005gf}. One can start with a Type~IIA
compactification on a ten-dimensional background given by the product
of a Calabi--Yau threefold and the four-dimensional space $\real^3
\times\sphere^1$. Then the charge one D6-brane background can be
transformed into a charge one Taub--NUT geometry by applying a $T \,S
\,T$ duality transformation, where the D2 and D0 branes become
respectively fundamental strings and Kaluza--Klein momentum
modes along the $\sphere^1$ of the Taub--NUT geometry. More precisely,
after a T-duality transformation along the $\sphere^1$ of the space
$\real^3 \times\sphere^1$, the Type~IIA D6-brane gets
mapped into the Type~IIB D5-brane. S-duality of the
physical Type~IIB string theory transforms this D5-brane into an
NS5-brane, and finally after the last T-duality transformation one
arrives at the Taub--NUT geometry. This configuration can then be
lifted to M-theory. See~\cite{Dijkgraaf:2006um} for further details.

It is reasonable to expect that this set of dualities continues to
hold for the nonabelian D6-brane configuration, though more work would
need to be done at the supergravity level to establish this. However,
if this is the case then it is interesting to speculate that our
computation predicts a relation between topological A-model amplitudes
on the Taub--NUT geometry and the partition function of $N$ Type~IIB
NS5-branes, as a consequence of S-duality. It was 
pointed out in~\cite{Kapustin:2004jm} that a ``mirror'' version of
this duality is already implied in~\cite{Dijkgraaf:2002ac}, where it
was argued that B-model amplitudes compute the partition function of
$N$ Type~IIA NS5-branes. In the setup
of~\cite{Dijkgraaf:2002ac} one starts with $N$ NS5-branes on a
Calabi--Yau threefold $X$ and relates this configuration through
T-duality with Type~IIB string theory on $X \times
\mathbb{M}^4$, where $\mathbb{M}^4$ is a Taub--NUT space (or better
an ALE space of type $A_{N-1}$ when we let the size of the
compactified circle grow to infinity). In concrete computations
they take $X = K3 \times \torus^2$ and its $\zed_2$ orbifold.

By a supergravity analysis and careful matching of F-terms on
both sides, one can compute the nonperturbative contribution to
the partition function of $N$ {\it well-separated} NS5-branes, which
turns out to be equal to the $(N-1)$-th power of the B-model
topological string amplitude.\footnote{The ``missing brane''
problem seems to be related to subtleties in the supergravity
analysis.} Reversing the statement, one may say that perturbative
topological string amplitudes only compute the partition function of
NS5-branes in the Coulomb branch. Thus even though it is not
completely clear how to explicitly translate this set of dualities to
our setup, our results are compatible with the findings
of~\cite{Dijkgraaf:2002ac}.

A similar picture arises
in~\cite{Gunaydin:2006bz} where a one-parameter generalization of
topological string theory was proposed. Although it seems natural
following the reasonings of~\cite{Gunaydin:2006bz} to identify this
parameter with the rank $N$ of the nonabelian Donaldson--Thomas
theory, it is not clear to us how to make precise contact with their
proposal. Other multiparameter extensions of topological string theory
are found in~\cite{Iqbal:2007ii} and~\cite{Saraikin:2007jc}. It would
also be interesting to compare our results with the $U(N)$ extension
of the construction of~\cite{Bonelli:2007hb} which considered the
equivariant reduction to lower dimension of the six-dimensional gauge
theory on a local surface.

\bigskip

\section*{Acknowledgments}

We thank G.~Bonelli, A.~Maciocia, N.~Nekrasov and A.~Tanzini for helpful
discussions and correspondence. This work was supported in part by the
Marie Curie Research Training Network Grants {\sl Constituents,
  Fundamental Forces and Symmetries of the Universe}
(Project~MRTN-CT-2004-005104) and {\sl Superstring Theory}
(Project~MRTN-CT-2004-512194) from the European Community's Sixth
Framework Programme, and by the INTAS Grant {\sl Strings, Branes and
  Higher Spin Fields} (Contract~03-51-6346). Part of this work was
carried out while R.J.S. was visiting the Isaac Newton Institute for
Mathematical Sciences under the auspices of the program {\sl Strong Fields,
  Integrability and Strings}, and the Erwin Schr\"odinger
International Institute for Mathematical Physics during the program
{\sl Applications of the Renormalization Group}. M.C. is thankful to
SISSA, Trieste and INFN Sezione di Trieste, the Department of Applied
Mathematics and Theoretical Physics, University of Cambridge and the
Department of Mathematics, Heriot--Watt University, Edinburgh
for warm hospitality and financial support during his visits over the
last year. M.C. acknowledges interesting discussions with the
participants of the workshop ``Black Holes and Topological
Strings'' in SISSA, Trieste 20--22 November 2006 and of the ``Fourth
Regional Meeting in String Theory'' in Patras 10--17 June 2007, and is
thankful to the organizers for providing a stimulating atmosphere.

\end{document}